\documentclass[letterpaper,11pt]{article}
\pdfoutput=1
\pdfoutput=1
\pdfoutput=1

\usepackage{jheppub}
\usepackage{multirow}

\usepackage{subfig}
\usepackage{xspace}
\usepackage[countmax]{subfloat}

\usepackage{amssymb}
\usepackage{amsmath}
\usepackage{cancel}
\usepackage{tabu,booktabs}
\usepackage{color}
\usepackage{braket}
\usepackage{graphicx}
\usepackage{multirow}
\usepackage{verbatim}
\usepackage{amsthm}
\usepackage{slashed}
\usepackage{wasysym}
\usepackage{simplewick}
\usepackage{mathtools}
\usepackage{soul}
\usepackage{xspace}

\definecolor{dancomment}{RGB}{0,159,0}

\def\cB{\mathcal{B}}
 
\def\cD{\mathcal{D}}

\def\cL{\mathcal{L}}
\def\cM{\mathcal{M}}

\def\cO{\mathcal{O}}

\def\cP{\mathcal{P}}

\def\cY{\mathcal{Y}}

\def\tr{{\rm tr}}

\def\nn{{\nonumber}}

\newcommand{\hard}{\mathrm{hard}}
\newcommand{\dyn}{\mathrm{dyn}}

\newcommand{\BPS}{\mathrm{BPS}}

\newcommand{\Eq}[1]{Equation~\eqref{#1}}

\DeclareRobustCommand{\Sec}[1]{Sec.~\ref{#1}}

\DeclareRobustCommand{\App}[1]{App.~\ref{#1}}
\DeclareRobustCommand{\Tab}[1]{Table~\ref{#1}}

\DeclareRobustCommand{\Eq}[1]{Eq.~(\ref{#1})}
\DeclareRobustCommand{\Eqs}[2]{Eqs.~(\ref{#1}) and (\ref{#2})}
\DeclareRobustCommand{\Ref}[1]{Ref.~\cite{#1}}
\DeclareRobustCommand{\Refs}[1]{Refs.~\cite{#1}}

\def\be{\begin{equation}}
\def\ee{\end{equation}}

\newcommand{\nbar}{{\bar n}}

\def\l{\langle}
\def\r{\rangle}

\def\bt{\beta}

\newcommand{\lotsdots}{{
+\cdot\cdot:\cdot\cdot 
(\cdot\cdot:\cdot\cdot\ldots\cdot\cdot:\cdot\cdot)
[\cdot\cdot:\cdot \cdot-]
}}

\newcommand{\Sl}[1]{\slashed{#1}}

\renewcommand{\arraystretch}{1.05}
\allowdisplaybreaks[3]

\newcommand{\eq}[1]{Eq.~\eqref{eq:#1}}
\newcommand{\eqs}[2]{Eqs.~\eqref{eq:#1} and \eqref{eq:#2}}

\newcommand{\app}[1]{App.~\ref{app:#1}}

\newcommand{\ord}[1]{\mathcal{O}(#1)}

\newcommand{\mae}[3]{\langle#1\rvert#2\rvert#3\rangle}

\newcommand{\df}{\mathrm{d}}

\newcommand{\sdt}{\!\cdot\!}

\newcommand{\al}{\alpha}

\newcommand\bn{{\bar n}}
\newcommand{\ga}{\gamma}

\newcommand{\de}{\delta}

\newcommand{\ve}{\varepsilon}
\newcommand{\la}{\lambda}

\newcommand{\w}{\omega}
\newcommand{\balpha}{{\bar \alpha}}
\newcommand{\bbeta}{{\bar \beta}}
\newcommand{\bgamma}{{\bar \gamma}}
\newcommand{\bdelta}{{\bar \delta}}

\newcommand{\fd}[2]{\parbox{#1}{\includegraphics[width=#1]{#2}}}

\newcommand{\hH}{\widehat{H}}
\newcommand{\hS}{\widehat{S}}

\newcommand{\vT}{\bar{T}}

\newcommand{\vC}{\vec{C}}
\newcommand{\vO}{\vec O}

\newcommand{\lp}{\tilde p}        

\newcommand{\bnP}{\overline {\mathcal P}}

  \newcount\hour \newcount\minute
  \hour=\time \divide \hour by 60 \minute=\time
  \newcommand{\todaytime}{\today \ -- \number\hour :\ifnum \minute<10 0\fi\number\minute}

\preprint{MIT--CTP 4837}

\title{A Subleading Operator Basis and Matching for $gg\to H$}

\author[1,2]{Ian Moult,}
\author[3]{Iain W. Stewart,}
\author[3]{and Gherardo Vita}

\affiliation[1]{Berkeley Center for Theoretical Physics, University of California, Berkeley, CA 94720, USA}
\affiliation[2]{Theoretical Physics Group, Lawrence Berkeley National Laboratory, Berkeley, CA 94720, USA}
\affiliation[3]{Center for Theoretical Physics, Massachusetts Institute of Technology, Cambridge, MA 02139, USA}

\emailAdd{ianmoult@lbl.gov}
\emailAdd{iains@mit.edu}
\emailAdd{vita@mit.edu}

\abstract{The Soft Collinear Effective Theory (SCET) is a powerful framework for studying factorization of amplitudes and cross sections in QCD. While factorization at leading power has been well studied, much less is known at subleading powers in the $\lambda\ll 1$ expansion. In SCET subleading soft and collinear corrections to a hard scattering process are described by power suppressed operators, which must be fixed case by case, and by well established power suppressed Lagrangians, which correct the leading power dynamics of soft and collinear radiation. Here we present a complete basis of power suppressed operators for $gg \to H$, classifying all operators which contribute to the cross section at $\mathcal{O}(\lambda^2)$, and showing how helicity selection rules significantly simplify the construction of the operator basis. We perform matching calculations to determine the tree level Wilson coefficients of our operators. These results are useful for studies of power corrections in both resummed and fixed order perturbation theory, and for understanding the factorization properties of gauge theory amplitudes and cross sections at subleading power.  As one example, our basis of operators can be used to analytically compute power corrections for $N$-jettiness subtractions for $gg$ induced color singlet production at the LHC.
}


\begin{document} 

\maketitle

\section{Introduction}\label{sec:intro}

Factorization theorems play an important role in understanding the all orders behavior of observables in Quantum Chromodynamics (QCD). While typically formulated at leading power, the structure of subleading power corrections is of significant theoretical and practical interest. A convenient formalism for studying factorization in QCD is the Soft Collinear Effective Theory (SCET) \cite{Bauer:2000ew, Bauer:2000yr, Bauer:2001ct, Bauer:2001yt}, an effective field theory describing the soft and collinear limits of QCD. SCET allows for a systematic power expansion in $\lambda \ll 1$ at the level of the Lagrangian, and simplifies many aspects of factorization proofs \cite{Bauer:2002nz}. SCET has been used to study power corrections at the level of the amplitude \cite{Larkoski:2014bxa} and to derive factorization theorems at subleading power for $B$ decays \cite{Lee:2004ja,Beneke:2004in,Hill:2004if,Bosch:2004cb,Beneke:2004rc,Paz:2009ut,Benzke:2010js}. More recently, progress has been made towards understanding subleading power corrections for event shape observables \cite{Freedman:2013vya,Freedman:2014uta,Moult:2016fqy,Feige:2017zci}.

In this paper, we focus on the power suppressed hard scattering operators describing the gluon initiated production (or decay) of a color singlet scalar. We present a complete operator basis to $\cO(\lambda^2)$ in the SCET power expansion using operators of definite helicity \cite{Moult:2015aoa,Kolodrubetz:2016uim,Feige:2017zci}, and discuss how helicity selection rules simplify the structure of the basis. We also classify all operators which can contribute at the cross section level at $\cO(\lambda^2)$, and discuss the structure of interference terms between different operators in the squared matrix element. We then perform the tree level matching onto our operators. These results can be used to study subleading power corrections either in fixed order, or resummed perturbation theory, and compliment our recent analysis for the case of $q\bar q$ initiated production \cite{Feige:2017zci}.

We will consider the production of a color singlet final state, which we take for concreteness to be the Higgs, with the underlying hard Born process
\begin{equation} \label{eq:interaction}
g_a(q_a)\, g_b(q_b) \to H(q_1)
\,,\end{equation}
where $g_{a,b}$ denote the colliding gluons, and $H$ the outgoing Higgs particle. We work in the Higgs effective theory, with an effective Higgs gluon coupling
\begin{align}\label{eq:heft_intro}
\cL_{\text{hard}}=\frac{C_1(m_t, \alpha_s)}{12\pi v}G^{\mu \nu}G_{\mu \nu} H\,,
\end{align}
obtained from integrating out the top quark. Here $v=(\sqrt{2} G_F)^{-1/2}=246$ GeV, and the matching coefficient is known to $\cO(\alpha_s^3)$ \cite{Chetyrkin:1997un}.

The active-parton exclusive jet cross section corresponding to \eq{interaction} can be proven to factorize for a variety of jet resolution variables. For concreteness we will take the case of beam thrust, $\tau_B$. The leading power factorized expression for the beam thrust cross section can be written schematically in the form \cite{Stewart:2010tn}
\begin{align} \label{eq:sigma}
\frac{\df\sigma^{(0)}}{\df \tau_B} &=
\int\!\df x_a\, \df x_b\, \df \Phi(q_a \!+ q_b; q_1)\, M(\{q_1\})\
 H_g^{(0)}(\{q_i\})\: 
\Bigl[ B_g^{(0)} B_g^{(0)}  \Bigr]\otimes S_g^{(0)}  
\,,\end{align}
where the $x_{a,b}$ denote the momentum fractions of the incoming partons, $\df \Phi$ denotes the Lorentz-invariant phase space for the Born process in \eq{interaction}, and $M(\{q_i\})$ denotes the measurement made on the color singlet final state.\footnote{By referring to active-parton factorization we imply that this formula ignores contributions from proton spectator interactions~\cite{Gaunt:2014ska} that occur through the Glauber Lagrangian of Ref.~\cite{Rothstein:2016bsq}. There are also perturbative corrections at ${\cal O}(\alpha_s^4)$ that are described by a single function $B_{gg}$ in place of $B_gB_g$~\cite{Zeng:2015iba,Rothstein:2016bsq}.} The dependence on the underlying hard interaction is encoded in the hard function $\hH(\{q_i\})$ and the trace is over color.  The soft function $\hS$ describes soft radiation, and the beam functions $B_i$ describe energetic initial-state radiation along the beam directions \cite{Stewart:2009yx}. The factorization theorem of \Eq{eq:sigma} allows logarithms of $\tau_B$ to be resummed to all orders through the renormalization group evolution of the hard, beam and soft functions.

The factorization formula in \Eq{eq:sigma} captures all terms in the cross section scaling as $\tau_B^{-1}$, including delta function terms. More generally the cross section can be expanded in powers of $\tau_B$ as,
\begin{align}\label{eq:cross_expand}
\frac{\df\sigma}{\df\tau_B} &=\frac{\df\sigma^{(0)}}{\df\tau_B} +\frac{\df\sigma^{(1)}}{\df\tau_B} +\frac{\df\sigma^{(2)}}{\df\tau_B}+\frac{\df\sigma^{(3)}}{\df\tau_B} +{\cal O}(\tau)\,.
\end{align}
Here the superscript refers to the suppression in powers of $\sqrt{\tau_B}$ relative to the leading power cross section. This particular convention is chosen due to the power expansion in SCET, where one typically takes the SCET power counting parameter $\lambda$ to scale like $\lambda^2 \sim \tau_B$. Odd orders in \Eq{eq:cross_expand} are expected to vanish, and we will show this explicitly for $\df\sigma^{(1)}/\df\tau_B$. The first non-vanishing power correction to the cross section then arises from $\df\sigma^{(2)}/\df\tau_B$, which contains all terms that scale like $\cO(\tau_B^0)$.  

It is generally expected that the power corrections in \Eq{eq:cross_expand} obey a factorization formula similar to that of \Eq{eq:sigma}. Schematically, 
\begin{align} \label{eq:sigma_sub}
&\hspace{-0.25cm}\frac{\df\sigma^{(n)}}{\df\tau_B} =
\int\!\df x_a\, \df x_b\, \df \Phi(q_a \!+ q_b; q_1)\,M(\{q_1\})\
\sum_{j}   H^{(n_{Hj})}_{j} \otimes 
  \Big[ B^{(n_{Bj})}_{j}  B^{(n'_{Bj})}_{j}\Big] \otimes S_j^{(n_{Sj})}  
,\end{align}
where $j$ sums over the multiple contributions that appear at each order, $n_{Hj}+n_{Bj}+n_{Bj}'+n_{Sj}=n$, and $\otimes$ denotes a set of convolutions, whose detailed structure has not been specified and is known to be more complicated than typical leading power factorization theorems. We also let $\otimes$ include nontrivial color contractions. The derivation of such a formula would enable for the resummation of subleading power logarithms using the renormalization group evolution of the different functions appearing in \Eq{eq:sigma_sub}, allowing for an all orders understanding of power corrections to the soft and collinear limits.

To derive a factorization theorem in SCET, QCD is matched onto SCET, which consists of hard scattering operators in $\cL_\hard$ and a Lagrangian $\cL_\dyn$ describing the dynamics of soft and collinear radiation
\begin{align} 
\cL_{\text{SCET}}=\cL_\hard+\cL_\dyn \,.
\end{align}
The dynamical Lagrangian can be divided into two parts
\begin{align}
\cL_\dyn=\cL_{\text{fact}}+\cL_{G}^{(0)} \,.
\end{align}
Here $\cL_{G}^{(0)}$ is the leading power Glauber Lagrangian determined in Ref.~\cite{Rothstein:2016bsq} which couples together soft and collinear fields in an apriori non-factorizable manner, and $\cL_{\text{fact}}$ includes both the leading interactions which can be factorized into independent soft and collinear Lagrangians, and subleading power interactions which are factorizable as products of soft and collinear fields. Our focus here is on determining the subleading power $\cL_\hard$ for $gg\to H$, and $\cL_\dyn$ only plays a minor role when we carry out explicit matching calculations (and $\cL_{G}^{(0)}$ does not play a role at all since these matching calculations are tree level). 

The hard scattering operators are process dependent, while the Lagrangian $\cL_\dyn$ is universal and the relevant terms for our analysis are known in SCET to $\cO(\lambda^2)$ in the power expansion \cite{Manohar:2002fd,Chay:2002vy,Beneke:2002ni,Beneke:2002ph,Pirjol:2002km,Bauer:2003mga}. A field redefinition can be performed in the effective theory \cite{Bauer:2002nz} which allows for the decoupling of leading power soft and collinear interactions in $\cL_{\text{fact}}$. If $\cL_{G}^{(0)}$ is proven to be irrelevant, then the Hilbert spaces for the soft and collinear dynamics are factorized, and a series of algebraic manipulations can be used to write the cross section as a product of squared matrix elements, each involving only collinear or soft fields. This provides a field theoretic definition of each of the functions appearing in \Eq{eq:sigma_sub} in terms of hard scattering operators and Lagrangian insertions in SCET.  Since the Lagrangian insertions are universal, the remaining ingredient which is required to derive a subleading power factorization theorem for the $gg\to H$ process is a complete basis of subleading power hard scattering operators. The derivation of a basis, which is the goal of this paper, provides the groundwork for a systematic study of power corrections for color singlet production through gluon fusion. 

An important application of the results presented in this paper is to the calculation of subleading power corrections to event shape observables for $gg\to H$, such as $0$-jettiness \cite{Stewart:2010tn}. Recently, there has been considerable interest in the use of event shape observables for performing NNLO fixed order subtractions using the $q_T$ \cite{Catani:2007vq} or $N$-jettiness \cite{Boughezal:2015aha,Gaunt:2015pea} subtraction schemes. These ideas have been applied to color singlet production \cite{Catani:2009sm,Ferrera:2011bk,Catani:2011qz,Grazzini:2013bna,Cascioli:2014yka,Ferrera:2014lca,Gehrmann:2014fva,Grazzini:2015nwa,Grazzini:2015hta,Campbell:2016yrh,Boughezal:2016wmq}, to the production of a single jet in association with a color singlet particle \cite{Boughezal:2015aha,Boughezal:2015dva,Boughezal:2016isb,Boughezal:2016dtm}, and to inclusive photon production \cite{Campbell:2016lzl}.  By analytically computing the power corrections for the subtractions, their stability and numerical accuracy can be significantly improved. This was shown explicitly in \cite{Moult:2016fqy} with the SCET based analytic calculation of the leading power corrections for $0$-jettiness  for $q\bar q$ initiated Drell Yan like production of a color singlet, and it would be interesting to extend this calculation to $gg\to H$. For a direct calculation of the power corrections in QCD, see \cite{Boughezal:2016zws}.

An outline of this paper is as follows. In \Sec{sec:review} we provide a brief review of SCET and of the helicity building blocks required for constructing subleading operators in SCET. In \Sec{sec:basis} we present a complete basis of operators to $\cO(\lambda^2)$ for the gluon initiated production of a color singlet, and carefully classify which operators can contribute to the cross section at $\cO(\lambda^2)$. In \Sec{sec:matching} we perform the tree level matching to the relevant operators. We conclude and discuss directions for future study in \Sec{sec:conclusions}.

\section{Helicity Operators in SCET}\label{sec:review}

In this section we briefly review salient features of SCET, as well as the use of helicity operators in SCET. Reviews of SCET can be found in \Refs{iain_notes,Becher:2014oda}, and more detailed discussions on the use of helicity operators can be found in \Refs{Moult:2015aoa,Kolodrubetz:2016uim,Feige:2017zci}.

\subsection{SCET}\label{sec:review_scet}

SCET is an effective field theory of QCD describing the interactions of collinear and soft particles in the presence of a hard interaction \cite{Bauer:2000ew, Bauer:2000yr, Bauer:2001ct, Bauer:2001yt, Bauer:2002nz}. Collinear particles  are characterized by a large momentum along a particular light-like direction, while soft particles are characterized by having a small momentum with homogenous scaling of all its components. For each jet direction present in the problem we define two light-like reference vectors $n_i^\mu$ and $\bn_i^\mu$ such that $n_i^2 = \bn_i^2 = 0$ and $n_i\sdt\bn_i = 2$. We can then write any four-momentum $p$ as
\begin{equation} \label{eq:lightcone_dec}
p^\mu = \bn_i\sdt p\,\frac{n_i^\mu}{2} + n_i\sdt p\,\frac{\bn_i^\mu}{2} + p^\mu_{n_i\perp}\
\,.\end{equation}
A particle with momentum $p$ close to the $\vec{n}_i$ direction will be referred to as $n_i$-collinear. In lightcone coordinates its momenta scale like $(n_i\!\cdot\! p, \bn_i \!\cdot\! p, p_{n_i\perp}) \sim \bn_i\!\cdot\! p$ $\,(\la^2,1,\la)$. Here $\la \ll 1$ is a formal power counting parameter determined by the measurements or kinematic restrictions imposed on the QCD radiation. The choice of reference vectors is not unique, and any two reference vectors, $n_i$ and $n_i'$, with $n_i\cdot n_i' \sim \ord{\lambda^2}$ describe the same physics. The freedom in the choice of $n_i$ is represented in the effective theory as a symmetry known as reparametrization invariance (RPI) \cite{Manohar:2002fd,Chay:2002vy}. More explicitly, there are three classes of RPI transformations under which the EFT is invariant
\begin{alignat}{3}\label{eq:RPI_def}
&\text{RPI-I} &\qquad &  \text{RPI-II}   &\qquad &  \text{RPI-III} \nn \\
&n_{i \mu} \to n_{i \mu} +\Delta_\mu^\perp &\qquad &  n_{i \mu} \to n_{i \mu}   &\qquad & n_{i \mu} \to e^\alpha n_{i \mu} \nn \\
&\bar n_{i \mu} \to \bar n_{i \mu}  &\qquad &  \bar n_{i \mu} \to \bar n_{i \mu} +\epsilon_\mu^\perp  &\qquad & \bar n_{i \mu} \to e^{-\alpha} \bar n_{i \mu}\,.
\end{alignat}
The transformation parameters are assigned the power counting $\Delta^\perp \sim \lambda$, $\epsilon^\perp \sim \lambda^0$, and $\alpha\sim \lambda^0$. Additionally, while $\alpha$ can be a finite parameter, the parameters $\Delta^\perp$ and $\epsilon^\perp$ are infinitesimal, and satisfy $n_i\cdot \Delta^\perp=\bar n_i\cdot \Delta^\perp=n_i \cdot \epsilon^\perp=\bar n_i \cdot \epsilon^\perp=0$. RPI symmetries can be used to relate operators at different orders in the power expansion, and will be used in this paper to relate the Wilson coefficients of several subleading power operators to the leading power Wilson coefficients for the $gg\to H$ process. Furthermore, the RPI-III symmetry will constrain the form of the Wilson coefficients of our subleading power operators. At tree level the Wilson coefficients are simply rational functions of the large momentum components of the fields appearing in the operator, which must satisfy the rescaling symmetries of RPI-III.

SCET is constructed by decomposing momenta into label and residual components
\begin{equation} \label{eq:label_dec}
p^\mu = \lp^\mu + k^\mu = \bn_i \sdt\lp\, \frac{n_i^\mu}{2} + \lp_{n_i\perp}^\mu + k^\mu\,.
\,\end{equation}
The momenta $\bn_i \cdot\lp \sim Q$ and $\lp_{n_i\perp} \sim \la Q$ are referred to as the label components, where $Q$ is a typical scale of the hard interaction, while $k\sim \la^2 Q$ is a small residual momentum describing fluctuations about the label momentum.  Fields with momenta of definite scaling are obtained by performing a multipole expansion. Explicitly, the effective theory consists of collinear quark and gluon fields for each collinear direction, as well as soft quark and gluon fields. Independent gauge symmetries are enforced for each set of fields, which have support for the corresponding momenta carried by that field \cite{Bauer:2003mga}. The leading power gauge symmetry is exact, and is not corrected at subleading powers.

In SCET, fields for $n_i$-collinear quarks and gluons, $\xi_{n_i,\lp}(x)$ and $A_{n_i,\lp}(x)$, are labeled by their collinear direction $n_i$ and their large momentum $\lp$. The collinear fields are written in a mixed representation, namely they are written in position space with respect to the residual momentum and in momentum space with respect to the large momentum components. Derivatives acting on collinear fields give the residual momentum dependence, which scales as $i \partial^\mu \sim k \sim \la^2 Q$, whereas the label momentum operator $\cP^\mu$ gives the label momentum component. It acts on a collinear field as $\cP^\mu\, \xi_{n_i,\lp} = \lp^\mu\, \xi_{n_i,\lp}$. Note that we do not need an explicit $n_i$ label on the label momentum operator, since it is implied by the field that the label momentum operator is acting on. We will use the shorthand notation $\bnP = \bn_i\sdt\cP$.  We will often suppress the explicit momentum labels on the collinear fields, keeping only the label of the collinear sector, ${n_i}$. Of particular relevance for the construction of subleading power operators is the $\cP_\perp$ operator, which identifies the $\cO(\lambda)$ perp momenta between two collinear fields within a collinear sector.

Soft degrees of freedom are described in SCET by quark and gluon fields $q_{us}(x)$ and $A_{us}(x)$. In this paper we will restrict ourselves to the SCET$_\text{I}$ theory where the soft degrees of freedom are referred to as ultrasoft so as to distinguish them from the soft modes of SCET$_\text{II}$ \cite{Bauer:2002aj}. The operators we construct are also applicable in the SCET$_\text{II}$ theory, but additional soft operators would be required. For a more detailed discussion see \Ref{Feige:2017zci}. The ultrasoft fields carry residual momenta, $i \partial^\mu \sim \la^2Q$, but do not carry label momenta, since they are not associated with any collinear direction. Correspondingly, they also do not carry a collinear sector label. The ultrasoft fields are able to exchange residual momenta between distinct collinear sectors while remaining on-shell.

SCET is constructed such that manifest power counting in the expansion parameter $\la$ is maintained at every stage of a calculation. All fields have a definite power counting \cite{Bauer:2001ct}, shown in \Tab{tab:PC}, and the SCET Lagrangian is expanded as a power series in $\lambda$
\begin{align} \label{eq:SCETLagExpand}
\cL_{\text{SCET}}=\cL_\hard+\cL_\dyn= \sum_{i\geq0} \cL_\hard^{(i)}+ 
 {\cal L}_G^{(0)} + \sum_{i\geq0} \cL^{(i)} \,.
\end{align}
Here $(i)$ denotes objects at ${\cal O}(\lambda^i)$ in the power counting. The Lagrangians $ \cL_\hard^{(i)}$ contain the hard scattering operators $O^{(i)}$, and are determined by an explicit matching calculation. The hard scattering operators encode all process dependence, while the $\cL^{(i)}$ describe the dynamics of ultrasoft and collinear modes in the effective theory, and are universal. The terms we need are explicitly known to $\mathcal{O}(\lambda^2)$, and can be found in a summarized form in \cite{iain_notes}. Finally, ${\cal L}_G^{(0)} $ is the leading power Glauber Lagrangian \cite{Rothstein:2016bsq}, which describes the leading power coupling of soft and collinear degrees of freedom through potential operators.

\begin{table}
\begin{center}
\begin{tabular}{| l | c | c |c |c|c| r| }
  \hline                       
  Operator & $\cB_{n_i\perp}^\mu$ & $\chi_{n_i}$& $\cP_\perp^\mu$&$q_{us}$&$D_{us}^\mu$ \\
  Power Counting & $\lambda$ &  $\lambda$& $\lambda$& $\lambda^3$& $\lambda^2$ \\
  \hline  
\end{tabular}
\end{center}
\caption{
Power counting for building block operators in $\text{SCET}_\text{I}$.
}
\label{tab:PC}
\end{table}

In this paper we will be interested in subleading power hard scattering operators, in particular, $\cL_\hard^{(1)}$ and $\cL_\hard^{(2)}$.  The hard effective Lagrangian at each power is given by a product of hard scattering operators and Wilson coefficients,
\begin{align} \label{eq:Leff_sub_explicit}
\cL^{(j)}_{\text{hard}} = \sum_{\{n_i\}} \sum_{A,\cdot\cdot} 
  \bigg[ \prod_{i=1}^{\ell_A} \int \! \! \df \omega_i \bigg] \,
& \vO^{(j)\dagger}_{A\,\lotsdots}\big(\{n_i\};
   \omega_1,\ldots,\omega_{\ell_A}\big) \nn\\
& \times
\vC^{(j)}_{A\,\lotsdots}\big(\{n_i\};\omega_1,\ldots,\omega_{\ell_A} \big)
\,.
\end{align}
The appropriate collinear sectors $\{n_i\}$ are determined by directions found in the collinear states of the hard process being considered. If there is a direction $n_1'$ in the state then we sum over the cases where each of $n_1$, $\ldots$, $n_4$ is set equal to this $n_1'$.\footnote{Technically the $n_i$ in $\{n_i\}$ are representatives of an equivalence class determined by demanding that distinct classes $\{n_i\}$ and $\{n_j\}$ have $n_i\cdot n_j\gg \lambda^2$.}  The sum over $A,\cdot\cdot$ in \eq{Leff_sub_explicit} runs over the full basis of operators that appear at this order, which are specified by either explicit labels $A$ and/or helicity labels $\cdot\cdot$ on the operators and coefficients.  The $\vC^{(j)}_{A}$ are also vectors in the color subspace in which the $\mathcal{O}(\lambda^j)$ hard scattering operators $\vec O_A^{(j)\dagger}$ are decomposed. Explicitly, in terms of color indices, we follow the notation of \Ref{Moult:2015aoa} and have
\begin{align} \label{eq:Opm_color}
\vO^\dagger_\lotsdots  &= O_\lotsdots^{a_1\dotsb \alpha_n}\, \vT^{\, a_1\dotsb \alpha_n}
 \,, \nn\\
C_{\lotsdots}^{a_1\dotsb\alpha_n}
 &= \sum_k C_{\lotsdots}^k T_k^{a_1\dotsb\alpha_n}
\equiv \vT^{ a_1\dotsb\alpha_n} \vC_{\lotsdots}
\,.\end{align}
Here $\vT^{\, a_1\dotsb\alpha_n}$ is a row vector of color structures that spans the color conserving subspace. The $a_i$ are adjoint indices and the $\alpha_i$ are fundamental indices.  The color structures do not necessarily have to be independent, but must be complete. 

Hard scattering operators involving collinear fields are constructed out of products of fields and Wilson lines that are invariant under collinear gauge transformations~\cite{Bauer:2000yr,Bauer:2001ct}. The field building blocks for these operators are collinear gauge-invariant quark and gluon fields, defined as
\begin{align} \label{eq:chiB}
\chi_{{n_i},\w}(x) &= \Bigl[\delta(\w - \bnP_{n_i})\, W_{n_i}^\dagger(x)\, \xi_{n_i}(x) \Bigr]
\,,\\
\cB_{{n_i}\perp,\w}^\mu(x)
&= \frac{1}{g}\Bigl[\delta(\w + \bnP_{n_i})\, W_{n_i}^\dagger(x)\,i  D_{{n_i}\perp}^\mu W_{n_i}(x)\Bigr]
 \,. \nn
\end{align}
For this particular definition of $\chi_{{n_i},\w}$, we have $\w > 0$ for an incoming quark and $\w < 0$ for an outgoing antiquark. For $\cB_{{n_i},\w\perp}$, $\w > 0$ ($\w < 0$) corresponds to outgoing (incoming) gluons. The covariant derivative in \eq{chiB} is given by,
\begin{equation}
i  D_{{n_i}\perp}^\mu = \cP^\mu_{{n_i}\perp} + g A^\mu_{{n_i}\perp}\,,
\end{equation}
and the collinear Wilson line is defined as
\begin{equation} \label{eq:Wn}
W_{n_i}(x) = \biggl[~\sum_\text{perms} \exp\Bigl(-\frac{g}{\bnP_{n_i}}\,\bn\sdt A_{n_i}(x)\Bigr)~\biggr]\,.
\end{equation}
The emissions summed in the Wilson lines are $\ord{\lambda^0}$ in the power counting. The square brackets indicate that the label momentum operators act only on the fields in the Wilson line. The collinear Wilson line, $W_{n_i}(x)$, is localized with respect to the residual position $x$, so that
$\chi_{{n_i},\w}(x)$ and $\cB_{{n_i},\w}^\mu(x)$ can be treated as local quark and gluon fields from the perspective of the ultrasoft degrees of freedom. 

All operators in the theory must be invariant under ultrasoft gauge transformations. Collinear fields transform under ultrasoft gauge transformations as background fields of the appropriate representation. 
Dependence on the ultrasoft degrees of freedom enters the operators through the ultrasoft quark field $q_{us}$, and the ultrasoft covariant derivative $D_{us}$, defined as 
\begin{equation}
i  D_{us}^\mu = i  \partial^\mu + g A_{us}^\mu\,.
\end{equation}
Other operators, such as the ultrasoft gluon field strength, can be constructed from the ultrasoft covariant derivative. The power counting for these operators is shown in \Tab{tab:PC}.

The complete set of collinear and ultrasoft building blocks is summarized in \Tab{tab:PC}. These can be combined, along with Lorentz and Dirac structures, to construct a basis of hard scattering operators at any order in the SCET power counting. All other field and derivative combinations can be reduced to this set by the use of equations of motion and operator relations~\cite{Marcantonini:2008qn}.  As shown in \Tab{tab:PC}, both the collinear quark and collinear gluon building block fields scale as ${\cal O}(\lambda)$. Therefore, while for most jet processes only a single collinear field appears in each sector at leading power, subleading power operators can involve multiple collinear fields in the same collinear sector, as well as $\cP_\perp$ insertions. The scaling of an operator is simply obtained by adding up the powers for the building blocks it contains. This implies that at higher powers hard scattering operators involve more and more fields, or derivative insertions, leading to any increasingly complicated structure. Furthermore, to ensure that the effective theory completely reproduces all IR limits of the full theory, as well as to guarantee that the renormalization group evolution of the operators is closed, it is essential that operator bases in SCET are complete, namely all operators consistent with the symmetries of the problem must be included. Enumerating a minimal basis of operators becomes difficult at subleading power, and it is essential to be able to efficiently identify independent operators, as well as to make manifest all symmetries of the problem.

\subsection{Helicity Operators}\label{sec:review_helicity}

An efficient approach to simplify operator bases in SCET is to use operators of definite helicity \cite{Moult:2015aoa,Kolodrubetz:2016uim,Feige:2017zci}. This general philosophy is well known from the study of on-shell scattering amplitudes, where it leads to compact expressions, removes gauge redundancies, and makes symmetries manifest. The use of helicities is also natural in SCET since the effective theory is formulated as an expansion about identified light like directions with respect to which helicities are naturally defined, and collinear fields carry these directions as labels. Furthermore, since SCET is formulated in terms of collinear gauge invariant fields, see \Eq{eq:chiB}, one can naturally project onto physical polarizations. SCET helicity operators were introduced in \cite{Moult:2015aoa} where they were used to study leading power processes with high multiplicities. This was extended to subleading power in \cite{Kolodrubetz:2016uim} where it was shown that the use of helicity operators is also convenient when multiple fields appear in the same collinear sector.  In this section we briefly review SCET helicity operators, since we will use them to simplify the structure of the subleading power basis for $gg\to H$. We will follow the notation and conventions of~\cite{Moult:2015aoa,Kolodrubetz:2016uim,Feige:2017zci}. A summary of the complete set of operators that we will use is given in \Tab{tab:helicityBB}. 

We define collinear gluon and quark fields of definite helicity as
\begin{subequations}
	\label{eq:cBpm_quarkhel_def}
\begin{align} 
\label{eq:cBpm_def}
\cB^a_{i\pm} &= -\ve_{\mp\mu}(n_i, \bn_i)\,\cB^{a\mu}_{n_i\perp,\w_i}
\,, \\
\label{eq:quarkhel_def}
 \chi_{i \pm}^\alpha &= \frac{1\,\pm\, \gamma_5}{2} \chi_{n_i, - \omega_i}^\alpha
\,,\qquad\quad
\bar{\chi}_{i \pm}^\balpha =  \bar{\chi}_{n_i, - \omega_i}^\balpha \frac{1\,\mp\, \gamma_5}{2}\,.
\end{align}
\end{subequations}
Here $a$, $\alpha$, and $\balpha$ are adjoint, $3$, and $\bar 3$ color indices respectively, and the $\omega_i$ labels on both the gluon and quark building blocks are taken to be outgoing, which is also used for our helicity convention.  Using the standard spinor helicity notation (see e.g. \cite{Dixon:1996wi} for an introduction) 
\begin{align} \label{eq:braket_def}
|p\rangle\equiv \ket{p+} &= \frac{1 + \ga_5}{2}\, u(p)
  \,,
 & |p] & \equiv \ket{p-} = \frac{1 - \ga_5}{2}\, u(p)
  \,, \\
\bra{p} \equiv \bra{p-} &= \mathrm{sgn}(p^0)\, \bar{u}(p)\,\frac{1 + \ga_5}{2}
  \,, 
 & [p| & \equiv \bra{p+} = \mathrm{sgn}(p^0)\, \bar{u}(p)\,\frac{1 - \ga_5}{2}
  \,, \nn 
\end{align}
with $p$ lightlike, the polarization vector of an outgoing gluon with momentum $p$ can be written
\begin{equation}
 \ve_+^\mu(p,k) = \frac{\mae{p+}{\ga^\mu}{k+}}{\sqrt{2} \langle kp \rangle}
\,,\qquad
 \ve_-^\mu(p,k) = - \frac{\mae{p-}{\ga^\mu}{k-}}{\sqrt{2} [kp]}
\,,\end{equation}
where $k\neq p$ is an arbitrary light-like reference vector, chosen to be $\bn_i$ in \eq{cBpm_def}.  

Since fermions always arise in pairs, we can define currents with definite helicities. Here we will restrict to the case of two back to back directions, $n$ and $\bar n$, as is relevant for $gg\to H$. A more general discussion can be found in \Refs{Kolodrubetz:2016uim,Feige:2017zci}. We define helicity currents where the quarks are in opposite collinear sectors,
 \begin{align} \label{eq:jpm_back_to_bacjdef}
 & h=\pm 1:
 & J_{n \bn \pm}^{\balpha\beta}
 & = \mp\, \sqrt{\frac{2}{\omega_n\, \omega_\bn}}\, \frac{   \ve_\mp^\mu(n, \bn) }{\langle \bn \mp | n \pm\rangle}   \, \bar{\chi}^\balpha_{n\pm}\, \gamma_\mu \chi^\beta_{\bn \pm}
 \,, \\
 & h=0:
 & J_{n \bn 0}^{\balpha\beta}
 & =\frac{2}{\sqrt{\vphantom{2} \omega_n \,\omega_\bn}\,  [n \bn] } \bar \chi^\balpha_{n+}\chi^\beta_{\bn-}
 \,, \qquad
 (J^\dagger)_{n \bn 0}^{\balpha\beta}=\frac{2}{\sqrt{ \vphantom{2} \omega_n \, \omega_\bn}  \langle n  \bn \rangle  } \bar \chi^\balpha_{n-}\chi^\beta_{\bn+}
 \,, \nn
 \end{align}
 as well as where the quarks are in the same collinear sector,
\begin{align}\label{eq:coll_subl}
 & h=0:
 & J_{i0}^{\balpha \beta} 
  &= \frac{1}{2 \sqrt{\vphantom{2} \omega_{\bar \chi} \, \omega_\chi}}
  \: \bar \chi^\balpha_{i+}\, \Sl{\bar n}_i\, \chi^\beta_{i+}
   \,,\qquad
   J_{i\bar 0}^{\balpha \beta} 
  = \frac{1}{2 \sqrt{\vphantom{2} \omega_{\bar \chi} \, \omega_\chi}}
  \: \bar \chi^\balpha_{i-}\, \Sl {\bar n}_i\, \chi^\beta_{i-}
 \,, \\[5pt]
  & h=\pm 1:
 & J_{i\pm}^{\balpha \beta}
  &= \mp  \sqrt{\frac{2}{ \omega_{\bar \chi} \, \omega_\chi}}  \frac{\epsilon_{\mp}^{\mu}(n_i,\bar n_i)}{ \big(\l n_i \mp | \bar{n}_i \pm \r \big)^2}\: 
   \bar \chi_{i\pm}^\balpha\, \gamma_\mu \Sl{\bar n}_i\, \chi_{i\mp}^\beta
 \,. \nn
\end{align}
Here $i$ can be either $n$ or $\bar n$. All of these currents are manifestly invariant under the RPI-III symmetry of SCET.   The Feynman rules for all currents are very simple, and are given in~\cite{Feige:2017zci}.  Note that the operators $J_{n \bn \pm}^{\balpha\beta}$, $J_{i0}^{\balpha \beta}$, and $J_{i\bar 0}^{\balpha \beta}$ have quarks of the same chirality, and hence are the ones that will be generated by vector gauge bosons.

\begin{table}
 \begin{center}
  \begin{tabular}{|c|c|cc|ccc|c|ccc|}
	\hline \phantom{x} & \phantom{x} & \phantom{x} 
	& \phantom{x} & \phantom{x} & \phantom{x} & \phantom{x} 
	& \phantom{x} & \phantom{x} & \phantom{x} & \phantom{x} 
	\\[-13pt]                      
 Field: & 
    $\cB_{i\pm}^a$ & $J_{ij\pm}^{\balpha\beta}$ & $J_{ij0}^{\balpha\beta}$ 
    & $J_{i\pm}^{\balpha \beta}$ 
	& $J_{i0}^{\balpha \beta}$ & $J_{i\bar 0}^{\balpha \beta}$  
    & $\cP^{\perp}_{\pm}$ 
	& $\partial_{us(i)\pm}$ & $\partial_{us(i)0}$ & $\partial_{us(i)\bar{0}}$
	\\[3pt] 
 Power counting: &	
    $\lambda$ &  $\lambda^2$ &  $\lambda^2$
	& $\lambda^2$ & $\lambda^2$& $\lambda^2$ & $\lambda$ 
    & $\lambda^2$ & $\lambda^2$  & $\lambda^2$
	\\
 Equation: & 
   (\ref{eq:cBpm_def}) & \multicolumn{2}{c|}{(\ref{eq:jpm_back_to_bacjdef})} 
     & \multicolumn{3}{c|}{(\ref{eq:coll_subl})} & (\ref{eq:Pperppm}) 
     & \multicolumn{3}{c|}{(\ref{eq:partialus})}
    \\
  \hline  
  \end{tabular}\\
\vspace{.3cm} 
  \begin{tabular}{|c|cc|}
	\hline  \phantom{x} &  \phantom{x} & \phantom{x} 
	\\[-13pt]                        
 Field: & 
 	$\cB^a_{us(i)\pm}$ & 
    \!\!$\cB^a_{us(i)0}$  
	\\[3pt] 
 Power counting: &
 	$\lambda^2$ & $\lambda^2$ 
	\\ 
 Equation: & 
     \multicolumn{2}{c|}{(\ref{eq:Bus})}  
    \\
	\hline
  \end{tabular}
 \end{center}
\vspace{-0.3cm}
\caption{The helicity building blocks in $\text{SCET}_\text{I}$ that will be used to construct a basis of hard scattering operators for $gg\to H$, together with their power counting order in the $\lambda$-expansion, and the equation numbers where their definitions may be found. The building blocks also include the conjugate currents $J^\dagger$ in cases where they are distinct from the ones shown.
} 
\label{tab:helicityBB}
\end{table}

At subleading power one must also consider insertions of the $\cP_{i\perp}^\mu$ operator. Note that we can drop the explicit $i$ index on the $\cP_\perp$ operator, as it is implied by the field that the operator is acting on.  The $\cP_{\perp}^\mu$ operator acts on the perpendicular subspace defined by the vectors $n_i, \bar n_i$, so  it is naturally written as
\begin{align} \label{eq:Pperppm}
\cP_{+}^{\perp}(n_i,\bar n_i)=-\epsilon^-(n_i,\bar n_i) \cdot \cP_{\perp}\,, \qquad \cP_{-}^{\perp}(n_i,\bar n_i)=-\epsilon^+(n_i,\bar n_i) \cdot \cP_{\perp}\,.
\end{align} 
The $\cP^\perp_\pm$ operator carry helicity $h=\pm 1$. 
We use square brackets to denote which fields are acted upon by the $\cP^{\perp}_{\pm}$ operator, for example
$\cB_{i+} \left [ \cP^{\perp}_{+}  \cB_{i-}  \right]  \cB_{i-}$,
indicates that the $\cP^{\perp}_{+}$ operator acts only on the middle field, whereas for currents, we use a curly bracket notation
\begin{align}\label{eq:p_perp_notation}
  \big\{ \cP^{\perp}_\lambda J_{i 0 }^{\balpha \beta} \big\}  
  & = \frac{1}{2 \sqrt{\vphantom{2}\omega_{\bar \chi} \, \omega_\chi }} \:
   \Big[  \cP^{\perp}_{\lambda}  \bar \chi^\balpha_{i +}\Big] \Sl {\bar n}_i \chi^\beta_{i+}
  \,, \\
 \big\{ J_{i0 }^{\balpha \beta} (\cP^{\perp}_{\lambda})^\dagger \big\}
  &=  \frac{1}{2\sqrt{\vphantom{2}\omega_{\bar \chi} \, \omega_\chi}} \:
  \bar \chi^\balpha_{i+} \Sl {\bar n}_i \Big[   \chi^\beta_{i+} (\cP^{\perp}_{\lambda})^\dagger \Big]
  \,, \nn
\end{align}
to indicate which of the fields within the current is acted on.

To work with gauge invariant ultrasoft gluon fields, we construct our basis post BPS field redefinition. The BPS field redefinition is defined by \cite{Bauer:2002nz}
\be \label{eq:BPSfieldredefinition}
\cB^{a\mu}_{n\perp}\to \cY_n^{ab} \cB^{b\mu}_{n\perp} , \qquad \chi_n^\alpha \to Y_n^{\alpha \bbeta} \chi_n^\beta,
\ee
and is performed in each collinear sector. Here $Y_n$, $\cY_n$ are fundamental and adjoint ultrasoft Wilson lines. For a general representation, r, the ultrasoft Wilson line is defined by
\be
Y^{(r)}_n(x)=\bold{P} \exp \left [ ig \int\limits_{-\infty}^0 ds\, n\cdot A^a_{us}(x+sn)  T_{(r)}^{a}\right]\,,
\ee
where $\bold P$ denotes path ordering.  The BPS field redefinition has the effect of decoupling ultrasoft and collinear degrees of freedom at leading power \cite{Bauer:2002nz}, and it accounts for the full physical path of ultrasoft Wilson lines~\cite{Chay:2004zn,Arnesen:2005nk}.

The BPS field redefinition introduces ultrasoft Wilson lines into the hard scattering operators. These Wilson lines can be arranged with the ultrasoft fields to define ultrasoft gauge invariant building blocks. In particular, the gauge covariant derivative in an arbitrary representation, $r$, can be sandwiched by Wilson lines and decomposed as
\begin{align}\label{eq:soft_gluon}
Y^{(r)\,\dagger}_{n_i} i D^{(r)\,\mu}_{us} Y^{(r)}_{n_i }=i \partial^\mu_{us} + [Y_{n_i}^{(r)\,\dagger} i D^{(r)\,\mu}_{us} Y^{(r)}_{n_i}]=i\partial^\mu_{us}+T_{(r)}^{a} g \cB^{a\mu}_{us(i)}\,.
\end{align}
Here we have defined the ultrasoft gauge invariant gluon field by
\begin{align} \label{eq:softgluondef}
g \cB^{a\mu}_{us(i)}= \left [   \frac{1}{in_i\cdot \partial_{us}} n_{i\nu} i G_{us}^{b\nu \mu} \cY^{ba}_{n_i}  \right] \,.
\end{align}
In the above equations the derivatives act only within the square brackets. Note from \eq{softgluondef}, that  $n_i\cdot \cB^{a}_{us(i)}= 0$.    The Wilson lines which remain after this procedure can be absorbed into a generalized color structure, $\vT_{\BPS}$ (see \cite{Kolodrubetz:2016uim} for more details). Determining a complete basis of color structures is straightforward, and detailed examples are given in~\cite{Feige:2017zci}.

Having defined gauge invariant ultrasoft gluon fields, we can now define ultrasoft gauge invariant gluon helicity fields and derivative operators which mimic their collinear counterparts. For the ultrasoft gluon helicity fields we define the three building blocks
\begin{equation} \label{eq:Bus}
\cB^a_{us(i)\pm} = -\ve_{\mp\mu}(n_i, \bn_i)\,\cB^{a\mu}_{us(i)},\qquad  \cB^a_{us(i)0} =\bar n_\mu  \cB^{a \mu}_{us(i)}   
\,,\end{equation}
and similarly for the ultrasoft derivative operators
\begin{equation}  \label{eq:partialus}
\partial_{us(i)\pm} = -\ve_{\mp\mu}(n_i, \bn_i)\,\partial^{\mu}_{us},\qquad   \partial_{us(i)0} =\bar n_{i\mu} \partial^{\mu}_{us}, \qquad \partial_{us(i)\bar 0} = n_{i \mu} \partial^{\mu}_{us}
\,.\end{equation}
Unlike for the gauge invariant collinear gluon fields, for the ultrasoft gauge invariant gluon field we use three building block fields to describe the two physical degrees of freedom because the ultrasoft gluons are not fundamentally associated with any direction. Without making a further gauge choice, their polarization vectors do not lie in the perpendicular space of any fixed external reference vector. When inserting ultrasoft derivatives into operators we will use the same curly bracket notation defined for the $\cP_\perp$ operators in \Eq{eq:p_perp_notation}. 

Gauge invariant ultrasoft quark fields can also appear explicitly in operator bases at subleading powers. From \Tab{tab:PC} we see that they power count as $\cO(\lambda^3)$, and are therefore not relevant for our construction of an $\cO(\lambda^2)$ operator basis. Details on the structure of subleading power helicity operators involving ultrasoft quarks can be found in~\cite{Feige:2017zci}. It is important to emphasize that although ultrasoft quarks do not appear in the hard scattering operators at $\cO(\lambda^2)$ they do appear in the calculation of cross sections or amplitudes at $\cO(\lambda^2)$ due to subleading power Lagrangian insertions in the effective (examples where they play an important role for factorization in $B$-decays include both exclusive decays~\cite{Bauer:2002aj,Mantry:2003uz,Beneke:2003pa} and inclusive decays~\cite{Bosch:2004cb,Lee:2004ja,Beneke:2004in}). Such ultrasoft quark contributions also played an important role in the recent subleading power perturbative SCET calculation of \Ref{Moult:2016fqy}.

Finally, we note that the helicity operator basis presented in this section only provides a complete basis in $d=4$, and we have not discussed evanescent operators \cite{Buras:1989xd,Dugan:1990df,Herrlich:1994kh}. An extension of our basis to include evanescent operators would depend on the regularization scheme. However, in general additional building block fields would be required, for example an $\epsilon$ scalar gluon $\cB^a_{\epsilon}$ to encode the $(-2\epsilon)$ transverse degrees of freedom of the gluon. As in standard loop calculations, we expect that the evanescent operators at each loop order could be straightforwardly identified and treated. Since we do not perform a one-loop matching to our operators, we leave a complete treatment of evanescent operators to future work.

\section{Operator Basis}\label{sec:basis}

In this section we enumerate a complete basis of power suppressed operators up to $\cO(\lambda^2)$  for the process $gg\to H$. The organization of the operator basis in terms of helicity operators will make manifest a number of symmetries arising from helicity conservation, greatly reducing the operator basis. Helicity conservation is particularly powerful in this case due to the spin-$0$ nature of the Higgs. The complete basis of field structures is summarized in \Tab{tab:summary}. In \Sec{sec:discussion} we will show which operators contribute to the cross section at $\cO(\lambda^2)$. These operators are indicated with a check mark in the table.

Examining \eq{Leff_sub_explicit} we see that the hard Lagrangian in SCET is written as a sum over label momenta of the hard operators.  For the special case of two back-to-back collinear sectors this reduces to
\begin{align}\label{eq:sum_dir}
\cL_{\text{hard}}^{(j)} = \sum_{n} \sum_{A,\cdot\cdot}  
\bigg[ \prod_{i=1}^{\ell_A} \int \! \! \df \omega_i \bigg] \,
& \vO^{(j)\dagger}_{A\,\lotsdots}\big(n,\bn;
   \omega_1,\ldots,\omega_{\ell_A}\big) \nn\\
& \times
\vC^{(j)}_{A\,\lotsdots}\big(n,\bn;\omega_1,\ldots,\omega_{\ell_A} \big)
\,.
\end{align}
When writing our basis, we therefore do not need to include operators which are identical up to the swap of $n\leftrightarrow \bar n$.  This means that when writing an operator with different field structures in the two collinear sectors we are free to make an arbitrary choice for which is labeled $n$ and which $\bar n$, and this choice can be made independently for each operator. When squaring matrix elements, all possible interferences are properly incorporated by the sum over directions in \Eq{eq:sum_dir}.

As discussed in \Sec{sec:intro}, we will work in the Higgs effective theory with a Higgs gluon coupling given by the effective Lagrangian in \Eq{eq:heft_intro}. We therefore do not consider operators generated by a direct coupling of quarks to the Higgs. All quarks in the final state are produced by gluon splittings. The extension to include operators involving quarks coupling directly to Higgs, as relevant for $H\to b\bar b$, is straightforward using the helicity building blocks given in \Sec{sec:review_helicity}.

{
\renewcommand{\arraystretch}{1.4}
\begin{table}[t!]
\hspace{-0.15cm} \scalebox{0.9}{
\begin{tabular}{| c | l | l | c | c | c | }
	\hline 
Order & $\!$Category &  Operators (equation number) 
 & \#$\!$  helicity & \#$\!$ of
 & $\!\sigma_{2j}^{\cO(\lambda^2)}\!\! \ne\! 0\!$
 \\[-8pt]
 & & & configs & \! color\! & 
 \\ \hline 
$\mathcal{O}(\lambda^0)$  
   & $\! H gg$ & $O_{\cB \lambda_1 \lambda_1}^{(0)ab}=\cB^a_{n  \lambda_1} \cB^a_{\bar n  \lambda_1} H$ \,(\ref{eq:hgg})
   & 2  & 1 & $\checkmark$
    \\ \hline
$\mathcal{O}(\lambda)$ 
   & $\! H q \bar{q} g$  
    & $O_{\cB n,\bar n \lambda_1 ( \lambda_i)}^{(1)a\,\balpha\bt}
	=  \cB_{n,\bar n  \lambda_1}^a\, J_{n\bar n\, \lambda_j}^{\balpha\bt}\,H$ \,(\ref{eq:H1_basis},\ref{eq:H1_basis2}) 
    & 4 &  1 &   $\checkmark$ 
    \\ \hline
$\mathcal{O}(\lambda^2)$
   & $\! H q \bar q Q \bar Q$  & $O_{qQ1(\lambda_1;\lambda_2)}^{(2)\balpha\bt\bgamma\delta}
	= J_{(q)n {\lambda_1}\, }^{\balpha\bt}\, J_{(Q) \bar n {\lambda_2}\, }^{\bgamma\delta}\,H$ \,(\ref{eq:Z2_basis_qQ}) 
    & 4  & 2 & 
    \\
	& & $O_{qQ2(\lambda_1;\lambda_1)}^{(2)\balpha\bt\bgamma\delta}
	= J_{(q \bar{Q}) n  \lambda_1\, }^{\balpha\bt}\, J_{(Q \bar{q}) \bar{n}  \, \lambda_1\, }^{\bgamma\delta}\,H $ \,(\ref{eq:Z2_basis_qQ_2})
    & 2  & 2 &  
    \\
    & & $O_{qQ3(\lambda_1;-\lambda_1)}^{(2)\balpha\bt\bgamma\delta}
	= J_{(q) n \bar n \lambda_1\, }^{\balpha\bt}\, J_{(Q) n\bar n -\lambda_1\, }^{\bgamma\delta}\,H$ \,(\ref{eq:Z2_basis_qQ_3})
    & 2 &  2 & 
    \\ \cline{2-6}
   & $\! H q \bar q q \bar q$  & $O_{qq1(\lambda_1;\lambda_2)}^{(2)\balpha\bt\bgamma\delta}
	= J_{(q)n {\lambda_1}\, }^{\balpha\bt}\, J_{(q) \bar n {\lambda_2}\, }^{\bgamma\delta}\,H$ \,(\ref{eq:Z2_basis_qq}) 
    & 3 & 2 &   
    \\
	& & $O_{qq3(\lambda_1;-\lambda_1)}^{(2)\balpha\bt\bgamma\delta}
	= J_{(q) n \bar n \lambda_1\, }^{\balpha\bt}\, J_{(q) n\bar n -\lambda_1\, }^{\bgamma\delta}\,H$ \,(\ref{eq:Z2_basis_qq_3})
    & 1 & 2 & 
    \\ \cline{2-6}
   & $\! H q \bar q g g$  & $O_{\cB1\lambda_1 \lambda_2(\lambda_3)}^{(2)ab\, \balpha\bt}
	=  \cB_{n \lambda_1}^a \cB_{\bar n \lambda_2}^b \, J_{n\,{\lambda_3} }^{\balpha\bt}   \,H $ \,(\ref{eq:Hqqgg_basis3}) 
    & 4  &  3 & $\checkmark$
    \\
	& & $O_{\cB2\lambda_1 \lambda_2(\lambda_3)}^{(2)ab\, \balpha\bt}
	=  \cB_{\bar n \lambda_1}^a \cB_{\bar n \lambda_2}^b \, J_{n\,{\lambda_3} }^{\balpha\bt}   \,H $ \,(\ref{eq:Hqqgg_basis4}) 
    & 2  & 3 & 
    \\ \cline{2-6}
   & $\! H gggg$  & $O_{4g1\lambda_1 \lambda_2 \lambda_3 \lambda_4}^{(2)a b c d}
	= S  \cB^a_{n \lambda_1} \cB^b_{n \lambda_2} \cB^c_{\bn \lambda_3} \cB^d_{\bn \lambda_4} H $ \,(\ref{eq:H_basis_gggg_1}) 
    & 3  & 9 &  
    \\
	&   & $O_{4g2\lambda_1 \lambda_2 \lambda_3 \lambda_4}^{(2)a b c d}
	= S  \cB^a_{n \lambda_1} \cB^b_{\bn \lambda_2} \cB^c_{\bn \lambda_3} \cB^d_{\bn \lambda_4} H$ \,(\ref{eq:H_basis_gggg_2}) 
    & 2  & 9 & $\checkmark$
    \\ \cline{2-6}
   &  $\cP_\perp$   & $O_{\cP\chi \lambda_1 (\lambda_2)[\lambda_{\cP}]}^{(2)a\,\balpha\bt}
	= \cB_{n\lambda_1}^a \, \{J_{\bar n\, {\lambda_2}    }^{\balpha\bt}(\cP_{\perp}^{\lambda_{\cP}})^\dagger\}\,H$  \,(\ref{eq:Hqqgpperp_basis_same})  
    & 4  & 1 & $\checkmark$
    \\
	& & $O_{\cP\cB \lambda_1 \lambda_2 \lambda_3[\lambda_{\cP}]}^{(2)abc}
	= S\, \cB_{n\lambda_1}^a\, \cB_{\bar n \lambda_2}^b \left [\cP_{\perp}^{\lambda_{\cP}} \cB_{\bar n \lambda_3}^c \right ]\! H\! $  \,(\ref{eq:Hgggpperp_basis}) 
    & 4  & 2 & $\checkmark$ 
    \\ \cline{2-6}
	&  	$\!$Ultrasoft $\!\!\!$    & $O_{\chi (us(n))0:(\lambda_1)}^{(2)a\,\balpha\bt} 
	= \cB_{us(n)0}^a \, J_{n\bar n\,\lambda_1}^{\balpha\bt}\,H $ \,(\ref{eq:soft_insert_basis}) 
    & 2 & 1 & 
    \\
	&   & $O_{\chi (us(\bar n))0:(\lambda_1)}^{(2)a\,\balpha\bt} 
	= \cB_{us(\bar n)0}^a \, J_{n\bar n\,\lambda_1}^{\balpha\bt}\,H $ \,(\ref{eq:soft_insert_basis2}) 
    & 2 &  1 & 
    \\
	& & $O_{\partial \chi (us(i))\lambda_1:(\lambda_2)}^{(2)\,\balpha\bt}
	= \{\partial_{us(i)\lambda_1} \, J_{n\bar n\,\lambda_2}^{\balpha\bt}\}\,H $ \,(\ref{eq:soft_derivative_basis}) 
    & 4 & 1 & 
    \\
	& & $O_{\cB (us(n))\lambda_1:\lambda_2 \lambda_3 }^{(2)abc}
	=\cB_{us(n) \lambda_1}^a \,\cB_{n\, \lambda_2}^{b}\,\cB_{\bn\, \lambda_3}^{c}\,H$ \,(\ref{eq:Hgggus}) 
    & 2  & 2 & $\checkmark$
    \\
	& & $O_{\cB (us(\bar n))\lambda_1:\lambda_2 \lambda_3 }^{(2)abc}
	=\cB_{us(\bar n) \lambda_1}^a \,\cB_{n\, \lambda_2}^{b}\,\cB_{\bn\, \lambda_3}^{c}\,H$ \,(\ref{eq:Hgggus_2}) 
	& 2  & 2 & $\checkmark$
    \\
	& & $O_{\partial \cB  (us(i))\lambda_1:\lambda_2 \lambda_3 }^{(2)ab}
	=\left[ \partial_{us(i) \lambda_1} \,\cB_{n\, \lambda_2}\right] \,\cB_{\bn\, \lambda_3}\,H $ \,(\ref{eq:Hdggus}) 
	& 4  & 1 & $\checkmark$ 
    \\ \hline
\end{tabular}}
\vspace{0.1cm}
\caption{Basis of hard scattering operators for $gg\to H$ up to ${\cal O}(\lambda^2)$. The $\lambda_i$ denote helicities, $S$ represents a symmetry factor present for some cases, and detailed lists of operators can be found in the indicated equation.  The number of allowed helicity configurations are summarized in the fourth column. The final column indicates which operators contribute to the cross section up to $\mathcal{O}(\lambda^2)$ in the power expansion, as discussed in Sec. \ref{sec:discussion}. Counting the helicity configurations there are a total of 53 operators, of which only 28 
contribute to the cross section at $\mathcal{O}(\lambda^2)$. Of those 28, only 24 have non zero Wilson coefficients at tree level since the operators in \eq{Hqqgpperp_basis_same} are absent at this order. These numbers do not include the number of distinct color configurations which are indicated in the 5th column. 
}
\label{tab:summary}
\end{table}
}

\subsection{Leading Power}\label{sec:lp}

The leading power operators for $gg\to H$ in the Higgs effective theory are well known. Due to the fact that the Higgs is spin zero, the only two operators are
\begin{align}
 \boldsymbol{g_n g_{\bn}:}   {\vcenter{\includegraphics[width=0.18\columnwidth]{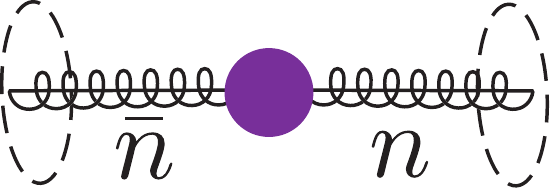}}} \nn
\end{align}
\vspace{-0.4cm}
\begin{alignat}{2}\label{eq:hgg}
 &O_{\cB++}^{(0)ab}
=\cB_{n +}^a\, \cB_{\bar n+}^b\,   H
\,, \qquad &&O_{\cB--}^{(0)ab}
=\cB_{n-}^a\, \cB_{ \bn -}^b\, H\,.
\end{alignat}
Here the purple circled denotes that this is a hard scattering operator in the effective theory, while the dashed circles indicate which fields are in each collinear sector.
Note that here we have opted not to include a symmetry factor at the level of the operator. We will include symmetry factors in the operator only when there is an exchange symmetry within a given collinear sector. We assume that overall symmetry factors which involve exchanging particles from different collinear sectors are taken into account at the phase space level. The color basis here is one-dimensional, and we take it to be
\begin{align} \label{eq:leading_color}
 \vT^{ab} = \de_{ab}\,, \qquad 
 \vT_{\BPS}^{ab} = \bigl( \cY_{n}^T \cY_{\bar n} \bigr)^{ab} 
                 = \bigl( \cY_{\bn}^T \cY_{n} \bigr)^{ba} 
\,.
\end{align}

\subsection{Subleading Power}\label{sec:nlp}

Due to the spin zero nature of the Higgs, the $\cO(\lambda)$ operators are highly constrained. To simplify the operator basis we will work in the center of mass frame and we will further choose our $n$ and $\bar n$ axes so that the total label $\perp$ momentum of each collinear sector vanishes. This is possible in an SCET$_\text{I}$ theory since the ultrasoft sector does not carry label momentum, and it implies that we do not need to include operators where the $\cP_\perp$ operator acts on a sector with a single collinear field. At $\cO(\lambda)$ the suppression in the operator must therefore come from an explicit collinear field. 

There are two possibilities for the collinear field content of the operators, either three collinear gluon fields, or two collinear quark fields and a collinear gluon field. Interestingly, the helicity selection rules immediately eliminate the possibility of $\cO(\lambda)$ operators with three collinear gluon fields, since they cannot sum to a zero helicity state. We therefore only need to consider operators involving two collinear quark fields and a collinear gluon field. The helicity structure of these operators is also constrained. In particular, to cancel the spin of the collinear gluon field, the collinear quark current must have helicity $\pm1$. Furthermore, the quark-antiquark pair arises from a gluon splitting, since we are considering gluon fusion in the Higgs EFT, and therefore both have the same chirality. Together this implies that the quarks are described by the current $J_{n\bar n\,\pm}^{\balpha\bt}$. The only two operators in the basis at  $\cO(\lambda)$ are
\begin{align}
 \boldsymbol{q_n (\bar qg)_{\bn}:}   {\vcenter{\includegraphics[width=0.18\columnwidth]{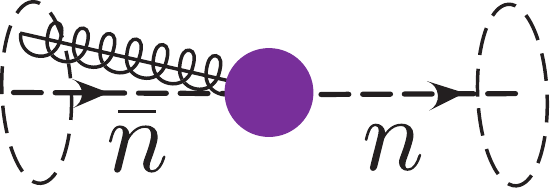}}} \nn
\end{align}
\vspace{-0.4cm}
\begin{alignat}{2} \label{eq:H1_basis}
&O_{\cB\bar n+(+)}^{(1)a\,\balpha\bt}
= \cB_{\bar n+}^a\, J_{n\bar n\,+}^{\balpha\bt} \,H
\,,\qquad &
&O_{\cB\bar n-(-)}^{(1)a\,\balpha\bt}
= \cB_{\bar n-}^a \, J_{n\bar n\,-}^{\balpha\bt}\,H
\,,
\end{alignat}
for the case that the gluon field is in the same sector as the antiquark field, which we have taken to be $\bar n$, and
\begin{align}
 \boldsymbol{(q g)_n \bar q_{\bn}:}   {\vcenter{\includegraphics[width=0.18\columnwidth]{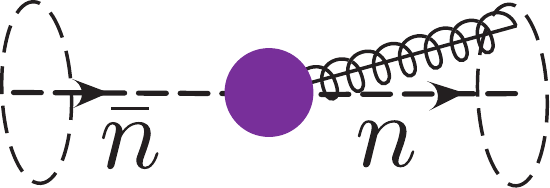}}} \nn
\end{align}
\vspace{-0.4cm}
\begin{alignat}{2} \label{eq:H1_basis2}
&O_{\cB n-(+)}^{(1)a\,\balpha\bt}
= \cB_{n-}^a\, J_{n\bar n\,+}^{\balpha\bt} \,H
\,,\qquad &
&O_{\cB n+(-)}^{(1)a\,\balpha\bt}
= \cB_{n+}^a \, J_{n\bar n\,-}^{\balpha\bt}\,H
\,,
\end{alignat}
for the case that the gluon field is in the same direction as the collinear quark field.  In both cases the color basis is one-dimensional $\vT^{a\, \al\bbeta} = T^a_{\al\bbeta}$. After the BPS field redefinition we have
\begin{align} \label{eq:nlp_color}
 \vT_{\BPS}^{ a \al\bbeta} 
    &= \left (Y_n^\dagger Y_{\bar n} T^a   \right )_{\alpha \bar \beta}
    \,,
& \vT_{\BPS}^{ a \al\bbeta} 
    &= \left (T^a Y_n^\dagger Y_{\bar n}  \right )_{\alpha \bar \beta}
    \,,
\end{align}
for \eqs{H1_basis}{H1_basis2} respectively.

\subsection{Subsubleading Power}\label{sec:nnlp}

At $\cO(\lambda^2)$ the allowed operators can include either additional collinear field insertions, insertions of the $\cP_\perp$ operator, or ultrasoft field insertions. We will treat each of these cases in turn.

\subsubsection{Collinear Field Insertions}\label{sec:nnlp_collinear}

We begin by considering operators involving only collinear field insertions. At $\cO(\lambda^2)$ the operator can have four collinear fields. These operators can be composed purely of collinear gluon fields, purely of collinear quark fields, or of two collinear gluon fields and a collinear quark current. In each of these cases helicity selection rules will restrict the possible helicity combinations of the operators.

\vspace{0.4cm}
\noindent{\bf{Two Quark-Two Gluon Operators:}}

We begin by considering operators involving two collinear quark fields and two collinear gluon fields, which are again severely constrained by the helicity selection rules. Since the two gluons fields can give either helicity $0$ or $2$, the only way to achieve a total spin zero is if the quark fields must be in a helicity zero configuration. Furthermore, since they arise from a gluon splitting they must have the same chirality. This implies that all operators must involve only the currents $J_{n\, 0}^{\balpha\bt}$ or $J_{n\, \bar 0}^{\balpha\bt}$, where we have taken without loss of generality that the two quarks are in the $n$-collinear sector, as per the discussion below \Eq{eq:sum_dir}. The gluons can then either be in opposite collinear sectors, or in the same collinear sector. The color basis before BPS field redefinition is identical for the two cases. It is three dimensional, and we take as a basis
\begin{equation} \label{eq:ggqqll_color}
\vT^{\, ab \alpha\bbeta}
= \Bigl(
(T^a T^b)_{\alpha\bbeta}\,,\, (T^b T^a)_{\alpha\bbeta} \,,\, \tr[T^a T^b]\, \delta_{\alpha\bbeta}
\Bigr)
\,.\end{equation} 

In the case that the two collinear gluons are in opposite collinear sectors a basis of helicity operators is given by
\begin{align}
 \boldsymbol{(g q\bar q)_n (g)_{\bn}:}   {\vcenter{\includegraphics[width=0.18\columnwidth]{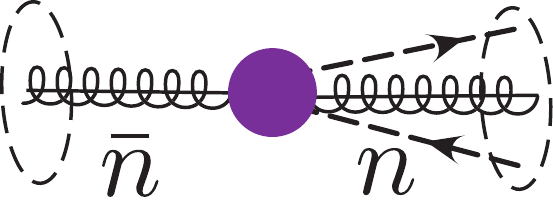}}} \nn
\end{align}
\vspace{-0.4cm}
\begin{alignat}{2} \label{eq:Hqqgg_basis3}
&O_{\cB1++(0)}^{(2)ab\, \balpha\bt}
=   \cB_{n+}^a\, \cB_{\bar n+}^b \, J_{n\,{0} }^{\balpha\bt}  \,H\, , \qquad 
&&O_{\cB1++(\bar 0)}^{(2)ab\, \balpha\bt}
=\cB_{n+}^a\, \cB_{\bar n+}^b  \, J_{n\,{\bar 0} }^{\balpha\bt}   \,H\, ,  \\
&O_{\cB1--(0)}^{(2)ab\, \balpha\bt}
=  \cB_{n-}^a\, \cB_{\bar n-}^b \, J_{n\,{0} }^{\balpha\bt}    \,H\, , \qquad 
&&O_{\cB1--(\bar 0)}^{(2)ab\, \balpha\bt}
= \cB_{ n-}^a\, \cB_{\bar n-}^b  \, J_{n\,{\bar 0} }^{\balpha\bt}   \,H\, . \nn
\end{alignat}
The color basis after BPS field redefinition is given by
\begin{equation}
\vT_{\BPS}^{\, ab \alpha\bbeta}
= \Bigl(
(\cY_n^T \cY_\bn)^{cb}  (T^a T^c)_{\alpha\bbeta} \,,\, 
(\cY_n^T \cY_\bn)^{cb}  (T^c T^a)_{\alpha\bbeta} \,,\, 
T_F (\cY_n^T \cY_\bn)^{ab} \, \delta_{\alpha\bbeta}
\Bigr)
\,,
\end{equation}
where we have used $\tr[T^a T^b] = T_F \delta^{ab}$.

In the case that the two gluons are in the same collinear sector a basis of helicity operators is given by
\begin{align}
& \boldsymbol{(q \bar q)_n (gg)_{\bn}:}{\vcenter{\includegraphics[width=0.18\columnwidth]{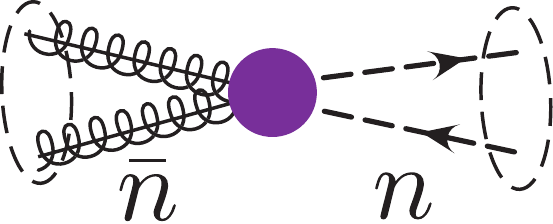}}}  \nn
\end{align}
\vspace{-0.4cm}
\begin{align}\label{eq:Hqqgg_basis4}
&O_{\cB2+-(0)}^{(2)ab\, \balpha\bt}
= \cB_{\bar n+}^a\, \cB_{ \bar n-}^b  \, J_{n\,{0} }^{\balpha\bt}\,   H\,, \qquad 
&O_{\cB2+-(\bar 0)}^{(2)ab\, \balpha\bt}
= \cB_{\bar n+}^a\, \cB_{ \bar n-}^b  \, J_{n\,{\bar 0} }^{\balpha\bt}\,   H\,.
\end{align}
The color basis after BPS field redefinition is
\begin{equation}
\vT_{\BPS}^{\, ab \alpha\bbeta}
= \Bigl(
(Y_n^\dagger Y_\bn T^a T^b Y^\dagger_{\bar n} Y_n )_{\alpha\bbeta}
  \,,\, 
(Y_n^\dagger Y_\bn T^b T^a Y^\dagger_{\bar n} Y_n )_{\alpha\bbeta}
   \,,\, 
\tr[T^a T^b]\, \delta_{\alpha\bbeta} \Bigr)
\,.
\end{equation}

\vspace{0.4cm}
\noindent{\bf{Four Gluon Operators:}}

Operators involving four collinear gluon fields can have either two collinear gluon fields in each sector, or three collinear gluon fields in one sector. A basis of color structures before BPS field redefinition is given by
\begin{equation} \label{eq:gggg_color}
\vT^{ abcd} =
\frac{1}{2}\begin{pmatrix}
\tr[abcd] + \tr[adcb] \\ \tr[acdb] + \tr[abdc] \\ \tr[adbc] + \tr[acbd] \\
\tr[abcd] - \tr[adcb] \\ \tr[acdb] - \tr[abdc] \\ \tr[adbc] - \tr[acbd] \\ 2\tr[ab]\, \tr[cd] \\ 2\tr[ac]\, \tr[db] \\ 2\tr[ad]\, \tr[bc]
\end{pmatrix}^{\!\!\!T}
.\end{equation}
Here we have used a simplified notation, writing only the adjoint indices of the color matrices appearing in the trace.  For example, $\tr[abcd] \equiv \tr[T^a T^b T^c T^d]$.
The color bases after BPS field redefinition will be given separately for each case.
For the specific case of SU($N_c$) with $N_c=3$ we could further reduce the color basis by using the relation
\begin{align}
&\tr[abcd+dcba] + \tr[acdb+bdca] + \tr[adbc+cbda]
\nn\\ & \qquad
= \tr[ab]\tr[cd] + \tr[ac]\tr[db] + \tr[ad]\tr[bc]
\,.\end{align}
We choose not to do this, as it makes the structure more complicated, and because it does not hold for $N_c>3$.

In the case that there are two collinear gluon fields in each collinear sector, a basis of helicity operators is given by
\begin{align}
& \boldsymbol{(gg)_n (gg)_{\bn}:}{\vcenter{\includegraphics[width=0.18\columnwidth]{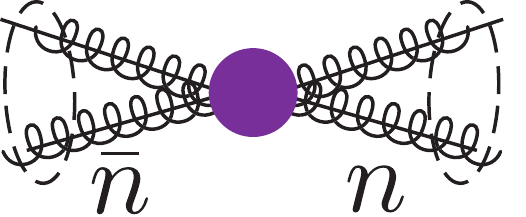}}}  \nn
\end{align}
\vspace{-0.4cm}
\begin{alignat}{2} \label{eq:H_basis_gggg_1}
&O_{4g1++++}^{(2)a b c d}
= \,\frac{1}{4}  \cB^a_{n +} \cB^b_{n +} \cB^c_{\bn +} \cB^d_{\bn +} \,H
\,,\qquad &
&O_{4g1+-+-}^{(2)a b c d}
= \, \cB^a_{n +} \cB^b_{n -} \cB^c_{\bn +} \cB^d_{\bn -} \,H
\,,\\
&O_{4g1----}^{(2)a b c d}
= \,\frac{1}{4}  \cB^a_{n -} \cB^b_{n -} \cB^c_{\bn -} \cB^d_{\bn -} \,H
\,.\qquad &\nn
\end{alignat}
The spin zero nature of the Higgs implies that a number of helicity configurations do not contribute, and therefore are not included in our basis operators here.
The color basis after BPS field redefinition is given by
\begin{equation}
\vT_{\text{BPS}}^{ abcd} =
\frac{1}{2}\begin{pmatrix}
(\tr[ T^{a'}   T^{b'}     T^{c'}   T^{d'}   ] + \tr[  T^{d'}   T^{c'}     T^{b'}   T^{a'}  ]) \cY^{a' a}_{n} \cY^{b' b}_{n} \cY^{c' c}_{\bn} \cY^{d' d}_{\bn}
\\
(\tr[ T^{a'}    T^{c'}   T^{d'}     T^{b'}  ] + \tr[ T^{b'}     T^{d'}   T^{c'}    T^{a'} ] ) \cY^{a' a}_{n} \cY^{b' b}_{n} \cY^{c' c}_{\bn} \cY^{d' d}_{\bn}
\\
(\tr[ T^{a'}    T^{d'}     T^{b'}     T^{c'}   ] + \tr[  T^{c'}     T^{b'}     T^{d'}    T^{a'}  ] ) \cY^{a' a}_{n} \cY^{b' b}_{n} \cY^{c' c}_{\bn} \cY^{d' d}_{\bn}
\\
(\tr[ T^{a'}   T^{b'}     T^{c'}   T^{d'}   ] - \tr[  T^{d'}   T^{c'}     T^{b'}   T^{a'}  ]) \cY^{a' a}_{n} \cY^{b' b}_{n} \cY^{c' c}_{\bn} \cY^{d' d}_{\bn}
\\
(\tr[ T^{a'}    T^{c'}   T^{d'}     T^{b'}  ] - \tr[ T^{b'}     T^{d'}   T^{c'}    T^{a'} ] ) \cY^{a' a}_{n} \cY^{b' b}_{n} \cY^{c' c}_{\bn} \cY^{d' d}_{\bn}
\\
(\tr[ T^{a'}    T^{d'}     T^{b'}     T^{c'}   ] - \tr[  T^{c'}     T^{b'}     T^{d'}    T^{a'}  ] ) \cY^{a' a}_{n} \cY^{b' b}_{n} \cY^{c' c}_{\bn} \cY^{d' d}_{\bn}
\\
\frac{1}{2}  \delta^{ab} \delta^{cd} 
\\
\frac{1}{2} (\cY^T_n \cY_\bn)^{ac} (\cY^T_n \cY_\bn)^{bd} 
\\
\frac{1}{2} (\cY^T_n \cY_\bn)^{ad} (\cY^T_n \cY_\bn)^{bc}
\end{pmatrix}^{\!\!\!T} 
.\end{equation}

The other relevant case has three gluons in one sector, which we take to be the $\bn$ collinear sector. The basis of operators is then given by
\begin{align}
& \boldsymbol{(g)_n (ggg)_{\bn}:}{\vcenter{\includegraphics[width=0.18\columnwidth]{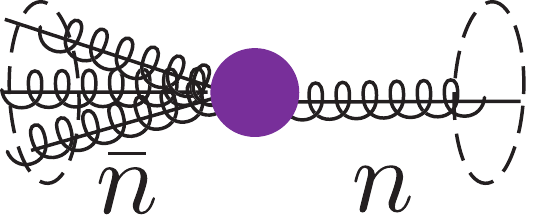}}}  \nn
\end{align}
\vspace{-0.4cm}
\begin{alignat}{2} \label{eq:H_basis_gggg_2}
&O_{4g2+++-}^{(2)a b c d}
= \,\frac{1}{2}  \cB^a_{n +} \cB^b_{\bn +} \cB^c_{\bn +} \cB^d_{\bn -} \,H
\,,\qquad &
&O_{4g2-+--}^{(2)a b c d}
= \,\frac{1}{2}  \cB^a_{n -} \cB^b_{\bn +} \cB^c_{\bn -} \cB^d_{\bn -} \,H
\,.
\end{alignat}
In this case, the post-BPS color basis is given by
\begin{equation}
\vT_{\text{BPS}}^{ abcd} =
\frac{1}{2}\begin{pmatrix}
(\tr[ T^{a'}   T^{b'}     T^{c'}   T^{d'}   ] + \tr[  T^{d'}   T^{c'}     T^{b'}   T^{a'}  ]) \cY^{a' a}_{n} \cY^{b' b}_{\bn} \cY^{c' c}_{\bn} \cY^{d' d}_{\bn}
\\
(\tr[ T^{a'}    T^{c'}   T^{d'}     T^{b'}  ] + \tr[ T^{b'}     T^{d'}   T^{c'}    T^{a'} ] ) \cY^{a' a}_{n} \cY^{b' b}_{\bn} \cY^{c' c}_{\bn} \cY^{d' d}_{\bn}
\\
(\tr[ T^{a'}    T^{d'}     T^{b'}     T^{c'}   ] + \tr[  T^{c'}     T^{b'}     T^{d'}    T^{a'}  ] ) \cY^{a' a}_{n} \cY^{b' b}_{\bn} \cY^{c' c}_{\bn} \cY^{d' d}_{\bn}
\\
(\tr[ T^{a'}   T^{b'}     T^{c'}   T^{d'}   ] - \tr[  T^{d'}   T^{c'}     T^{b'}   T^{a'}  ]) \cY^{a' a}_{n} \cY^{b' b}_{\bn} \cY^{c' c}_{\bn} \cY^{d' d}_{\bn}
\\
(\tr[ T^{a'}    T^{c'}   T^{d'}     T^{b'}  ] - \tr[ T^{b'}     T^{d'}   T^{c'}    T^{a'} ] ) \cY^{a' a}_{n} \cY^{b' b}_{\bn} \cY^{c' c}_{\bn} \cY^{d' d}_{\bn}
\\
(\tr[ T^{a'}    T^{d'}     T^{b'}     T^{c'}   ] - \tr[  T^{c'}     T^{b'}     T^{d'}    T^{a'}  ] ) \cY^{a' a}_{n} \cY^{b' b}_{\bn} \cY^{c' c}_{\bn} \cY^{d' d}_{\bn}
\\
\frac{1}{2}  (\cY^T_n \cY_\bn)^{ab} \delta^{cd} 
\\
\frac{1}{2} (\cY^T_n \cY_\bn)^{ac} \delta^{bd} 
\\
\frac{1}{2} (\cY^T_n \cY_\bn)^{ad} \delta^{bc}
\end{pmatrix}^{\!\!\!T} 
.\end{equation}

The helicity basis has made extremely simple the task of writing down a complete and minimal basis of four gluon operators, which would be much more difficult using traditional Lorentz structures. The helicity operators also make it simple to implement the constraints arising from the spin zero nature of the Higgs.

\vspace{0.4cm}
\noindent{\bf{Four Quark Operators:}}

We now consider the case of operators involving four collinear quark fields. These operators are again highly constrained by the helicity selection rules and chirality conservation, since each quark-antiquark pair was produced from a gluon splitting. In particular, these two constraints imply that there are no operators with non-vanishing Wilson coefficients with three quarks in one collinear sector. Therefore, we need only consider the cases where there are two quarks in each collinear sector.

When constructing the operator basis we must also treat separately the case of identical quark flavors $H q \bar q q\bar q$ and distinct quark flavors $H  q \bar q Q\bar Q $. For the case of distinct quark flavors $H q \bar q Q\bar Q $ we will have a $q\leftrightarrow Q$ symmetry for the operators. Furthermore the two quarks of flavor $q$, and the two quarks of flavor $\bar Q$, are necessarily of the same chirality.  In the case that both quarks of the same flavor appear in the same current, the current will be labeled by the flavor. Otherwise, the current will be labeled with ($q \bar{Q}$) or ($Q\bar{q}$) appropriately. For all these cases, the color basis is
\begin{equation} \label{eq:qqqq_color}
\vT^{\, \al\bbeta\ga\bdelta} =
\Bigl(
\de_{\al\bdelta}\, \de_{\ga\bbeta}\,,\, \delta_{\al\bbeta}\, \de_{\ga\bdelta}
\Bigr)
\,.\end{equation}
We will give results for the corresponding $\bar T_{\rm BPS}^{\, \al\bbeta\ga\bdelta}$ basis as we consider each case below.

For the case of operators with distinct quark flavors $H q \bar q Q\bar Q $ and two collinear quarks in each of the $n$ and $\bn$ sectors there are three possibilities. There is either a quark anti-quark pair of the same flavor in each sector (e.g. $(q \bar q)_n(Q \bar Q)_\bn$), a quark and an antiquark of distinct flavors in the same sector (e.g. $(q \bar Q)_n(Q \bar q)_\bn$), or two quarks with distinct flavors in the same sector(e.g. $(q Q)_n(\bar q \bar Q)_\bn$). In the case that there is a quark anti-quark pair of the same flavor in each sector, the basis of helicity operators is
\vspace{0.3cm}\begin{align}
 \boldsymbol{(q \bar q)_n(Q \bar Q)_\bn:}    {\vcenter{\includegraphics[width=0.18\columnwidth]{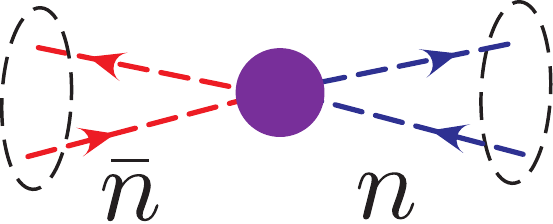}}}
\nn
\end{align}
\vspace{-0.4cm}
\begin{alignat}{2} \label{eq:Z2_basis_qQ}
&O_{qQ1(0;0)}^{(2)\balpha\bt\bgamma\delta}
= \, J_{(q)n 0\, }^{\balpha\bt}\, J_{(Q) \bar n 0\, }^{\bgamma\delta} H
\,,\qquad &
&O_{qQ1(0;\bar 0)}^{(2)\balpha\bt\bgamma\delta}
= \, J_{(q) n  0\, }^{\balpha\bt}\, J_{(Q) \bar n \bar 0\,}^{\bgamma\delta} H
\,,\\
&O_{qQ1(\bar 0;0)}^{(2)\balpha\bt\bgamma\delta}
= \,  J_{(q)n  \bar 0\, }^{\balpha\bt}\, J_{(Q) \bar n 0\, }^{\bgamma\delta} H
\,,\qquad &
&O_{qQ1(\bar 0;\bar 0)}^{(2)\balpha\bt\bgamma\delta}
=\, J_{(q) n \bar 0\, }^{\balpha\bt}\, J_{(Q) \bar n \bar 0\, }^{\bgamma\delta} H
\,,\nn
\end{alignat}
where we have chosen the $q$ quark to be in the $n$ sector. Since all the operators have total helicity $0$ along the $\hat n$ direction, there are only chirality constraints here and no constraints  from angular momentum conservation.  In the case that there is a quark anti-quark of distinct flavors in the same sector, chirality and angular momentum conservation constrains the basis to be 
\begin{align}
\boldsymbol{(q \bar Q)_n(Q \bar q)_\bn:}     {\vcenter{\includegraphics[width=0.18\columnwidth]{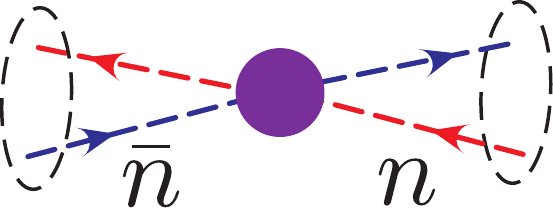}}}
\nn
\end{align}
\vspace{-0.4cm}
\begin{align} \label{eq:Z2_basis_qQ_2}
&O_{qQ2(0;0)}^{(2)\balpha\bt\bgamma\delta}
=\, J_{(q \bar{Q}) n 0\, }^{\balpha\bt}\, J_{(Q\bar{q} ) \bar{n} 0\, }^{\bgamma\delta} H
\,,\qquad 
O_{qQ2(\bar 0;\bar 0)}^{(2)\balpha\bt\bgamma\delta}
=\, J_{(q \bar{Q}) n \bar 0\, }^{\balpha\bt}\, J_{(Q\bar{q} )\bar{n} \bar 0\, }^{\bgamma\delta} H
\,.
\end{align}
For the operators in \eqs{Z2_basis_qQ}{Z2_basis_qQ_2} the color basis after BPS field redefinition is
\begin{align}  \label{eq:TBPS_OqQ12}
\vT_{\BPS}^{ \al\bbeta\ga\bdelta} &=
\left( \Big[ Y_{n}^\dagger Y_{\bar n}  \Big]_{\al\bdelta}\,\Big[ Y_{\bar n}^\dagger Y_{n}  \Big]_{\ga\bbeta}\,,\, \delta_{\al\bbeta}\, \delta_{\ga\bdelta}
\right)
\,.
\end{align}
When there are two quarks of distinct flavors in the same sector the basis of helicity operators is constrained by chirality and reduced further to just two operators by angular momentum conservation, giving 
\begin{align}
\boldsymbol{(q Q)_n(\bar q \bar Q)_\bn:}    {\vcenter{\includegraphics[width=0.18\columnwidth]{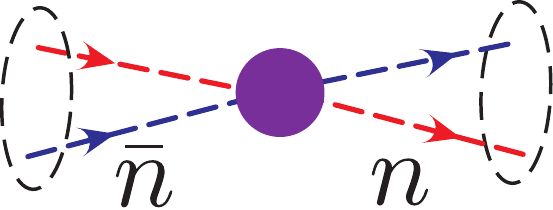}}}
\nn
\end{align}
\vspace{-0.4cm}
\begin{align}\label{eq:Z2_basis_qQ_3}
&O_{qQ3(+;-)}^{(2)\balpha\bt\bgamma\delta}
=\, J_{(q) n \bar n +\, }^{\balpha\bt}\, J_{(Q) n\bar n -\, }^{\bgamma\delta} H
\,,\qquad &
O_{qQ3(-;+)}^{(2)\balpha\bt\bgamma\delta}
= \, J_{(q)n \bar n -\, }^{\balpha\bt}\, J_{(Q) n\bar n +\, }^{\bgamma\delta} H
\,.
\end{align}
For the operators in \eq{Z2_basis_qQ_3} the color basis after BPS field redefinition is
\begin{align} \label{eq:TBPS_OqQ3}
\vT_{\BPS}^{ \al\bbeta\ga\bdelta} &=
\left( \left[ Y_{n}^\dagger Y_{\bar n}  \right ]_{\al\bdelta}\,\left[ Y_{n}^\dagger Y_{\bar n}  \right ]_{\ga\bbeta}
 \,,\, 
 \left[ Y_{n}^\dagger Y_{\bar n}  \right ]_{\al\bbeta}\,\left[ Y_{n}^\dagger Y_{\bar n}  \right ]_{\ga\bdelta}
\right)
\,.
\end{align}

In the cases considered in \eqs{Z2_basis_qQ}{Z2_basis_qQ_2} where there is a quark and antiquark field in the same collinear sector, we have chosen to work in a basis using $J_{i0}^{\balpha\beta}$ and $J_{i\bar 0}^{\balpha\beta}$ which contain only fields in a single collinear sector. One could also construct an alternate form for the basis, for example using the currents $J_{n\bn\lambda}^{\balpha\beta}$. From the point of view of factorization our basis is the most convenient since the fields in the $n$ and $\bar n$-collinear sectors are only connected by color indices, which will simplify later steps of factorization proofs. In the following, we will whenever possible use this logic when deciding between equivalent choices for our basis. 

For identical quark flavors the operators are similar to those in Eqs.~(\ref{eq:Z2_basis_qQ},\ref{eq:Z2_basis_qQ_3}). The distinct operators include
\begin{align}
 \boldsymbol{(q \bar q)_n(q \bar q)_\bn:}    {\vcenter{\includegraphics[width=0.18\columnwidth]{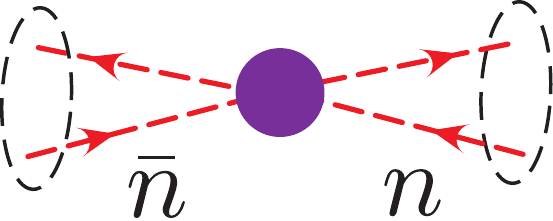}}}
\nn
\end{align}
\vspace{-0.4cm}
\begin{alignat}{2} \label{eq:Z2_basis_qq}
&O_{qq1(0;0)}^{(2)\balpha\bt\bgamma\delta}
= \,\frac14\, J_{(q)n 0\, }^{\balpha\bt}\, J_{(q) \bar n 0\, }^{\bgamma\delta} H
\,,\qquad &
&\phantom{O_{qq1(0;\bar 0)}^{(2)\balpha\bt\bgamma\delta}
= \, J_{(q) n  0\, }^{\balpha\bt}\, J_{(q) \bar n \bar 0\,}^{\bgamma\delta} H
\,,}\\
&O_{qq1(\bar 0;0)}^{(2)\balpha\bt\bgamma\delta}
= \,  J_{(q)n  \bar 0\, }^{\balpha\bt}\, J_{(q) \bar n 0\, }^{\bgamma\delta} H
\,,\qquad &
&O_{qq1(\bar 0;\bar 0)}^{(2)\balpha\bt\bgamma\delta}
=\, \frac14\, J_{(q) n \bar 0\, }^{\balpha\bt}\, J_{(q) \bar n \bar 0\, }^{\bgamma\delta} H
\,,\nn
\end{alignat}
\begin{align}
\boldsymbol{(q q)_n(\bar q \bar q)_\bn:}    {\vcenter{\includegraphics[width=0.18\columnwidth]{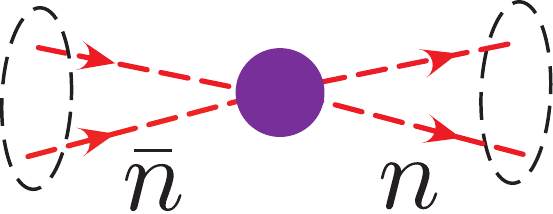}}}
\nn
\end{align}
\vspace{-0.4cm}
\begin{align}\label{eq:Z2_basis_qq_3}
&O_{qq3(+;-)}^{(2)\balpha\bt\bgamma\delta}
=\, J_{(q) n \bar n +\, }^{\balpha\bt}\, J_{(q) n\bar n -\, }^{\bgamma\delta} H
\,.
\end{align}
Note that in \Eq{eq:Z2_basis_qq} there are only three operators due to the equivalence between the two operators
\begin{align}
 \sum_n J_{(q) n  0\, }^{\balpha\bt}\, J_{(q) \bar n \bar 0\,}^{\bgamma\delta} H 
\equiv \,  \sum_n J_{(q)n  \bar 0\, }^{\balpha\bt}\, J_{(q) \bar n 0\, }^{\bgamma\delta} H
\,,
\end{align}
due to the fact that the $n$ label is summed over, as in \Eq{eq:sum_dir}.
We also have the same color bases as in \eqs{TBPS_OqQ12}{TBPS_OqQ3} for $O_{qq1}^{(2)}$ and $O_{qq3}^{(2)}$ respectively.

\subsubsection{$\cP_\perp$ Insertions}\label{sec:nnlp_perp}

Since we have chosen to work in a frame where the total $\perp$ momentum of each collinear sector vanishes, operators involving explicit insertions of the $\cP_\perp$ operator first appear at $\cO(\lambda^2)$. The $\cP_\perp$ operator can act only in a collinear sector composed of two or more fields. At $\cO(\lambda^2)$, there are then only two possibilities, namely that the $\cP_\perp$ operator is inserted into an operator involving two quark fields and a gluon field, or it is inserted into an operator involving three gluon fields. 

In the case that the $\cP_\perp$ operator is inserted into an operator involving two quark fields and a gluon field, the helicity structure of the operator is highly constrained. In particular, the quark fields must be in a helicity zero configuration. Combined with the fact that they must have the same chirality, this implies that all operators must involve only the currents $J_{\bar n\, 0}^{\balpha\bt}$ or $J_{\bar n\, \bar 0}^{\balpha\bt}$. Here we have again taken without loss of generality that the two quarks are in the $\bar n$-collinear sector. A basis of operators is then given by
\begin{align}
&   \boldsymbol{(g)_n (q\bar q\, \cP_\perp)_{\bn}:}{\vcenter{\includegraphics[width=0.18\columnwidth]{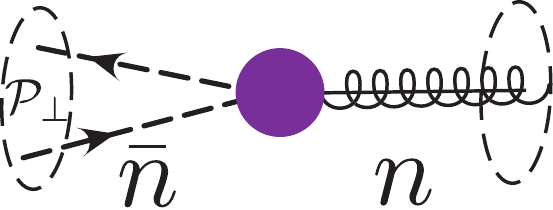}}}  \nn
\end{align}
\vspace{-0.4cm}
\begin{alignat}{2}\label{eq:Hqqgpperp_basis_same}
&O_{\cP \chi + (0)[+]}^{(2)a\,\balpha\bt}
= \cB_{n+}^a\, \big\{ \cP_{\perp}^{+} J_{\bar n\, 0}^{\balpha\bt} \big\}\,  H
\,,\qquad &
&O_{\cP\chi - (0)[-]}^{(2)a\,\balpha\bt}
= \cB_{n-}^a\, \big\{ \cP_{\perp}^{-} J_{\bar n\, 0    }^{\balpha\bt} \big\} \, H 
\,,\\
&O_{\cP\chi + (\bar 0)[+]}^{(2)a\,\balpha\bt}
=  \cB_{n+}^a\,  \big\{ \cP_{\perp}^{+} J_{\bar n\, \bar0    }^{\balpha\bt} \big\}\,  H
\,,\qquad &
&O_{\cP\chi - (\bar 0)[-]}^{(2)a\,\balpha\bt}
= \cB_{n-}^a \, \big\{ \cP_{\perp}^{-} J_{\bar n\, \bar0    }^{\balpha\bt} \big\}  \, H
\,.\nn
\end{alignat}
Since we have assumed that the total $\cP_\perp$ in each collinear sector is zero, integration by parts can be used to make the $\cP_\perp$ operator act only on either the quark, or the antiquark field, which has been used in \Eq{eq:Hqqgpperp_basis_same}. (The additional operators that are needed when we relax this assumption are discussed in \app{gen_pt}.)
The color basis is one-dimensional
\begin{equation} \label{eq:nnlp_color_quark_perp}
\vT^{a\, \al\bbeta} = T^a_{\al\bbeta}\,.
\end{equation}
After BPS field redefinition the structure is given by
\begin{align} \label{eq:nnlp_color_quark_perpBPS}
\vT_{\BPS}^{ a \al\bbeta} 
=\left ( Y^\dagger_{\bar n} T^b \cY_n^{ba}   Y_{\bar n}  \right )_{\alpha \bar \beta}
= \bigl( \cY_n^T \cY_\bn \bigr)^{ac}\: T^c_{\alpha \bar \beta} 
\,.
\end{align}

In the case that the $\cP_\perp$ operator is inserted into an operator involving three gluon fields, the helicity selection rules simply imply that the helicities must add to zero.
A basis of operators involving three collinear gluon fields and a $\cP_\perp^\pm$ insertion is given by
\begin{align}
&  \boldsymbol{(g)_n (gg\, \cP_\perp)_{\bn}:}{\vcenter{\includegraphics[width=0.18\columnwidth]{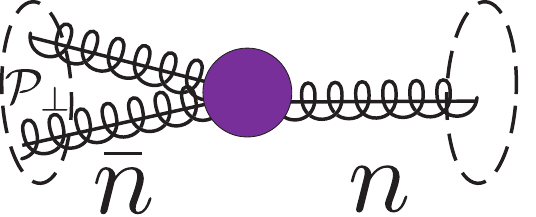}}}  \nn
\end{align}
\vspace{-0.4cm}
\begin{alignat}{2} \label{eq:Hgggpperp_basis}
&O_{\cP\cB +++[-]}^{(2)abc}
=  \cB_{n+}^a\, \cB_{\bar n+}^b\, \left [\cP_{\perp}^{-} \cB_{\bar n+}^c \right ] \,H
\,,\qquad && O_{\cP\cB ---[+]}^{(2)abc}
= \cB_{n-}^a\, \cB_{\bar n-}^b\, \left [\cP_{\perp}^{+} \cB_{\bar n-}^c \right ] \,H\,,
 \nn \\
&O_{\cP\cB ++-[+]}^{(2)abc}
= \, \cB_{n+}^a\, \cB_{ \bar n+}^b\, \left [\cP_{\perp}^{+} \cB_{\bar n-}^c \right ] \,H
\,,\qquad 
&&O_{\cP\cB --+[-]}^{(2)abc}
= \cB_{n-}^a\, \cB_{ \bn -}^b\, \left [\cP_{\perp}^{-} \cB_{\bar n +}^c \right ] \,H
\,.
\end{alignat}
Note that the analogous operators with the helicities $O_{\cP\cB +-+[+]}^{(2)abc}$ and $O_{\cP\cB -+-[-]}^{(2)abc}$ are not eliminated, but instead are equivalent to those in the last row by integrating the $\cP_\perp^\pm$  by parts onto the other $\bn$-collinear field since the total $\cP_\perp$ in each collinear sector is zero. (The additional operators that are needed when we relax this assumption are discussed in \app{gen_pt}.)

The basis of color structures here is two dimensional,
\begin{equation} \label{eq:ggg_perp_color}
\vT^{abc} =
\begin{pmatrix}
  i  f^{abc} \\ d^{abc}
\end{pmatrix}
\,,
\qquad
\vT_{\BPS}^{abc} =
\begin{pmatrix}
  i  f^{a'b'c'}\, {\cal Y}_n^{a'a} {\cal Y}_\bn^{b'b} {\cal Y}_{\bn}^{c'c}  \\  
  d^{a'b'c'}\, {\cal Y}_n^{a'a} {\cal Y}_\bn^{b'b} {\cal Y}_{\bn}^{c'c} 
\end{pmatrix}
 =
\begin{pmatrix}
   i f^{bcd} {\cal Y}_{\bn}^{a'd} {\cal Y}_n^{a'a}   \\  
  d^{bcd}\, {\cal Y}_\bn^{a'd}  {\cal Y}_n^{a'a}  
\end{pmatrix}
\,.
\end{equation}
In the BPS redefined color structure we have written it both in a form that makes the structure of the Wilson lines appearing from the field redefinition clear, as well as in a simplified form.

\subsubsection{Ultrasoft Insertions}\label{sec:nnlp_soft}

At $\cO(\lambda^2)$ we have the possibility of operators with explicit ultrasoft insertions. To have label momentum conservation these operators must have a collinear field in each collinear sector. Interestingly, despite the fact that the leading power operator has two collinear gluon fields, for the operators involving an ultrasoft insertion one can have either two collinear quark fields, or two collinear gluon fields.

The construction of an operator basis involving ultrasoft gluons is more complicated due to the fact that they are not naturally associated with a given lightcone direction. There are therefore different choices that can be made when constructing the basis. We will choose to work in a basis where all ultrasoft derivatives acting on ultrasoft Wilson lines are absorbed into $\cB_{us}$ fields. To understand why it is always possible to make this choice, we consider two pre-BPS operators involving two collinear quark fields, and an ultrasoft derivative
\begin{align}
O^\mu_1=\bar \chi_{\bar n} (i D_{us}^\mu) \chi_n\,, \qquad O^\mu_2=\bar \chi_{\bar n} (-i \overleftarrow D_{us}^\mu) \chi_n\,,
\end{align}
where $(-i \overleftarrow D_{us}^\mu)=(i D_{us}^\mu)^\dagger$ and we have not made the contraction of the $\mu$ index explicit, as it is irrelevant to the current discussion. Performing the BPS field redefinition, we obtain 
\begin{align}
O^\mu_{1\text{BPS}}= i \bar \chi_{\bar n}Y_{\bar n}^\dagger D_{us}^\mu Y_n \chi_n\,, \qquad
 O^\mu_{2\text{BPS}}=-i \bar \chi_{\bar n} Y_{\bar n}^\dagger \overleftarrow D_{us}^\mu Y_n \chi_n\,.
\end{align}
If we want to absorb all derivatives acting on Wilson lines into $\cB_{us}$ fields, we must organize the Wilson lines in the operators as
\begin{align}
O^\mu_{1\text{BPS}}= i\bar \chi_{\bar n}Y_{\bar n}^\dagger Y_n (Y_n^\dagger  D_{us}^\mu Y_n) \chi_n\,, \qquad 
O^\mu_{2\text{BPS}}=-i \bar \chi_{\bar n} (Y_{\bar n}^\dagger \overleftarrow D_{us}^\mu Y_{\bar n}) Y_{\bar n}^\dagger  Y_n \chi_n\,.
\end{align}
Using \Eq{eq:soft_gluon} we see that this can be written entirely in terms of $\partial_{us}$ operators acting on collinear fields, and the two ultrasoft gauge invariant gluon fields $\cB_{us(n)}$ and $\cB_{us(\bar n)}$ for $O^\mu_{1\text{BPS}}$ and $O^\mu_{2\text{BPS}}$ respectively. Note, however, that ultrasoft gluon fields defined with respect to both lightcone directions are required. Alternatively, it is possible to work only with $\cB_{us(n)}$, for example, but in this case we see that the ultrasoft derivative must also be allowed to act explicitly on pairs of ultrasoft Wilson lines, for example $[\partial_{us}^\mu (Y_n^\dagger Y_\bn)]$. In constructing our complete basis we will choose to avoid this so that ultrasoft derivatives acting on soft Wilson lines occur only within the explicit $\cB_{us}$ fields. This choice also makes our basis more symmetric.

For the operators involving one ultrasoft gluon and two collinear quarks we have the basis
\begin{align}
&  \boldsymbol{g_{us}(q)_n (\bar q)_{\bn}:}{\vcenter{\includegraphics[width=0.18\columnwidth]{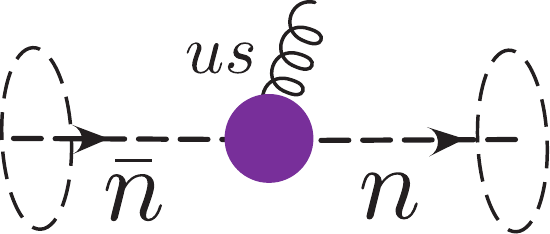}}}  \nn
\end{align}
\vspace{-0.4cm}
\begin{alignat}{2} \label{eq:soft_insert_basis}
&O_{\chi(us(n))-:(+)}^{(2)a\,\balpha\bt}
=  \cB_{us(n)-}^a\, J_{n\bar n\,+}^{\balpha\bt}\, H
\,,\qquad &
&O_{\chi(us(n))+:(-)}^{(2)a\,\balpha\bt}
=\cB_{us(n)+}^a \, J_{n\bar n\, -}^{\balpha\bt}\, H
\,, 
\end{alignat}
with the unique color structure
\begin{equation} 
\vT_{\BPS}^{\,a\, \al\bbeta} = \left ( T^a Y^\dagger_{n} Y_{\bn} \right )_{\alpha \bar\beta}
\,,\end{equation}
and
\begin{alignat}{2} \label{eq:soft_insert_basis2}
&O_{\chi(us(\bar n))+:(+)}^{(2)a\,\balpha\bt}
=  \cB_{us(\bar n)+}^a\, J_{n\bar n\,+}^{\balpha\bt}\, H
\,,\qquad &
&O_{\chi(us)(\bar n))-:(-)}^{(2)a\,\balpha\bt}
=\cB_{us(\bar n)-}^a \, J_{n\bar n\, -}^{\balpha\bt}\, H
\,, 
\end{alignat}
with the unique color structure
\begin{equation} 
\vT_{\BPS}^{\,a\, \al\bbeta}=(Y_n^\dagger Y_\bn T^a)_{\alpha\bbeta}
\,.\end{equation}
Note that the color structures associated with the two different projections of the $\cB_{us}$ field are distinct. All other helicity combinations vanish due to helicity selection rules. The helicity selection rules differ between \Eq{eq:soft_insert_basis} and \Eq{eq:soft_insert_basis2} due to the different choice of reference vector for the ultrasoft field in the two cases.

We also have operators involving two collinear quark fields and a single ultrasoft derivative, 
\begin{align}
&  \boldsymbol{\partial_{us} (q)_n (\bar q)_{\bn}:}{\vcenter{\includegraphics[width=0.18\columnwidth]{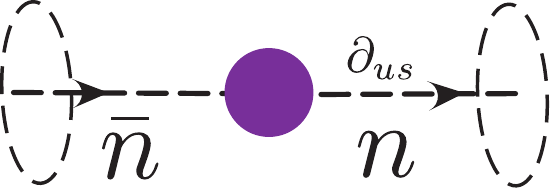}}}  \nn
\end{align}
\vspace{-0.4cm}
\begin{alignat}{2} \label{eq:soft_derivative_basis}
&O_{\partial \chi (us(n))-:(+)}^{(2)\,\balpha\bt}
=  \{\partial_{us(n)-}\, J_{n\bar n\,+}^{\balpha\bt}\}\, H
\,,\qquad &
&O_{\partial \chi (us(n))+:(-)}^{(2)\,\balpha\bt}
=  \{\partial_{us(n)+} \, J_{n\bar n\, -}^{\balpha\bt}\}\, H\,,
\end{alignat}
with the unique color structure given before and after BPS field redefinition by
\begin{align} \label{eq:leading_color_deriv}
 \vT^{\al\bbeta} = (\de_{\al\bbeta})\,, \qquad 
 \vT_{\BPS}^{\al\bbeta} =  \big[Y_n^\dagger Y_{\bar n} \big]_{\al\bbeta} 
\,,
\end{align}
and
\begin{alignat}{2} \label{eq:soft_derivative_basis2}
&O_{\partial^\dagger \chi (us(\bar n))+:(+)}^{(2)\,\balpha\bt}
=  \{ J_{n\bar n\,+}^{\balpha\bt}\,     (i\partial_{us(\bar n)+})^\dagger\}\, H
\,,~ &
&O_{\partial^\dagger \chi (us(\bar n))-:(-)}^{(2)\,\balpha\bt}
=  \{  J_{n\bar n\, -}^{\balpha\bt}\, (i\partial_{us(\bar n)-})^\dagger\}\, H\,,
\end{alignat}
with the unique color structure given before and after BPS field redefinition by
\begin{align} \label{eq:leading_color_deriv2}
 \vT^{\al\bbeta} = (\de_{\al\bbeta})\,, \qquad 
 \vT_{\BPS}^{\al\bbeta} =  \big[Y_n^\dagger Y_{\bar n} \big]_{\al\bbeta} 
\,.
\end{align}
Although the color structure happens to be the same in both cases, we have separated them to highlight the different decompositions of the ultrasoft derivatives in the two cases. Note that the form of the ultrasoft derivatives which appear is constrained by the helicity constraints.

Similarly, we have the corresponding operators involving two collinear gluons. A basis of helicity operators involving two collinear gluons and a single ultrasoft gluon field is given by
\begin{align}
& \boldsymbol{g_{us}(g)_n (g)_{\bn}:}{\vcenter{\includegraphics[width=0.18\columnwidth]{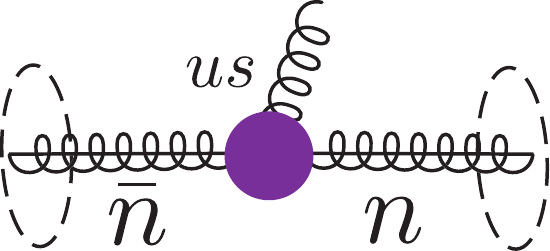}}}  \nn
\end{align}
\vspace{-0.4cm}
\begin{align}\label{eq:Hgggus}
  &O_{\cB (us(n))0:++}^{(2)abc}
  =  \cB_{us(n)0}^a\, \cB_{n+}^b\, \cB_{\bar n+}^c\,  H
  \,, \qquad && O_{\cB (us(n))0:--}^{(2)abc} 
  =   \cB_{ us(n)0}^a\, \cB_{n-}^b\, \cB_{\bar n-}^c\,  H
  \,, 
\end{align}
with the basis of color structures,\footnote{In order to see how the Wilson line structure in \Eq{eq:Z2g_colorus} arises, we look at the object $D_{us}^{ab} \cB_{n}^c \cB_{\bn}^d$ pre-BPS field redefinitions. This object must be contracted with a tensor to make it a singlet under ultrasoft gauge transformations. Each of these resulting forms can be mapped onto the color structures of \Eq{eq:Z2g_colorus} after performing the BPS field redefinition} 
\begin{equation} \label{eq:Z2g_colorus}
\vT_{\BPS}^{abc} =
\begin{pmatrix}
i  f^{abd}\, \big({\cal Y}_n^T {\cal Y}_{\bar n}\big)^{dc} \\
 d^{abd}\, \big({\cal Y}_n^T {\cal Y}_{\bar n}\big)^{dc} 
\end{pmatrix}^T
\,,
\end{equation}
and
\begin{align}\label{eq:Hgggus_2}
  &O_{\cB (us(\bar n))0:++}^{(2)abc}
  =  \cB_{us(\bar n)0}^a\, \cB_{n+}^b\, \cB_{\bar n+}^c\,  H
  \,, \qquad && O_{\cB (us(\bar n))0:--}^{(2)abc} 
  =   \cB_{ us(\bar n)0}^a\, \cB_{n-}^b\, \cB_{\bar n-}^c\,  H
  \,, 
\end{align}
with the basis of color structures,
\begin{equation} \label{eq:Z2g_colorus_2}
\vT_{\BPS}^{abc} =
\begin{pmatrix}
i  f^{abd}\, \big({\cal Y}_{\bar n}^T {\cal Y}_{n}\big)^{dc} \\
 d^{abd}\, \big({\cal Y}_{\bar n}^T {\cal Y}_{n}\big)^{dc} 
\end{pmatrix}^T
\,.
\end{equation}
We have only included the $\vT_{\BPS}^{abc}$ version of the color structure here because the $\cB_{us(n)\lambda}^a$ are generated by BPS field redefinition. 

The Wilson coefficients of the operators that include $\cB_{us(n)0}$ can be related to the Wilson coefficients of the leading power operators using RPI symmetry (see \cite{Larkoski:2014bxa}). In particular, we have 
\begin{align} \label{eq:usRPIrelation}
C^{(2)}_{\cB n(us)0:\lambda_1, \lambda_1}&=-\frac{\partial C^{(0)}_{\lambda_1, \lambda_1} }{\partial \omega_1}  
\,, 
\end{align}
where $C^{(0)}_{\lambda_1, \lambda_1}$ is the Wilson coefficient for the leading power operator of \Eq{eq:hgg}. We will explicitly verify this at the level of tree level matching in \Sec{sec:matching}.

We must also consider operators with an insertion of $\partial_{us(n)}$ with two collinear gluons in different collinear sectors. The gluon equations of motion allow us to eliminate the operators $in\cdot \partial \cB_{n\perp}$ and $i\bar n\cdot \partial \cB_{\bar n\perp}$, which can be rewritten purely in terms of collinear objects~\cite{Marcantonini:2008qn}. Furthermore, we again choose to organize our basis of operators such that ultrasoft derivatives act on ultrasoft Wilson lines only within the $\cB_{us}$ fields, as was done in the quark case. (We also do not include operators where the ultrasoft derivative acts on the Higgs field, since this is moved to the other fields by integration by parts.) The basis of operators involving ultrasoft derivatives is then given by
\begin{align}
& \boldsymbol{\partial_{us}(g)_n (g)_{\bn}:}{\vcenter{\includegraphics[width=0.18\columnwidth]{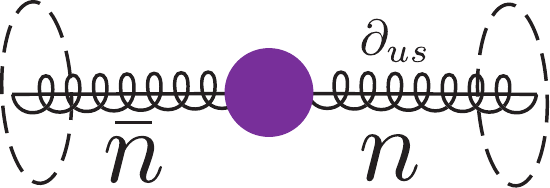}}}  \nn
\end{align}
\vspace{-0.4cm}
\begin{alignat}{2}\label{eq:Hdggus}
&O_{\partial \cB (us(n))\bar 0:++}^{(2)ab}
=  \cB_{n+}^a  \, \left[ \partial_{us(n)\bar 0} \cB_{\bar n+}^b\right] \,H
\,, \,~~ && O_{\partial \cB (us(n))\bar 0:--}^{(2)ab} 
=    \cB_{n-}^a \left[\partial_{us(n)\bar 0} \cB_{\bar n-}^b\right]\, H
\,,
\end{alignat}
with the basis of color structures
\begin{equation} \label{eq:Z2gd_colorus}
\vT_{\BPS}^{ab} =
\big({\cal Y}_n^T {\cal Y}_{\bar n}\big)^{ab} 
\,.
\end{equation}
and
\begin{alignat}{2}\label{eq:Hdggus2}
&O_{\partial \cB  (us(\bar n))0:++}^{(2)ab}
=  \left[ \partial_{us(\bar n)0}\, \cB_{n+}^a\right]\, \cB_{\bar n+}^b\,  H
\,, \,~~ && O_{\partial \cB  (us(\bar n))0:--}^{(2)ab} 
=  \left [ \partial_{us(\bar n)0}\, \cB_{n-}^a \right] \, \cB_{\bar n-}^b\,  H
\,,
\end{alignat}
with the basis of color structures
\begin{equation} \label{eq:Z2gd_colorus2}
\vT_{\BPS}^{ab} =
\big({\cal Y}_{\bar n}^T {\cal Y}_{n}\big)^{ab} 
\,.
\end{equation}

The Wilson coefficients of the operators that include $\partial_{us(n)0}$ can also be related to the Wilson coefficients of the leading power operators using RPI symmetry (see \cite{Larkoski:2014bxa}). In particular, we have 
\begin{align} \label{eq:usRPIrelationb}
C^{(2)}_{\partial \cB (us(\bar n)) 0:\lambda_1 \lambda_1}&=-\frac{\partial C^{(0)}_{\lambda_1,\lambda_1} }{\partial \omega_1}  
\,, 
\end{align}
where $C^{(0)}_{\lambda_1,\lambda_1}$ is the Wilson coefficient for the leading power operator of \Eq{eq:hgg}. We will explicitly show how this arises in the tree level matching in \Sec{sec:matching}.

\subsection{Cross Section Contributions and Factorization}\label{sec:discussion}

{
\renewcommand{\arraystretch}{1.6}
	\begin{center}
\begin{table}[t!]
	\begin{center}
		\begin{tabular}{| c | l | c | c | c | c | }
			\hline
			& Operators & Factorization  & Beam $n$ &  Beam $\bar n$ & Soft  \\
			\hline 
			$\!\!\mathbf{\mathcal{O}(\lambda^0)}\!\!$ 
			&$O_{\cB}^{(0)}  O_{\cB}^{(0)} $ 
			& $H_g^{(0)} B_{g}^{(0)} B_{g}^{(0)} S_g^{(0)}$
			& $\cB_n \,\hat\delta\, \cB_n$ 
			& $\cB_\bn \,\hat\delta\, \cB_\bn$
			& $\cY_n^T \cY_\bn \widehat\cM^{(0)}\, \cY_\bn^T \cY_n$ 
			\\
			\hline
			$\!\!\mathbf{\mathcal{O}(\lambda^2)}\!\!$ 
			&$O_{\cB\bar n}^{(1)}  O_{\cB\bar n}^{(1)}$
			& $H_{g1}^{(0)} B_q^{(0)} B_{qgg}^{(2)} S_{q}^{(0)}$  
			& $\bar \chi_n \,\hat\delta\,\chi_n $ 
			& $\bar \chi_\bn \cB_\bn \hat\delta\, \cB_\bn\chi_\bn   $   
			& $Y_\bn^\dagger Y_n \widehat\cM^{(0)}\,   Y_n^\dagger Y_\bn$   
			\\
			\cline{2-6}
			&$O^{(0)}  O_{\cB1}^{(2)} $
			& $H_{g2}^{(0)} B_{gqq}^{(2)} B_{g}^{(0)} S_{g}^{(0)}$  
			& $\!\!\bar \chi_n \cB_n  \chi_n \hat\delta\, \cB_n\!\!  $ 
			& $ \cB_\bn \,\hat\delta\, \cB_\bn $   
			& $\cY_n^T \cY_\bn \widehat\cM^{(0)}\, \cY_\bn^T \cY_n$    
			\\
			\cline{2-6}
			&$O^{(0)}  O_{\cP\chi}^{(2)} $
			& $H_{g3}^{(0)} B_g^{(0)} B_{gq P}^{(2)} S_{g}^{(0)}$  
			& $\cB_n \,\hat\delta\,\cB_n $ 
			& $\!\!\bar \chi_\bn [\cP_\perp \chi_\bn] \hat\delta\, \cB_\bn\!\!   $   
			& $\cY_n^T \cY_\bn \widehat\cM^{(0)}\, \cY_\bn^T \cY_n$   
			\\
			\cline{2-6}
			&$O^{(0)} O_{\cP \cB}^{(2)} $
			& $H_{g4}^{(0)} B_{g}^{(0)} B_{gg P}^{(2)} S_{g}^{(0)}$  
			& $\bar \cB_n \,\hat\delta\, \cB_n  $ 
			& $\!\!\cB_\bn [\cP_\perp \cB_\bn] \hat\delta\, \cB_\bn  \!\! $   
			& $\cY_n^T \cY_\bn \widehat\cM^{(0)}\, \cY_\bn^T \cY_n$   
			\\
			\cline{2-6}
			&$O^{(0)}  O_{4g2}^{(2)}$
			& $H_{g5}^{(0)} B_g^{(0)} B_{gg}^{(2)} S_{g}^{(0)}$  
			& $\cB_n \,\hat\delta\,\cB_n $ 
			& $\!\! \cB_\bn \cB_\bn \cB_\bn \hat\delta\, \cB_\bn \!\! $   
			& $\cY_n^T \cY_\bn \widehat\cM^{(0)}\, \cY_\bn^T \cY_n$
			\\
			\cline{2-6}
			&$\!\! O^{(0)}  O_{\cB (us)0}^{(2)} \!\!\!$
			& $H_{g6}^{(0)} B_g^{(0)} B_{g}^{(0)} S_{g\cB}^{(2)}$  
			&  $\cB_n \,\hat\delta\,\cB_n   $	 
			&$\cB_\bn \,\hat\delta\,\cB_\bn   $  
			& $\!\cB_{us(n) 0}\, \cY_n\cY_\bn \widehat\cM^{(0)}\, \cY_\bn \cY_n\!\!$     
			\\
			\cline{2-6}
			& $\!\!O^{(0)}  O_{\partial(us)0}^{(2)} \!\!\!$
			& $H_{g7}^{(0)} B_g^{(0)} B_{g}^{(0)} S_{g\partial 0}^{(2)}$  
			&  $\cB_n \,\hat\delta\,\cB_n   $	 
			&$\cB_\bn \,\hat\delta\,\cB_\bn   $  
			& $\!\partial_{us(n) 0}\, \cY_n\cY_\bn \widehat\cM^{(0)}\, \cY_\bn \cY_n\!\!$     
			\\	
			\cline{2-6}
			& $\!\!O^{(0)}  O_{\partial(us)\bar 0}^{(2)} \!\!\!$
			& $H_{g8}^{(0)} B_g^{(0)} B_{g}^{(0)} S_{g\partial \bar 0}^{(2)}$  
			&  $\cB_n \,\hat\delta\,\cB_n   $	 
			&$\cB_\bn \,\hat\delta\,\cB_\bn   $  
			& $\! \partial_{us(n) \bar 0}\, \cY_n\cY_\bn \widehat\cM^{(0)}\, \cY_\bn \cY_n\!\!$     
			\\
			\hline
		\end{tabular}
		\caption{Subleading beam and soft functions arising from products of hard scattering operators in the factorization of Higgs with a jet veto, and their field content. Helicity and color structures have been suppressed. We have  not included products of operators whose beam and soft functions are identical to those shown by charge conjugation or $n\leftrightarrow \bn$.}		\label{tab:fact_func}
	\end{center}
\end{table}
	\end{center}
}

While the basis of operators presented in this section is quite large, many of the operators will not contribute to a physical cross section at $\cO(\lambda^2)$. In this section we briefly discuss the helicity operator basis, focusing in particular on understanding which operators can contribute to the cross section for an SCET$_\text{I}$ event shape observable, $\tau_B$, measured on $gg\to H$. In \Sec{sec:contribs_lam}, we show that there are no contributions to the cross section from hard scattering operators at $\cO(\lambda)$, which would correspond to power corrections of $\sqrt{\tau_B}$. Then in \Sec{sec:contribs}, we use helicity selection rules to determine which operators can contribute at $\cO(\lambda^2)=\cO(\tau_B)$. The results are summarized in \Tab{tab:summary}.

Given the set of contributing operators, one can then determine the full subleading power factorization theorem for the related observables with Higgs production. Here we restrict ourselves to determining the structure of the factorization theorem terms arising purely from our subleading hard scattering operators, written in terms of hard, beam and soft functions. A summary of these results is given in \Tab{tab:fact_func}.  In many cases the beam and soft functions which appear in the subleading power factorization formula are identical to those at leading power. For the case of the soft functions this simplification arises due to color coherence, allowing a simplification to the Wilson lines in the soft functions that appear. For gluon-gluon and quark-quark color channels the leading power soft functions are
\begin{align}\label{eq:soft_func_def}
S_g^{(0)}=\frac{1}{(N_c^2-1)}  \tr \big\langle 0 \big| \cY^T_{\bar n} \cY_n \widehat{\cM}^{(0)}\cY_n^T \cY_{\bar n} \big|0 \big\rangle\,, \qquad
S_q^{(0)}=\frac{1}{N_c}  \tr \big\langle 0 \big| Y^\dagger_{\bar n} Y_n \widehat{\cM}^{(0)}Y_n^\dagger Y_{\bar n} \big|0 \big\rangle\,, 
\end{align}
and depend on the kinematic variables probed by the measurement operator $\widehat{\cM}^{(0)}$.
For the beam functions, this simplification occurs since the power correction is often restricted to a single collinear sector. The other collinear sector is then described by the leading power beam functions (incoming jet functions) for gluons and quarks~\cite{Stewart:2009yx,Stewart:2010qs}
\begin{align}\label{eq:beam_func_def}
\frac{\delta^{ab}}{N_c^2-1}\, B_g^{(0)}
  &=- \frac{\omega \, \theta(\omega)}{2\pi}  \int \!\!\frac{dx^-}{2|\omega|}\:  e^{\frac{i}{2} \ell^+ x^-}  \Big\langle p \Big|\, \cB^{\mu a}_{n\perp} \big(x^- \text{\small $\frac{n}{2}$}\big)\, \hat{\delta}\, \left[ \delta(\omega-\bar \cP) \cB^{b}_{n\perp \mu}(0) \right] \,\Big|p\Big\rangle
   \,, \\
\frac{\delta^{\alpha \bbeta}}{N_c}\, B_q^{(0)}
  &= \frac{\theta(\omega)}{2\pi} \int \!\!\frac{dx^-}{2|\omega|}\: e^{\frac{i}{2} \ell^+ x^-} 
  \Big\langle p\Big|\, \chi_{n}^{\alpha} \big(x^- \text{\small $\frac{n}{2}$}\big) \frac{\Sl{\bar n}}{2} \,\hat{\delta}\,  \left[ \delta(\omega-\bar \cP)     \bar \chi_{n}^{\bar\beta}(0) \right]
  \,\Big| p \Big\rangle
\,, \nn
\end{align}
where we take $\ell^+ \gg \Lambda_{\rm QCD}^2/\omega$.
The result for the leading power measurement function $\hat{\delta}$ appearing in these beam functions depends on the factorization theorem being treated. Often the beam functions are inclusive in which case $\hat{\delta}=1$, giving functions of the momentum fraction of the struck parton $x$ and a single invariant mass momentum variable,  $B_{g}^{(0)}(x,\omega \ell^+)$ and $B_{q}^{(0)}(x,\omega \ell^+)$. Here we assume an SCET$_{\text{I}}$ type measurement that does not fix the $\cP_\perp$ of the measured particle. This assumption has been explicitly used in writing the form of the beam functions in \Eq{eq:beam_func_def}, as well as in our construction of the operator basis.

\subsubsection{Vanishing at $\cO(\lambda)$}\label{sec:contribs_lam}

We begin by considering possible contributions to the cross section at $\cO(\lambda)$.
While we will not discuss the factorization of the cross section in detail, the contribution of the hard scattering operators to the cross section at $\cO(\lambda)$ can be written schematically as
{\begin{small}
\begin{align}\label{eq:xsec_lam}
&\frac{d\sigma^{(1)} }{d\tau_B} \supset N \sum_{X,i}  \tilde \delta^{(4)}_q  \bra{P_1 P_2} C_i^{(1)} O_i^{(1)}(0) \ket{X}\bra{X} C^{(0)} O^{(0)}(0) \ket{P_1 P_2}   \delta\big( \tau_B - \tau_B^{(0)}(X) \big) +\text{h.c.} 
\,. 
\end{align}
\end{small}}
Here $N$ is a normalization factor, $P_1, P_2$ denote the incoming hadronic states, and we use the shorthand notation $\tilde \delta^{(4)}_q=(2\pi)^4\delta^4(q-p_X)$ for the momentum conserving delta function. This expression should merely be taken as illustrative of the operator contributions, and in particular, we have not made explicit any color or Lorentz index contractions, nor the treatment of the initial state. The summation over all final states, $X$, includes phase space integrations. The measurement of the observable is enforced by $ \delta\big( \tau_B - \tau_B^{(0)}(X) \big) $, where $\tau_B^{(0)}(X)$, returns the value of the observable $\tau_B$ as measured on the final state $X$. The explicit superscript $(0)$ indicates that the measurement function is expanded to leading power, since here we focus on the power suppression due to the hard scattering operators. 

From \Eq{eq:xsec_lam} we see that hard scattering operators contribute to the $\cO(\lambda)$ cross section through their interference with the leading power operator. The $\cO(\lambda)$ basis of operators is given in \Eqs{eq:H1_basis}{eq:H1_basis2}, each of which involves a single collinear quark field in each collinear sector. Conservation of fermion number then immediately implies that these operators cannot have non-vanishing matrix elements with the leading power operator which consists of a single collinear gluon field in each sector. Therefore, all contributions from hard scattering operators vanish at $\cO(\lambda)$. Although we do not consider them in this paper, using similar arguments one can show that all other sources of power corrections, such as Lagrangian insertions, also vanish at $\cO(\lambda)$.

\subsubsection{Relevant Operators at $\cO(\lambda^2)$}\label{sec:contribs}

Unlike the $\cO(\lambda)$ power corrections, the power corrections at $\cO(\lambda^2)=\cO(\tau_B)$ will not vanish. Contributions to the cross section at $\cO(\lambda^2)$ whose power suppression arises solely from hard scattering operators take the form either of a product of two $\cO(\lambda)$ operators or as a product of an $\cO(\lambda^2)$ operator and an $\cO(\lambda^0)$ operator
{\begin{small}
\begin{align}\label{eq:xsec_lam2}
\frac{d\sigma^{(2)}}{d\tau_B} &\supset N \sum_{X,i}  \tilde \delta^{(4)}_q  \bra{P_1 P_2} C_i^{(2)} O_i^{(2)}(0) \ket{X}\bra{X} C^{(0)} O^{(0)}(0) \ket{P_1 P_2} \delta\big( \tau_B - \tau_B^{(0)}(X) \big) +\text{h.c.}\nn \\
&+ N \sum_{X,i,j}  \tilde \delta^{(4)}_q  \bra{P_1 P_2} C_i^{(1)} O_i^{(1)}(0) \ket{X}\bra{X} C_j^{(1)} O_j^{(1)}(0) \ket{P_1 P_2}   \delta\big( \tau_B - \tau_B^{(0)}(X) \big)+\text{h.c.} \,.
\end{align}
\end{small}}
For $gg\to H$ the operator basis has only a single operator at $\cO(\lambda)$ (up to helicities and $n\leftrightarrow \bar n$), which was given in \Eqs{eq:H1_basis}{eq:H1_basis2}. This operator will contribute to the cross section at $\cO(\lambda^2)$, as indicated in \Tab{tab:fact_func}.

The contributions from $\cO(\lambda^2)$ hard scattering operators are highly constrained since they must interfere with the leading power operator. We will discuss each possible contribution in turn, and the summary of all operators which can contribute to the $\cO(\lambda^2)$ cross section is given in \Tab{tab:summary}. The schematic structure of the beam and soft functions arising from each of the different operator contributions is shown in \Tab{tab:fact_func}. The subleading beam and soft functions enumerated in this table are universal objects that will appear in processes initiated by different Born level amplitudes (such as $q\bar q$ annihilation), unless forbidden by symmetry. In this initial investigation, we content ourselves with only giving the field content of the beam and soft functions. In \Tab{tab:fact_func}, to save space, we do not write the external vacuum states for the soft functions, or the external proton states for the beam functions, nor do we specify the space-time positions of the fields. We do not present here the full definitions analogous to the leading power definitions given in \Eqs{eq:soft_func_def}{eq:beam_func_def}, but using the field content given in \Tab{tab:fact_func}. Deriving full definitions goes hand in hand with presenting the complete factorization theorems for these contributions, which will be given in future work.

\vspace{0.6cm}
\noindent{\bf{Two Quark-One Gluon Operators:}}

The two quark-one gluon operators, $O_{\cB\bar n}^{(1)}$ can contribute to the cross section by interfering with themselves. These operators are interesting since they effectively have a quark like cusp, instead of a gluon like cusp as is true of the leading power operators.  They contribute with a leading power quark channel soft function $S_q^{(0)}$, a quark beam function $B_q^{(0)}$ and a subleading power beam function $B_{qgg}^{(2)}$ that has fermion number crossing the cut (as indicated by its first $q$ subscript). 

\vspace{0.4cm}
\noindent{\bf{Two Quark-Two Gluon Operators:}}\\
\indent In the case of the two quark-two gluon operators, the only operators that will have a non-vanishing contribution are those that have the two gluons in different collinear sectors, namely $O_{\cB1}^{(2)}$. This gives a gluon beam function $B_g^{(0)}$, soft function $S_g^{(0)}$, and a subleading power beam function with gluon quantum numbers crossing the cut $B_{gqq}^{(2)}$ (with three color contractions). The operator $O_{\cB2}^{(2)}$, which has two quarks in a helicity $0$ configuration in one collinear sector, and two gluons in a helicity $0$ configuration in the other collinear sector does not contribute, since rotational invariance implies that its interference with the leading power operator vanishes. 

\vspace{0.4cm}
\noindent{\bf{Four Gluon Operators:}}

To give a non-vanishing interference with the leading power operator the four gluon operators must have an odd number of collinear gluon fields in each sector. This implies that $O_{4g1}^{(2)}$ does not contribute, while  $O_{4g2}^{(2)}$ does. Once again we can prove that $O_{4g2}^{(2)}$ generates a contribution that enters with simply the leading power gluon soft function $S_g^{(0)}$ (the direct proof of this requires some fairly extensive color algebra). This happens despite the fact that the subleading power beam function $B_{gg}^{(2)}$ has six color contractions.  The contribution from this four gluon operator first enters the cross section at $\cO(\alpha_s^2)$.

\vspace{0.4cm}
\noindent{\bf{Four Quark Operators:}}

For a four quark operator to interfere with the leading power operator, it must have both zero fermion number and a helicity 1 projection in each collinear sector. This eliminates all four quark operators from contributing to the cross section at $\cO(\lambda^2)$.

\vspace{0.4cm}
\noindent{\bf{$\cP_\perp$ Operators:}}

Both the operators involving $\cP_\perp$ insertions have the correct symmetry properties and therefore both $O_{\cP \chi }^{(2)}$ and $O_{\cP \cB }^{(2)}$ can contribute to the $\cO(\lambda^2)$ cross section. Both contributions have a leading power gluon beam function $B_g^{(0)}$ and soft function $S_g^{(0)}$. The operator $O_{\cP \cB }^{(2)}$ has a similar structure to the operator $\cO^{(2)}_{\cP1}$ found in the quark case in \cite{Feige:2017zci}, which contributes a leading log to the thrust (beam thrust) cross section \cite{Moult:2016fqy}. It involves a subleading power beam function $B_{ggP}^{(2)}$ (with two color contractions). On the other hand, we find in \Sec{sec:match_qqg_lam2} that the operator $O_{\cP \chi }^{(2)}$ has a vanishing Wilson coefficient at tree level, so its factorized contribution starts at least at ${\cal O}(\alpha_s^2)$ for the cross section. It has a subleading beam function $B_{gqP}^{(2)}$ with a single color contraction.

\vspace{0.4cm}
\noindent{\bf{Ultrasoft Operators:}}

The ultrasoft operators involving quark fields cannot contribute to the cross section through interference with the leading power operator due to fermion number conservation. Therefore, only the gluon operators of \Eqs{eq:Hgggus}{eq:Hdggus} contribute. They have leading power gluon beam functions $B_g^{(0)}$.

\subsubsection{Comparison with $\bar q\, \Gamma q$}\label{sec:compare}

It is interesting to briefly compare the structure of the operator basis, as well as the contributions to the $\cO(\lambda^2)$ cross section, to the basis for a process with two collinear sectors initiated by the $\bar q \Gamma q$ current as discussed in \cite{Feige:2017zci}. The leading power factorization theorems for the two cases are essentially identical, with simply a replacement of quark and gluon jet (beam) functions, as well as the color charges of the Wilson lines in the soft functions. However, at subleading power there are interesting differences arising both from the helicity structure of the currents, as well as from the form of the leading power Wilson coefficient.

An interesting feature of $gg\to H$ is that the Wilson coefficient for the leading power operator, which is given in \Sec{sec:matching_lp}, depends explicitly on the large label momenta of the gluons at tree level. This is not the case for the $\bar q\, \Gamma q$ current, whose leading power operator has a Wilson coefficient that is independent of the large label momenta at tree level. As discussed in \Sec{sec:nnlp_soft}, the Wilson coefficients of hard scattering operators involving insertions of $n \cdot \partial$, $\bar n \cdot \partial$, or $\cB_{us(n)0}$ are related to the derivatives of the leading power Wilson coefficients by RPI. This implies that these particular operators vanish at tree level for a $\bar q\, \Gamma q$ current, but are present at tree level for $gg\to H$. For the $\bar q \, \Gamma q$ current the power corrections from the ultrasoft sector arise instead only from subleading power Lagrangian insertions. Therefore, the nature of power corrections in the two cases is quite different in terms of the organization of the effective theory in the ultrasoft sector. However, this does not say anything about their numerical size which would require a full calculation. Furthermore, the organization of the collinear hard scattering operators is nearly identical in the two cases.

Despite this difference in the organization of the particular corrections within the ultrasoft sector of the effective theory, there is also much similarity in the way that the subleading power operators contribute to the cross section at $\cO(\lambda^2)$. In particular, in both cases, operators involving an additional ultrasoft or collinear gluon field as compared with the leading power operator contribute as an interference of the form $\cO(\lambda^2) \cO(1)$, see \Tab{tab:fact_func}. This is guaranteed by the Low-Burnett-Kroll theorem \cite{Low:1958sn,Burnett:1967km}. However, the subleading hard scattering operators that have a different fermion number in each sector than the leading power operators contribute as $\cO(\lambda) \cO(\lambda)$. For the $gg\to H$ case, this is the $O_{\cB\bar n}^{(1)}$ operator, while for a $q\Gamma \bar q$ current considered in \cite{Feige:2017zci}, it was a hard scattering operator involving two collinear quarks recoiling against a collinear gluon. In the NNLO calculation of power corrections for the $q\Gamma \bar q$ case \cite{Moult:2016fqy}, this operator played an important role, as it gave rise to a leading logarithmic divergence not predicted by a naive exponentiation of the one-loop result, and it is expected that the same will be true here. We plan to consider this calculation in a future work, and to understand in more detail the relation between the leading logarithmic divergences for the $q\Gamma \bar q$ current, compared with a $gg$ current.

\section{Matching}\label{sec:matching}

In this section we perform the matching to the operators relevant for the calculation of the $\cO(\lambda^2)$ cross section, which were enumerated in \Sec{sec:contribs} and summarized in \Tab{tab:summary}. As discussed in \Sec{sec:intro}, we will work in the context of an effective Higgs gluon coupling
\begin{align}\label{eq:heft}
\cL_{\text{hard}}=\frac{C_1(m_t, \alpha_s)}{12\pi v}G^{\mu \nu}G_{\mu \nu} H\,,
\end{align}
obtained from integrating out the top quark. Here $v=(\sqrt{2} G_F)^{-1/2}=246$ GeV, and the matching coefficient is known to $\cO(\alpha_s^3)$ \cite{Chetyrkin:1997un}. Corrections to the infinite top mass can be included in the matching coefficient $C_1$. We use the sign convention
\begin{align}
G_{\mu \nu}^a=\partial_\mu A_\nu^a-\partial_\nu A_\mu^a+g f^{abc} A_{\mu}^b A_\nu^c\,,\qquad iD^\mu=i\partial^\mu+gA^\mu\,.
\end{align}
In the matching, we take all particles as outgoing. However, to avoid a cumbersome notation we use $\epsilon$ instead of $\epsilon^*$ for the polarization of an outgoing gluon. We also restrict to Feynman gauge although we check gauge invariance by enforcing relevant Ward identities. For operators involving collinear gluon fields gauge invariance is guaranteed through the use of the $\cB_\perp$ fields. 

The Higgs effective Lagrangian has Feynman rules for $2$, $3$, and $4$ gluons which are summarized in \App{app:expand_gluon}.  Due to the non-negative powers of momenta appearing in these Feynman rules they give rise to Wilson coefficients which are less singular than those arising from power corrections to the ultrasoft and collinear dynamics of SCET. This will be seen explicitly in the subleading power matching calculations. To simplify the notation throughout this paper we will suppress the factor of $C_1(m_t, \alpha_s)/(12\pi v)$, and simply write the Feynman rules and matching relations for the operator 
\begin{align}
O^{\text{hard}}=G^{\mu \nu}G_{\mu \nu} H\,.
\end{align}
The dependence on $C_1(m_t, \alpha_s)/(12\pi v)$ is trivially reinstated. 

Throughout the matching, collinear gluons in the effective theory will be indicated in Feynman diagrams as a spring with a line drawn through them, collinear quarks will be indicated by dashed lines, and ultrasoft gluons will be indicated with an explicit ``us". This will distinguish them from their full theory counterparts for which standard Feynman diagram notation for quarks and gluons is used. Furthermore, for the full theory diagrams, we will use the $\otimes$ symbol to denote the vertex of the Higgs effective theory, as compared with the purple circle used to denote a hard scattering operator in the effective theory.

Due to the large number of operators present in our basis, we find it most convenient to express the results of the tree-level matching in the form of the Wilson coefficient multiplying the relevant operator. For this purpose we define a shorthand notation with a caligraphic {\cal O},
\begin{align}
\cO_X^{(i)} = C_X^{\text{tree}} O_X^{(i)}\,,
\end{align}
where as before, the superscript indicates the power suppression, and the subscript is a label that denotes the field and helicity content.
We will write results for $\cO_X^{(i)}$ in a form such that it is trivial to identify the tree level Wilson coefficient $C_X^{\text{tree}}$, so that higher order corrections can be added as they become available.

\subsection{Leading Power Matching}\label{sec:matching_lp}

The leading power matching is of course well known, however, we reproduce it here for completeness and to illustrate the matching procedure. The matching can be performed using a two gluon external state. Since the leading power operator is independent of any $\perp$ momenta, in performing the matching we can take the momenta 
\begin{align}
p_1^\mu =\omega_1 \frac{n^\mu}{2}, \qquad  p_2 =\omega_2 \frac{\bar n^\mu}{2}\,,
\end{align}
and the polarizations to be purely $\perp$, namely $\epsilon_{i}^\mu=\epsilon_{i\perp}^\mu$. All of the operators in \Sec{sec:lp} give a non-vanishing contribution to the two-gluon matrix element for this choice of polarization.

In the two gluon matrix element, this choice of polarization does not remove overlap with any of the operators in \Sec{sec:lp}.
Expanding the QCD result, we find
\begin{align}
\left.\fd{2.0cm}{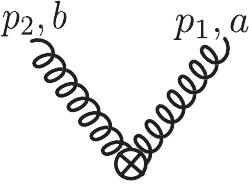} \right |_{\cO(\lambda^0)} &= -2i \delta^{ab} \omega_1 \omega_2 \epsilon_{1\perp} \cdot \epsilon_{2\perp}\,.
\end{align}
This is reproduced by the leading power operator
\begin{align}
\cO^{(0)}_{\cB}= -2\omega_1 \omega_2 \delta^{ab} \cB^a_{\perp \bar n, \omega_2} \cdot \cB^b_{\perp n, \omega_1} H\,,
\end{align}
or in terms of the helicity basis of \Eq{eq:hgg}, we have
\begin{align}
\cO^{(0)}_{\cB++}&=2\omega_1 \omega_2 \delta^{ab} \cB^a_{ \bar n +, \omega_2} \cdot \cB^b_{n +, \omega_1} H\,, \qquad
\cO^{(0)}_{\cB--}=2\omega_1 \omega_2 \delta^{ab} \cB^a_{ \bar n -, \omega_2} \cdot \cB^b_{n -, \omega_1} H\,.
\end{align}

While we focus here on the case where there is zero perp momentum in each collinear sector, we also give the Feynman rule in the case that each sector has a non-zero perp momentum. This will allow us to illustrate the gauge invariance properties of the collinear gluon field $\cB_\perp$. The expansion of the collinear gluon field with an incoming momentum $k$ is given by
\begin{align}
\cB^\mu_{n\perp}&=  A^{\mu a}_{\perp k} T^a -k^\mu_\perp \frac{\bar n \cdot A^a_{nk} T^a}{\bar n \cdot k} +\cdots \,,
\end{align}
where the dots represent terms with multiple gluon fields. The two gluon terms are given in \App{app:expand_gluon}. Gauge invariance therefore dictates the Feynman rule of our operator in the case of generic perp momenta for the two gluon fields,
\begin{align}
\fd{2.0cm}{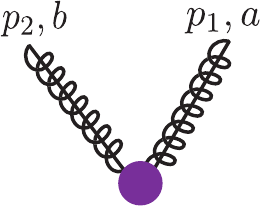} &= -2i \delta^{ab} \omega_1 \omega_2 \left( \epsilon_{1\perp}^{\mu} -p_{1\perp}^\mu \frac{\bar n \cdot \epsilon_1}{ \bar n \cdot p_1}  \right) \left( \epsilon_{2\perp}^{\mu} -p_{2\perp}^\mu \frac{\bar n \cdot \epsilon_2}{ \bar n \cdot p_2}  \right)  \,.
\end{align}
We note that the additional terms are essential to enforce that the required Ward identities are satisfied, and the result is gauge invariant. While this is trivial in this simple leading power example, for the more complicated matching calculations considered in the remainder of the paper we will often perform the matching for particular kinematic configurations, and the gauge invariance of the collinear gluon fields is an important ingredient to uniquely obtain the full result.

\subsection{Subleading Power Matching}\label{sec:matching_nlp}

We now consider the matching at $\cO(\lambda)$. In \Sec{sec:nlp} we argued that the only $\cO(\lambda)$ operator which can contribute to the cross section at $\cO(\lambda^2)$ has two collinear quark fields in opposite collinear sectors and a collinear gluon field. We can therefore perform the matching using this external state. For concreteness we start with the case with a quark in the $n$-collinear sector, and a gluon and antiquark in the $\bar n$-collinear sector, $(q)_n (\bar q g)_\bn$. Since the power suppression arises from the explicit fields, and all propagators are off shell, we can use the kinematics
\begin{align}
p_1^\mu =\omega_1 \frac{n^\mu}{2}, \qquad  p_2 =\omega_2 \frac{\bar n^\mu}{2}, \qquad p_3 =\omega_3 \frac{\bar n^\mu}{2}\,,
\end{align}
and take the polarization of the gluon to be purely $\perp$, $\epsilon_{i}^\mu=\epsilon_{i\perp}^\mu$. This choice suffices to obtain non-zero matrix elements for all the operators we want to probe, and to distinguish them from one another.

For the matching calculations, we will use the notation
\begin{align}
u_n(i)=P_n u(p_i)\,, \qquad \text{and} \qquad v_n(i)=P_n v(i)\,, \qquad \text{with} \qquad P_n=\frac{\Sl{n} \Sl{\bar n}}{4}\,,
\end{align}
for the projected SCET spinors. Here we have taken the momentum $p_i$ to be $n$-collinear, but similar relations exist for the case that it is $\bar n$-collinear. 
The spinors obey
\begin{align}
u(p_i) = \Big( 1+ \frac{\Sl{p}_{i\perp}}{\bar{n} \cdot p_i} \frac{\Sl{\bar{n}}}{2} \Big)  u_n(i) \,, 
 \qquad
u(p_i) = \Big( 1+ \frac{\Sl{p}_{i\perp}}{n \cdot p_i} \frac{\Sl{n}}{2} \Big) u_{\bar n}(i) \,,
\end{align}
for the $n$-collinear and $\bar n$-collinear cases respectively, with direct analogs for the $v(p_i)$ spinors.

Expanding the QCD diagram to $\cO(\lambda)$, we find
\begin{align}
\left. \fd{2.5cm}{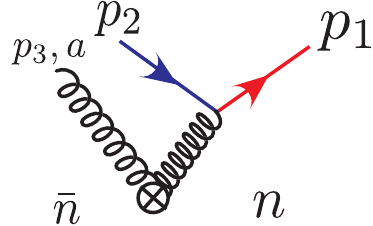}\right |_{\cO(\lambda)} &= \frac{-2ig\omega_3}{\omega_2}  \bar u_n (p_1) \Sl{\epsilon}_{3\perp} T^a v_{\bar n}(p_2)\,.
\end{align}
There are no contributions from time ordered products in the effective theory to this particular matrix element used in the matching. This is due to the fact that there are no $\cO(\lambda^0)$ or $\cO(\lambda^1)$ operators involving just two quark fields, and the collinear Lagrangian insertions in each section preserve the fermion number of each sector, so this particular matrix element can not be obtained from Lagrangian insertions starting from the leading power operator involving two collinear gluons.   Therefore, the result must be reproduced entirely by a hard scattering operator in SCET. This operator is given by
\begin{align}
\cO^{(1)}_{\cB \bar n}= -2g\, \frac{\omega_3}{\omega_2}\,
  \bar \chi_{n,\omega_1} \Sl{\cB}_{\perp \bar n, \omega_3} 
    \chi_{\bar n,-\omega_2} H\,.
\end{align}
or, in terms of the helicity operators of \Eq{eq:H1_basis}
\begin{align}
\cO^{(1)}_{\cB \bar n+(+)}&=4g\frac{\omega_3}{\omega_2} T^a_{\alpha \bbeta} \sqrt{\frac{\omega_1 \omega_2}{2}}  \langle \bar n n \rangle \cB^a_{\bar n+,\omega_3} J^{\balpha \beta}_{n\bar n+} H\,, \nn \\
\cO^{(1)}_{\cB \bar n-(-)}&=-4g\frac{\omega_3}{\omega_2}T^a_{\alpha \bbeta}  \sqrt{\frac{\omega_1 \omega_2}{2}}  [ \bar n n ] \cB^a_{\bar n-,\omega_3} J^{\balpha \beta}_{n\bar n-} H\,.
\end{align}
The Wilson coefficient has a singularity as the energy fraction of the quark in the $\bar n$-collinear sector becomes soft. This operator will therefore contribute to the leading logarithmic divergence at the cross section level at $\cO(\lambda^2)$.  Note that this operator is explicitly RPI-III invariant, with its Wilson coefficient taking the form of a ratio of the large momentum components of the two $\bar n$ collinear fields.

For convenience, we also give the full Feynman rule for this operator
\begin{align}
\fd{3cm}{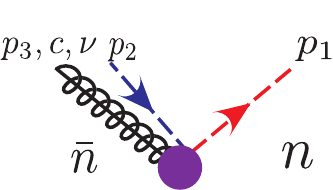}
  &=-2ig T^c \frac{\omega_3 }{\omega_2}   \left( \gamma^\nu_\perp-\frac{\Sl{p}_{3\perp} n^\nu}{\omega_3} \right)\,.
\end{align}
Note that this Feynman rule contains terms that were not present in the matching calculation due to the special choice of kinematics used there. These additional terms are determined by the gauge invariant gluon field, $\cB_{\perp \bar n}$, and it is easy to see that they ensure that this operator satisfies the required Ward identities.

The matching for the operators in the case  $(\bar q)_n (q g)_\bn$ can be easily obtained from the above results by exploiting charge conjugation. This gives
\begin{align}
\cO^{(1)}_{\cB n}= -2g \frac{\omega_3}{\omega_1} \bar \chi_{n,\omega_1} \Sl{\cB}_{\perp n, \omega_3} \chi_{\bn,-\omega_2} H\,,
\end{align}
and for the helicity operators in \eq{H1_basis2} we obtain
\begin{align}
\cO^{(1)}_{\cB n+(-)}
  &=4g\frac{\omega_3}{\omega_1} T^a_{\alpha \bbeta} 
   \sqrt{\frac{\omega_1 \omega_2}{2}}  \langle \bar n n \rangle \cB^a_{n-,\omega_3} J^{\balpha \beta}_{n\bar n+} H
  \,, \nn \\
\cO^{(1)}_{\cB n-(+)}
  &=-4g\frac{\omega_3}{\omega_1}T^a_{\alpha \bbeta}  
  \sqrt{\frac{\omega_1 \omega_2}{2}}  [ \bar n n ] 
  \cB^a_{n+,\omega_3} J^{\balpha \beta}_{n\bar n-} H
  \,.
\end{align}

This concrete matching calculation at subleading power also clearly illustrates the distinction between subleading power hard scattering operators, and the standard picture of leading power factorization in terms of splitting functions. In the leading power factorization for $H\to gq\bar q$, when the $q\bar q$ pair become collinear, the amplitude factorizes into $H\to gg$ multiplied by a universal $g\to q\bar q$ splitting function. This gives rise to a leading power contribution, due to the nearly on-shell propagator of the intermediate gluon that undergoes the splitting. For the operator considered here, the gluon which splits into the $q\bar q$ pair is far off-shell, due to the fact that the $q$ and $\bar q$ are in distinct collinear sectors. Because of this, it is represented in the effective theory by a local contribution (namely a hard scattering operator), and this operator is power suppressed. The hard scattering operators therefore describe precisely the contributions that are not captured by a splitting function type factorization. While this is particularly clear for the operator considered here, this picture remains true for the subleading power hard scattering operators for the more complicated partonic contributions considered at subsubleading power in \Sec{sec:matching_nnlp}. The hard scattering operators describe local contributions, which do not factorize in standard splitting function type picture, and therefore in general have no relation to known splitting functions which appear in the literature.

\subsection{Subsubleading Power Matching}\label{sec:matching_nnlp}

In this section we perform the tree level matching to the $\cO(\lambda^2)$ operators, considering only those which contribute at the cross section level at $\cO(\lambda^2)$, as discussed in \Sec{sec:contribs}. Since there are a number of operators, each with different field content, we will consider each case separately.

\subsubsection{Ultrasoft Derivative}\label{sec:us_deriv}

We begin by performing the matching to the ultrasoft derivative operators of \Sec{sec:nnlp_soft}. To perform the matching we can use a state consisting of two perpendicularly polarized collinear gluons, and we take our momenta as
\begin{align}
p_1^\mu=(\omega_1 +k_1)\frac{n^\mu}{2} +p_{1r}\frac{\bar n^\mu}{2}+p_{1\perp}^\mu\,, \qquad p_2^\mu=(\omega_2 +k_2)\frac{\bar n^\mu}{2} +p_{2r}\frac{n^\mu}{2}+p_{2\perp}^\mu\,. 
\end{align}
Since we have taken non-zero label perp momentum to keep the particles on shell we will have operators contributing that involving the $\cP_\perp$ operator. These operators were not included in our basis, since we assumed zero total perp momentum in each sector. (See \App{app:gen_pt} for the additional operators required in the case that the collinear sectors have non-vanishing perp momentum.) However, these terms are easy to identify. Dropping these terms involving the label perp momentum to identify the contributions relevant for the matching, we find
\begin{align}
\left. \fd{2.0cm}{figures/matching_gg_low}\right |_{\cO(\lambda^2)}&=-2i \delta^{ab} \omega_1 k_2 \epsilon_{1\perp} \cdot \epsilon_{2\perp} -2i\delta^{ab} \omega_2 k_1 \epsilon_{1\perp} \cdot \epsilon_{2\perp}\,.
\end{align}
This result must be completely reproduced by hard scattering operators in the effective theory, since the relevant subleading propagator insertions are proportional to residual components of the $\perp$ momentum, which we have taken to be zero in the matching (see \App{app:expand_gluon}, and in particular \Eq{eq:lam2_gluon_prop}).

The operators given in \Sec{sec:nnlp_soft} were defined post BPS field redefinition, in which case the partial derivative operator $\partial^\mu$ acts on gauge invariant building blocks. While the distinction between pre- and post-BPS field redefinition is not relevant for the calculation of the matrix elements in this particular case, since there are no ultrasoft emissions, it of course determines the form that the operators are written in. For convenience, we give the operators both before and after BPS field redefinition. Note that the collinear gluon field transforms as an adjoint matter field under ultrasoft gauge transformations since the ultrasoft gauge field acts as a background field.

Matching onto pre-BPS field redefinition operators, we find
\begin{align}\label{eq:preBPS_softgluon}
\cO^{(2)}_{n\cdot D}=4\omega_1 \tr \left[ \cB^\mu_{\perp n, \omega_1}   [in \cdot  D_{us}, \cB^\mu_{\perp \bar n, \omega_2} ] \right ] H\,, \quad \cO^{(2)}_{\bar n\cdot D}=4\omega_2 \tr \left[ \cB^\mu_{\perp \bar n, \omega_2}  [i\bar n \cdot D_{us},  \cB^\mu_{\perp n, \omega_1}] \right] H\,,
\end{align}
where the trace is over color. This color structure will be fixed by matching with an additional ultrasoft gluon in \Sec{sec:us_gluon}. To determine the operators post-BPS field redefinition, we can either directly apply the BPS field redefinition, or simply match to the operators of \Sec{sec:nnlp_soft}. We find that the operators where the ultrasoft derivative acts on the gluon fields are given by
\begin{align}
\cO^{(2)ab}_{\partial \cB (us)(0)}=-2\omega_1 \cB^{\mu a}_{\perp n, \omega_1}   i n \cdot  \partial \cB^{\mu b}_{\perp \bar n, \omega_2} H\,, \quad \cO^{(2)ab}_{\partial \cB (us) (\bar 0)}=-2\omega_2 \cB^{\mu a}_{\perp \bar n, \omega_2}  i\bar n \cdot  \partial   \cB^{\mu b}_{\perp n, \omega_1} H\,,
\end{align}
or expanded in terms of the helicity operator basis
\begin{align}
\cO^{(2)ab}_{\partial \cB (us(n)) \bar 0:++}
  &=-2\omega_1 \cB^{\mu a}_{n +, \omega_1}   i  \partial_{us(n)\bar 0} 
  \cB^{\mu b}_{\bar n +, \omega_2} H
  \,, \quad \\
\cO^{(2)ab}_{\partial \cB (us(\bar n))0:++}
  &=-2\omega_2  \cB^{\mu a}_{\bar n +, \omega_2}  i\partial_{us(\bar n) 0}  \cB^{\mu b}_{n +, \omega_1} H
  \,, \nn \\
\cO^{(2)ab}_{\partial \cB (us(n)) \bar 0:--}
  &=-2\omega_1  \cB^{\mu a}_{n -, \omega_1}   i  \partial_{us(n)\bar 0} \cB^{\mu b}_{\bar n -, \omega_2} H
  \,, \quad \nn\\
\cO^{(2)ab}_{\partial \cB (us(\bar n))0:--}
  &=-2\omega_2  \cB^{\mu a}_{\bar n -, \omega_2}  i\partial_{us(\bar n) 0}  \cB^{\mu b}_{n -, \omega_1} H
   \,. \nn
\end{align}
Here the color indices are contracted against the basis of color structures given in \Eq{eq:Z2gd_colorus}.
These operators also give rise, after BPS field redefinition to operators involving $\cB_{us}$. These will be discussed in \Sec{sec:us_gluon}.

As mentioned above \Eq{eq:Hdggus}, using the gluon equations of motion we can eliminate operators involving $\bar n\cdot \partial\cB_{\bar n}$ and $n\cdot \partial \cB_n$ from our basis to all orders in perturbation theory. This structure of the ultrasoft derivative operators is important for the matching at $\cO(\lambda^2)$. In particular, only the $\bar n \cdot \partial$ acts on the $n$-collinear sector, and only the $n\cdot \partial$ acts on the $\bar n$-collinear sector. These correspond to the residual components of the label momenta. In a graph consisting of only collinear particles (i.e. no ultrasoft particles) the residual components of the label momenta can be chosen to vanish, so that these operators do not contribute. In all purely collinear graphs computed in the remainder of this paper, we will always make this choice, and therefore, these operators will not contribute. However, these operators will contribute, and will play an important role, when ultrasoft particles are present in the graph.

\subsubsection{qqg}\label{sec:match_qqg_lam2}

We now consider the case of the  $\cO(\lambda^2)$ operators involving two collinear quark fields, a collinear gluon field, and a $\cP_\perp$ insertion. In \Sec{sec:nnlp} we argued that the only such operators have both quark fields in the same collinear sector, which we will take to be the $n$-collinear sector. To perform the matching we take the kinematics
\begin{align}\label{eq:qqg_momentum}
p_1^\mu=\omega_1 \frac{n^\mu}{2}+p_\perp^\mu +p_{1r} \frac{\bar n^\mu}{2}\,, \qquad  p_2^\mu=\omega_2 \frac{n^\mu}{2}-p_\perp^\mu +p_{2r} \frac{\bar n^\mu}{2}\,, \qquad p_3^\mu=\omega_3 \frac{\bar n^\mu}{2}\,.
\end{align}
With this choice all subleading Lagrangian insertions in SCET vanish. This can be seen from the explicit subleading Lagrangians and Feynman rules given in \App{app:expand_gluon} by noting that these give contributions to this matrix element the involve residual components of $\perp$ momentum, or residual components of the large label momentum, which are zero for the choice of momentum in \Eq{eq:qqg_momentum}.   The result must therefore be entirely reproduced by hard scattering operators. Expanding the QCD result we find that it vanishes at $\cO(\lambda^2)$
\begin{align}
\left. \fd{2.5cm}{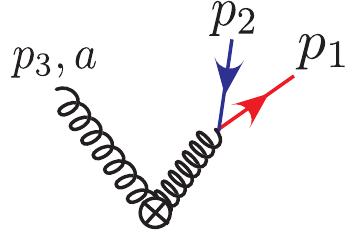} \right|_{\cO(\lambda^2)}&= 0\,.
\end{align}
This is expected since this diagram involves only collinear dynamics in a single collinear sector, and non-trivial terms will be reproduced by power suppressed Lagrangians.
Therefore, at tree level, the hard scattering operators involving two quarks in the same sector along with a $\cP_\perp$ insertion have  vanishing Wilson coefficients. We do not have an argument that the Wilson coefficients of these operators would continue to vanish at higher orders in perturbation theory, and therefore we do not expect this to be the case.

\subsubsection{ggg}\label{subsec:ggg_match_lam2}

We now consider matching to the $\cO(\lambda^2)$ three gluon operators which have a single $\cP_\perp$. In \Sec{sec:contribs} we have argued that the only such operators that contribute to the cross section at $\cO(\lambda^2)$ have two gluons in the same collinear sector, which we take to be $\bar n$ for concreteness. To perform the matching, we take the kinematics as
\begin{align}
p_1=\omega_1 \frac{n^\mu}{2}\,, \qquad p_2=\omega_2 \frac{\bar n^\mu}{2}+p_\perp^\mu+p_{2r} \frac{n^\mu}{2}\,, \qquad p_3=\omega_3 \frac{\bar n^\mu}{2}-p_\perp^\mu+p_{3r} \frac{n^\mu}{2}\,.
\end{align}
As a further simplification, we can take the polarization vector of the gluon in the $n$-collinear sector to be purely $\perp$,
$\epsilon_{1}^\mu=\epsilon_{1\perp}^\mu\,.$  All of the three gluon operators in our basis give a non-vanishing contribution to the three-gluon matrix element for this choice of polarizations.

In performing the expansion of the QCD diagrams we will obtain all three projections of the polarization vectors, namely $\bar n \cdot \epsilon_{2,3}$, $n \cdot \epsilon_{2,3}$, and $p_{\perp} \cdot \epsilon_{2,3\perp}$. However, all of the operators in our basis are formed from $\cB_{\perp}$, and therefore contain only the $n \cdot \epsilon_{2,3}$ and $p_{\perp} \cdot \epsilon_{2,3\perp}$ components. From the on-shell conditions for the gluon we have the relation
\begin{align}
\omega_2 \frac{\bar n \cdot \epsilon_2}{2}=\frac{p_\perp^2 n\cdot \epsilon_2}{2\omega_2}-p_\perp \cdot \epsilon_\perp\,,
\end{align}
and similarly for $\epsilon_3$. Note that we always use the Minkowski signature for the $\perp$ momenta, i.e. $p_\perp^2=-\vec p_\perp^{~2}$. In performing the matching one can therefore keep track of only the $\perp$ polarizations,  as long as the $\bar n \cdot \epsilon$ polarizations are converted into $n \cdot \epsilon$ and $p_\perp\cdot \epsilon_\perp$ using the above equation. This allows one to simplify the structure of the matching while keeping enough terms to reconstruct operators formed from $\cB_\perp$ gluon fields.

Expanding the QCD diagrams, and keeping only the $\perp$ terms of the polarizations we find
\begin{align}
&\left. \left(\fd{2.25cm}{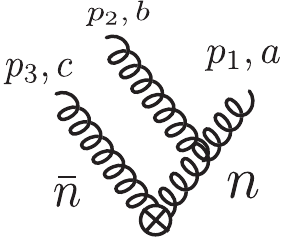} + \fd{2.25cm}{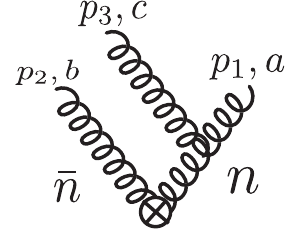}\right) \right |_{\cO(\lambda^2)} \nn \\
&\hspace{0cm}= -4gf^{abc} \frac{\omega_3}{\omega_2}\left( \epsilon_{1\perp} \cdot \epsilon_{2\perp} p_{\perp}\cdot \epsilon_{3\perp} -\epsilon_{2\perp} \cdot \epsilon_{3\perp} p_{\perp} \cdot \epsilon_{1\perp}    \right ) -4g f^{abc} \epsilon_{1\perp}\cdot \epsilon_{2\perp} p_{\perp}\cdot \epsilon_{3\perp}  +[(2,b)\leftrightarrow (3,c)]\,, \nn \\
&\left. \fd{2.25cm}{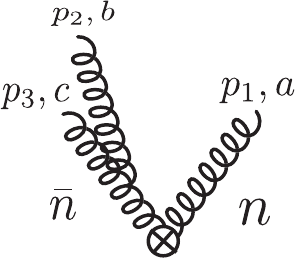}\right|_{\cO(\lambda^2)} =0\,, \nn \\
&\left.\fd{2.25cm}{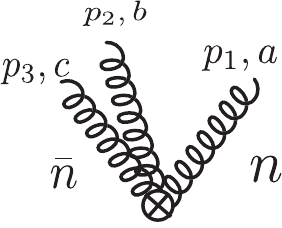}\right|_{\cO(\lambda^2)} =-4g f^{abc} \left(p_\perp \cdot \epsilon_{3\perp} \epsilon_{1\perp}\cdot \epsilon_{2\perp} -p_\perp \cdot \epsilon_{1\perp} \epsilon_{2\perp}\cdot \epsilon_{3\perp} +\frac{\omega_2}{\omega_3} p_\perp \cdot \epsilon_{3\perp} \epsilon_{1\perp} \cdot \epsilon_{2\perp}   \right) \nn \\
&\hspace{4cm}+[(2,b)\leftrightarrow (3,c)]\,,
\end{align}
We have shown results for the individual diagrams to emphasize the structure of the contributions, namely that only the diagrams involving an off-shell propagator or the Higgs EFT three gluon vertex contribute.
Simplifying this result, we find that the sum of the QCD diagrams is given by
\begin{align}
&\left. \left(\fd{2.25cm}{figures/matching_lam2_ggg4_low} + \fd{2.25cm}{figures/matching_lam2_ggg1_low}+\fd{2.25cm}{figures/matching_lam2_ggg2_low}+\fd{2.25cm}{figures/matching_lam2_ggg3_low}\right) \right |_{\cO(\lambda^2)} \nn \\
= &4gf^{abc} \left (2+\frac{\omega_3}{\omega_2}+\frac{\omega_2}{\omega_3} \right) \epsilon_{2\perp}\cdot \epsilon_{3\perp} p_\perp\cdot \epsilon_{1\perp} -4gf^{abc} \left( 2+\frac{\omega_3}{\omega_2}+\frac{\omega_2}{\omega_3}  \right)\epsilon_{1\perp} \cdot \epsilon_{2\perp} p_\perp \cdot \epsilon_{3\perp}\nn \\
&  -4gf^{abc} \left( 2+\frac{\omega_3}{\omega_2}+\frac{\omega_2}{\omega_3}  \right)\epsilon_{1\perp} \cdot \epsilon_{3\perp} p_\perp \cdot \epsilon_{2\perp} \,.
\end{align}
For the choice of kinematics and polarizations used in the matching there are no SCET subleading Lagrangian contributions at this power, for similar reasons to the case of $gq\bar q$ discussed above. Therefore, the hard scattering operators must exactly reproduce the QCD result.

We write the operators and their Wilson coefficients both in the helicity basis of \Eq{eq:Hgggpperp_basis}, as well as in a more standard Lorentz structures, as the two may prove useful for different purposes. In terms of standard Lorentz structures the tree level matching gives
\begin{align}
\cO^{(2)}_{\cP \cB1}&=-\left( \frac{1}{2}\right)4g \left(  2+\frac{\omega_3}{\omega_2}+ \frac{\omega_2}{\omega_3}  \right)i f^{abc} \cB^a_{n\perp,\omega_1}\cdot \left[  \cP_\perp \cB^b_{\bar n \perp,\omega_2}\cdot  \right] \cB_{\bar n \perp,\omega_3}^c    H\,, \nn \\
\cO^{(2)}_{\cP \cB2}&=4g\left( 2+\frac{\omega_3}{\omega_2}  + \frac{\omega_2}{\omega_3}\right) if^{abc}\left[ \cP_\perp \cdot \cB_{\bar n \perp,\omega_3}^a \right] \cB^b_{n\perp,\omega_1} \cdot \cB_{\perp \bar n, \omega_2}^c    H\,.
\end{align}
We have written the first operator in this form to incorporate the symmetry factor.
In the helicity basis, we have
\begin{align}\label{eq:3g_hel_match}
&\cO_{\cP\cB +++[-]}^{(2)}
= 4g if^{abc} \left(2+\frac{\omega_3}{\omega_2}+ \frac{\omega_2}{\omega_3}\right) \cB_{n+,\omega_1}^a\, \cB_{\bar n+,\omega_3}^b\, \left [\cP_{\perp}^{-} \cB_{\bar n+,\omega_2}^c \right ] \,H
\,,\nn \\
&\cO_{\cP\cB ---[+]}^{(2)}
=4g if^{abc}\left(2+\frac{\omega_3}{\omega_2}+ \frac{\omega_2}{\omega_3}\right) \cB_{n-,\omega_1}^a\, \cB_{\bar n-,\omega_3}^b\, \left [\cP_{\perp}^{+} \cB_{\bar n-,\omega_2}^c \right ] \,H\,,
 \nn \\
&\cO_{\cP\cB ++-[+]}^{(2)}
= -2g if^{abc}\left(2+\frac{\omega_3}{\omega_2}+ \frac{\omega_2}{\omega_3}\right)\, \cB_{n+,\omega_1}^a\, \cB_{ \bar n-,\omega_3}^b\, \left [\cP_{\perp}^{+} \cB_{\bar n+,\omega_2}^c \right ] \,H
\,,\nn \\
&\cO_{\cP\cB -+-[-]}^{(2)}
= -2g if^{abc}\left(2+\frac{\omega_3}{\omega_2}+ \frac{\omega_2}{\omega_3}\right)\cB_{n-,\omega_1}^a\, \cB_{ \bn -,\omega_3}^b\, \left [\cP_{\perp}^{-} \cB_{\bar n +,\omega_2}^c \right ] \,H
\,.
\end{align}
We therefore see explicitly that the helicity selection rules are realized in the tree level matching. Furthermore, the Wilson coefficient is formed from Bose symmetric combinations of ratios of the large momentum components of the $\bar n$ collinear fields, as required by RPI-III invariance.
For convenience, we also give the Feynman rule of the combined operator with three external gluons
\begin{align}\label{eq:feynrule_ggg3}
\fd{2.75cm}{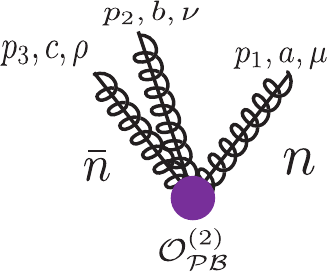}&\\%
	&\hspace{-2.5cm}=4g f^{abc}\Bigl( 2  + \frac{ \omega_2}{\omega_3} + \frac{ \omega_3}{\omega_2}\Bigr)\biggl[  p_\perp^\mu g_\perp^{\nu \rho}  - p_\perp^\nu g_\perp^{\mu \rho} - p_\perp^\rho g_\perp^{\mu \nu} +\frac{p^2_\perp}{\omega_2 \omega_3}\left( \omega_3 n^\nu g_\perp^{\mu \rho}- \omega_2 n^\rho g_\perp^{\mu \nu}  + p_\perp^\mu n^\nu n^\rho\right) \biggr]
    . \nn
\end{align}
This contains additional terms not present in the earlier matching calculation, due to the particular choice of $\perp$ polarizations used to simplify the matching. One can explicitly check that this operator satisfies the Ward identity, which is gauranteed by the fact that it is written in terms of $\cB_\perp$ fields. 
It is also interesting to  note that the Wilson coefficient of this operator has a divergence as either $\omega_2$, or $\omega_3$ become soft, so that it will give rise to a leading logarithmic divergence in the cross section at $\cO(\lambda^2)$.

\subsubsection{Ultrasoft Gluon}\label{sec:us_gluon}

The operators involving a single ultrasoft insertion were given in \Sec{sec:nnlp_soft}, and it was argued that they were related by RPI to the leading power operator involving two collinear gluons.  In this section we will explicitly perform the tree level matching to verify that this relation holds. The operators in \Sec{sec:nnlp_soft} were given after BPS field redefinition, since it is more convenient when enumerating a complete basis to work with a gauge invariant ultrasoft gluon field. While it is possible to directly match to the post-BPS operators, we will first perform the matching to pre-BPS field redefinition operators involving ultrasoft covariant derivatives, and verify the color structure given in \Eq{eq:preBPS_softgluon}. We will then give the operators after BPS field redefinition.

We perform the matching to a three particle external state, with one collinear gluon in each sector, and a single ultrasoft gluon.
To simplify the matching we take the momenta of the collinear particles as
\begin{align}
p_1^\mu=\omega_1 \frac{n^\mu}{2}\,, \qquad p_2^\mu =\omega_2 \frac{\bar n^\mu}{2}\,,
\end{align}
and the momentum of the ultrasoft particle as
\begin{align}
p_3^\mu=\bar n \cdot p_3 \frac{n^\mu}{2} +n\cdot p_3 \frac{\bar n^\mu}{2} +p_{3\perp}^\mu\,,
\end{align}
where $(n\cdot p_3, \bar n \cdot p_3, p_{3\perp})\sim (\lambda^2, \lambda^2,\lambda^2)$.
The full theory QCD diagrams expanded to $\cO(\lambda^2)$ are given by
\begin{align}
\left. \fd{2.25cm}{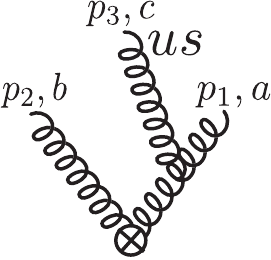}\right|_{\cO(\lambda^2)}  &= 2g \omega_2 f^{abc} \epsilon_1 \cdot \epsilon_2 \frac{\bar n\cdot p_3}{n\cdot p_3}n\cdot \epsilon_3+4g f^{abc} \omega_2 \epsilon_1 \cdot \epsilon_3 \frac{\epsilon_{2\perp} \cdot p_{3\perp}}{n \cdot p_3} \nn \\
&-4g f^{abc} \omega_2 \epsilon_2\cdot  \epsilon_3 \frac{p_{3\perp} \cdot \epsilon_{1\perp}}{ n \cdot p_3}\,, \nn \\
\left.\fd{2.25cm}{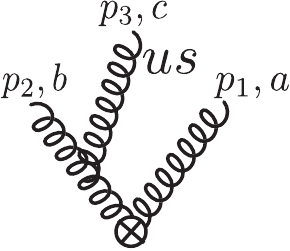} \right|_{\cO(\lambda^2)} &=-2g \omega_1 f^{abc} \epsilon_1 \cdot \epsilon_2 \frac{n\cdot p_3}{\bar n\cdot p_3}\bar n\cdot \epsilon_3-4g f^{abc} \omega_1 \epsilon_2 \cdot \epsilon_3 \frac{\epsilon_{1\perp} \cdot p_{3\perp}}{\bar n \cdot p_3} \nn \\
&+4g f^{abc} \omega_1 \epsilon_1\cdot \epsilon_3 \frac{p_{3\perp} \cdot \epsilon_{2\perp}}{\bar n \cdot p_3}\,, \nn \\
\left.\fd{2.25cm}{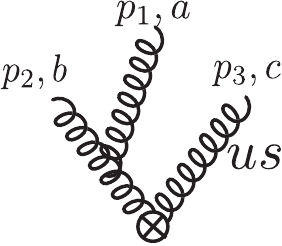} \right|_{\cO(\lambda^2)} &= 0\,,\nn \\
\left.\fd{2.25cm}{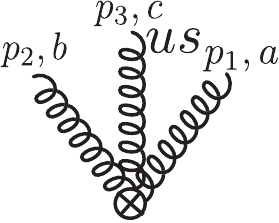} \right|_{\cO(\lambda^2)} &=2g f^{abc} \omega_1 \epsilon_1 \cdot \epsilon_2 n\cdot \epsilon_3-2g f^{abc} \omega_2 \epsilon_1 \cdot \epsilon_2 \bar n \cdot \epsilon_3 \,. 
\end{align}
In this case there are also contributions from $T$ product diagrams in SCET correcting the emission of an ultrasoft gluon. Once we subtract these terms from the full theory result, the remainder will be localized at the hard scale.
The $\cO(\lambda^2)$ Feynman rule for the emission of a ultrasoft gluon from a collinear gluon is given by (see \App{app:expand_gluon} and e.g.  \cite{Larkoski:2014bxa} for the explicit Feynman rule)
\begin{align}
\fd{3cm}{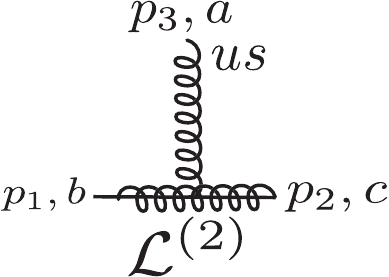}&=\langle | T \cB^\nu_{n\perp}(0) \cL^{(2)}_{A_n} | \epsilon_n, p_n; \epsilon_s,p_s \rangle=-i f^{abc} \epsilon_{n\mu} \frac{2\epsilon_{s\rho} p_{s\sigma}}{p_n^- n\cdot p_s} \left(  g^{\mu \rho}_\perp g^{\sigma \nu}_\perp -    g^{\mu \sigma}_\perp g^{\rho \nu}_\perp  \right)\,.
\end{align}
The two SCET diagrams involving this Lagrangian insertion are given by
\begin{align}
\fd{2.25cm}{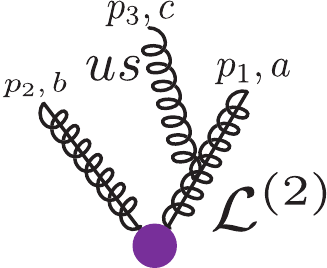}  &=\frac{4\omega_2 f^{abc} }{n\cdot p_3} (\epsilon_1 \cdot \epsilon_3 p_{3\perp} \cdot \epsilon_{2\perp} -\epsilon_{1\perp} \cdot p_{3\perp} \epsilon_2 \cdot \epsilon_3)\,, \nn \\
\fd{2.25cm}{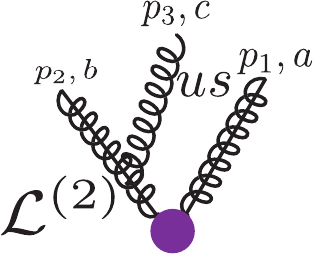}  &=\frac{4\omega_1 f^{abc} }{\bar n\cdot p_3} (\epsilon_1 \cdot \epsilon_3 p_{3\perp} \cdot \epsilon_{2\perp} -\epsilon_{1\perp} \cdot p_{3\perp} \epsilon_2 \cdot \epsilon_3)\,.
\end{align}
Finally we also have contributions from the ultrasoft derivative operators of \Sec{sec:us_deriv}, with a leading power emission of a ultrasoft gluon. For these diagrams we find
\begin{align}
\fd{2.25cm}{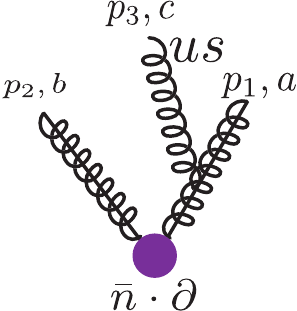}&= 2g \omega_2 f^{abc} \epsilon_1 \cdot \epsilon_2 \frac{\bar n\cdot p_3}{n\cdot p_3}n\cdot \epsilon_3\,, \quad
\fd{2.25cm}{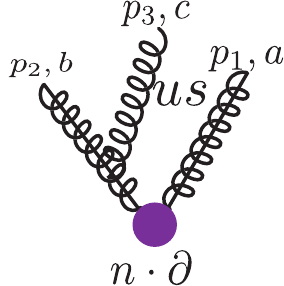}&= -2g \omega_1 f^{abc} \epsilon_1 \cdot \epsilon_2 \frac{n\cdot p_3}{\bar n\cdot p_3}\bar n\cdot \epsilon_3\,.
\end{align}
The SCET $T$-products therefore exactly reproduce the QCD diagrams, with the exception of the contribution from the three gluon vertex of the Higgs effective theory.
Subtracting the SCET contributions from the expansion of the QCD diagrams, we find that the hard scattering operators are given by
\begin{align}\label{eq:preBPS_softgluon2}
\cO^{(2)}_{n\cdot D}=4\omega_1 \tr \left[ \cB^\mu_{\perp n, \omega_1}   [n \cdot  D_{us}, \cB^\mu_{\perp \bar n, \omega_2} ] \right ] H\,, \quad \cO^{(2)}_{\bar n\cdot D}=4\omega_2 \tr \left[ \cB^\mu_{\perp \bar n, \omega_2}  [\bar n \cdot D_{us},  \cB^\mu_{\perp n, \omega_1}] \right] H\,,
\end{align}
as stated in \Eq{eq:preBPS_softgluon}.
In terms of gauge invariant ultrasoft gluon fields we have
\begin{align}
\cO^{(2)}_{\cB(us(n))}&=\left(i  f^{abd}\, \big({\cal Y}_n^T {\cal Y}_{\bar n}\big)^{dc}\right)  \left (-2g \omega_2  \cB^a_{n\perp, \omega_1} \cdot \cB^b_{\bar n \perp, \omega_2} \cB^c_{us(n)0} \right)\,,  \nn \\
\cO^{(2)}_{\cB(us(\bar n))}&=\left(i  f^{abd}\, \big({\cal Y}_{\bar n}^T {\cal Y}_{n}\big)^{dc}\right)  \left (-2g \omega_1  \cB^a_{n\perp, \omega_1} \cdot \cB^b_{\bar n \perp, \omega_2} \cB^c_{us(\bar n)0} \right)\,,
\end{align}
where the color structures that appear at tree level are the first components of the color basis of \Eqs{eq:Z2g_colorus}{eq:Z2g_colorus_2}.
In terms of helicity operators, 
\begin{align}
\hspace{-0.5cm}\cO_{\cB(us(n)) 0:++}^{(2)}&=-2g \left(i  f^{abd}\, \big({\cal Y}_n^T {\cal Y}_{\bar n}\big)^{dc}\right)  \omega_2  \cB^a_{n+,\omega_1}  \cB^b_{\bar n+,\omega_2}  \cB^c_{us(n)0} H\,,\nn \\
\cO_{\cB(us(n)) 0:--}^{(2)}&=-2g \left(i  f^{abd}\, \big({\cal Y}_n^T {\cal Y}_{\bar n}\big)^{dc}\right)   \omega_2  \cB^a_{n-,\omega_1}  \cB^b_{\bar n-,\omega_2} \cB^c_{us(n) 0} H\,, \nn \\
\hspace{-0.5cm}\cO_{\cB(us(\bar n)) 0:++}^{(2)}&=-2g \left(i  f^{abd}\, \big({\cal Y}_{\bar n}^T {\cal Y}_{n}\big)^{dc}\right)  \omega_1  \cB^a_{n+,\omega_1}  \cB^b_{\bar n+,\omega_2}  \cB^c_{us(\bar n)0} H\,,\nn \\
\cO_{\cB(us(\bar n)) 0:--}^{(2)}&=-2g \left(i  f^{abd}\, \big({\cal Y}_{\bar n}^T {\cal Y}_{n}\big)^{dc}\right)  \omega_1  \cB^a_{n-,\omega_1}  \cB^b_{\bar n-,\omega_2} \cB^c_{us(\bar n) 0} H\,.
\end{align}
This agrees with the relation derived from RPI symmetry, given in \Eq{eq:usRPIrelation}. For convenience, we also give the Feynman rule for the contribution of the hard scattering operators to a single ultrasoft emission both before BPS field redefinition
\begin{align}
\fd{2.75cm}{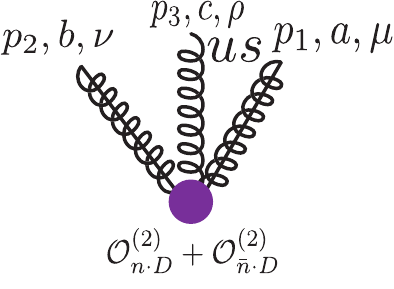}&=2g f^{abc} \omega_1 g^{\mu \nu}_\perp n^\rho -2g f^{abc} \omega_2 g^{\mu \nu}_\perp \bar n^\rho\,,
\end{align}
as well as after BPS field redefinition
\begin{align}
\fd{2.75cm}{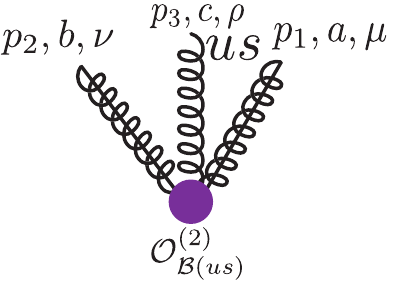}&=2g f^{abc} \left[ \omega_1 \left( n^\rho - \frac{ n \cdot p_3}{\bar n \cdot p_3} \bar n^\rho \right) - \omega_2  \left(\bar n^\rho - \frac{\bar n \cdot p_3}{n \cdot p_3} n^\rho \right) \right]  \nn \\
&=2g f^{abc} \left[  
	  n^\rho \left( \omega_1 + \frac{\bar n \cdot p_3}{n \cdot p_3} \omega_2  \right) 
-\bar n^\rho \left( \omega_2 + \frac{ n \cdot p_3}{\bar n \cdot p_3} \omega_1 \right) \right] \,.
\end{align}
Note that the contribution from hard scattering operators before the BPS field redefinition is local, but not gauge invariant, since before BPS field redefinition there are also SCET $T$-product diagrams involving. After BPS field redefinition, the contribution from the hard scattering operators is gauge invariant, but at the cost of locality. However, as emphasized in \cite{Feige:2017zci}, the form of the non-locality is dictated entirely by the BPS field redefinition, and is therefore not problematic. It is therefore advantageous to work in terms of the ultrasoft gauge invariant building blocks, so that the contributions from the hard scattering operators alone are gauge invariant.
Note also that here we have restricted the $\perp$ momentum of the two collinear particles to vanish for simplicity. Furthermore, because of the ultrasoft wilson lines in the color structure of \Eq{eq:Z2g_colorus}, there are also Feynman rules with multiple ultrasoft emissions. This is analogous to the familiar case of the $\cB_\perp$ operator which has Feynman rules for the emission of multiple collinear gluons.

\subsubsection{qqgg}

A basis for the operators involving two collinear quark and two collinear gluon fields was given in \Sec{sec:nnlp_collinear}. In \Sec{sec:contribs} it was argued that the only non-vanishing contributions to the cross section at $\cO(\lambda^2)$ arise from operators with the two collinear quarks and a collinear gluon in one sector, recoiling against a collinear gluon in the other sector.  

In performing the matching to these operators there are potentially $T$-product terms from the three gluon $\cO(\lambda^2)$ operator of \Sec{subsec:ggg_match_lam2}, where one of the gluons splits into a $q\bar q$ pair.
By choosing the momentum
\begin{align}
p_1^\mu&=\omega_1 \frac{n^\mu}{2}+p_\perp^\mu +p_{1r} \frac{\bar n^\mu}{2}\,, \quad p_2^\mu=\omega_2 \frac{n^\mu}{2}-p_\perp^\mu +p_{2r} \frac{\bar n^\mu}{2}\,, \quad
p_3^\mu=\omega_3 \frac{\bar n^\mu}{2}\,, \quad p_4^\mu=\omega_4 \frac{n^\mu}{2}\,,
\end{align}
we see from \Eq{eq:feynrule_ggg3} that all SCET $T$-product contributions vanish, so that the result must be reproduced by hard scattering operators in SCET. 
Expanding the QCD diagrams to $\cO(\lambda^2)$, we find that all the contributions from the two gluon vertex in the Higgs effective theory vanish
\begin{align}
\left. \fd{2.5cm}{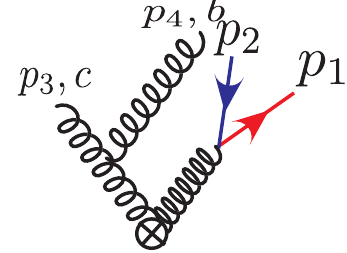}\right|_{\cO(\lambda^2)}  &=0\,, \quad \left.\fd{2.5cm}{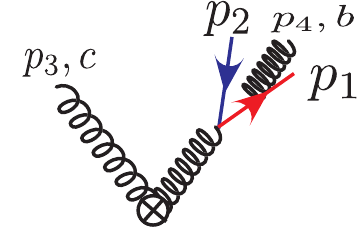}\right|_{\cO(\lambda^2)}  = 0\,, \quad \left.\fd{2.5cm}{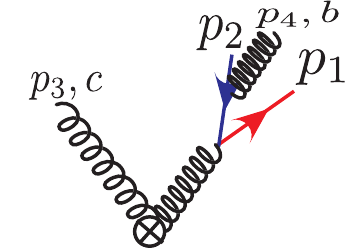}\right|_{\cO(\lambda^2)}   = 0\nn \\
\left.\fd{2.25cm}{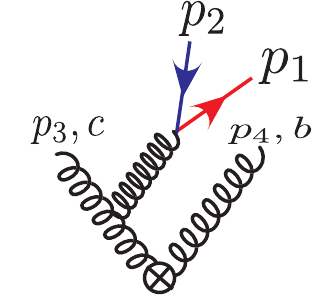}\right|_{\cO(\lambda^2)}  &=0\,, \quad \left.\fd{2.5cm}{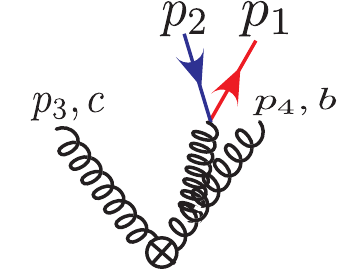}\right|_{\cO(\lambda^2)}  =0\,.
\end{align}
This result might be anticipated from the structure of the diagrams. However, there is a non-vanishing contribution from the three-gluon vertex in the Higgs effective theory
\begin{align}
\left. \fd{2.5cm}{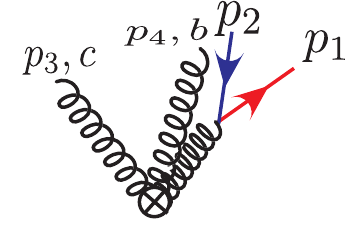} \right|_{\cO(\lambda^2)}  &=-\frac{4g^2 f^{abc} \omega_4 \epsilon_{3\perp}\cdot \epsilon_{4\perp}}{(\omega_1+\omega_2)^2}  \bar u_n(p_1) T^a \frac{\Sl{\bar n}}{2} v_n(p_2)\,. 
\end{align}
In terms of standard Lorentz and Dirac structures the corresponding hard scattering operator is given by
\begin{align}
\cO^{(2)}_{\cB1}=\frac{4g^2 if^{abc} \omega_4 }{(\omega_1+\omega_2)^2}\cB^b_{n\perp,\omega_4 }\cdot \cB^c_{\bar n \perp,\omega_3} \bar \chi_{n,\omega_1} T^a \frac{\Sl{\bar n}}{2} \chi_{n,-\omega_2}H\,.
\end{align}
Projected onto the helicity operator basis of \Eq{eq:Hqqgg_basis3}, and using the color basis of \Eq{eq:ggqqll_color}, we find 
\begin{align}
\cO^{(2)}_{\cB1++(0)}&=-\frac{4g^2 \omega_4 }{(\omega_1+\omega_2)^2}2\sqrt{\omega_1 \omega_2} \left(  (T^a T^b)_{\alpha \bbeta} -(T^b T^a)_{\alpha \bbeta}  \right) \cB^a_{n+,\omega_4} \cB^b_{\bar n+,\omega_3}  J_{n0}^{\balpha \beta} H\,,\nn \\
\cO^{(2)}_{\cB1--(0)}&=-\frac{4g^2 \omega_4 }{(\omega_1+\omega_2)^2}2\sqrt{\omega_1 \omega_2} \left(  (T^a T^b)_{\alpha \bbeta} -(T^b T^a)_{\alpha \bbeta}  \right) \cB^a_{n-,\omega_4} \cB^b_{\bar n-,\omega_3}  J_{n0}^{\balpha \beta} H\,, \nn\\
\cO^{(2)}_{\cB1++(\bar0)}&=-\frac{4g^2  \omega_4 }{(\omega_1+\omega_2)^2}2\sqrt{\omega_1 \omega_2} \left(  (T^a T^b)_{\alpha \bbeta} -(T^b T^a)_{\alpha \bbeta}  \right) \cB^a_{n+,\omega_4} \cB^b_{\bar n+,\omega_3}  J_{n\bar0}^{\balpha \beta}H\,, \nn \\
\cO^{(2)}_{\cB1--(\bar0)}&=-\frac{4g^2  \omega_4 }{(\omega_1+\omega_2)^2}2\sqrt{\omega_1 \omega_2} \left(  (T^a T^b)_{\alpha \bbeta} -(T^b T^a)_{\alpha \bbeta}  \right) \cB^a_{n-,\omega_4} \cB^b_{\bar n-,\omega_3}  J_{n\bar0}^{\balpha \beta}H\,.
\end{align}
For convenience, we also give the Feynman rule for the operator
\begin{align}
\fd{2.5cm}{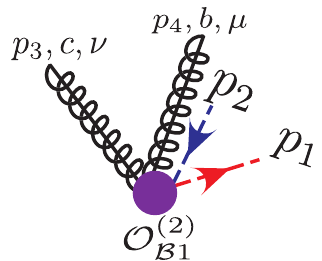}&=-\frac{4g^2 f^{abc} T^a \omega_4}{(\omega_1+\omega_2)^2} \left(  g^{\mu \nu}_\perp -\frac{p^\nu_{4\perp} \bar n^\mu}{\omega_4} \right) \frac{\Sl{\bar n}}{2}\,.
\end{align}
Again, this contains additional terms not present in the matching calculation, and it is straightforward to check that they are necessary to satisfy the required Ward identities.

\subsubsection{gggg}

Finally, we consider the matching to the operators involving four collinear gluon fields. A basis of such operators was given in \Eq{eq:H_basis_gggg_2}. In \Sec{sec:contribs} it was argued that to contribute to the cross section at $\cO(\lambda^2)$, there must be three collinear gluons in the same sector. For concreteness, we take this to be the $\bar n$ sector. The operators with three gluons in the $n$ sector can be obtained by crossing $\bar n \leftrightarrow n$.

To perform the matching we choose the momenta as
\begin{align}\label{eq:momentagggg}
p_1^\mu&=\omega_1 \frac{ n^\mu}{2}\,, \quad p_2^\mu=\omega_2 \frac{\bar n^\mu}{2}\,, \quad p_3^\mu=\omega_3 \frac{\bar n^\mu}{2}-p_\perp^\mu +p_{3r} \frac{n^\mu}{2}\,, \quad p_4^\mu=\omega_4 \frac{\bar n^\mu}{2}+p_\perp^\mu +p_{4r} \frac{ n^\mu}{2}\,.
\end{align}
With this choice, each particle in the $\bar n$ sector is on-shell, but the sum of any two of their momenta is off-shell,
\be
	p_i^2 = 0\,, \qquad (p_1 + p_j)^2 \sim \cO(1)\,, \qquad (p_j + p_k)^2 \sim \cO(\lambda^2) \,, \qquad j,k=2,3,4 \,;~ j\neq k\,,
\ee
which regulates all propagators. This particular choice of momenta is convenient since it simplifies $T$-product contributions from SCET. Furthermore, we take the external polarizations to be purely perpendicular, i.e. $\epsilon_i^\mu = \epsilon_{i\perp}^\mu$. All of the four gluon operators give a non-vanishing contribution to the four-gluon matrix element for this choice of polarization, allowing their Wilson coefficients to be obtained.

In computing the full theory diagrams for the matching it is convenient to separate the diagrams into those involving on-shell propagators, which will be partially reproduced by $T$-product terms in SCET, and diagrams involving only off-shell propagators. Since the four gluon operators obtain their power suppression entirely from the fields, for diagrams involving only off-shell propagators the residual momenta in \Eq{eq:momentagggg} can be ignored, as they contribute only power suppressed contributions. Diagrams with on-shell propagators are regulated by the residual momenta in \Eq{eq:momentagggg}.

We begin by considering the expansion of the full theory diagrams that don't involve any on-shell propagators. In this case, all $\perp$ momenta can be set to zero, and the result will be purely local. The relevant QCD diagrams expanded to $\cO(\lambda^2)$ arise from the four gluon vertex in the Higgs effective theory,\\
\noindent\begin{minipage}{.3\linewidth}
\begin{equation}
\hspace{2cm}\left.\fd{2.25cm}{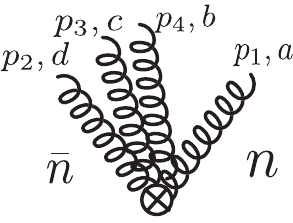}\right|_{\cO(\lambda^2)} \nn
\end{equation}
\end{minipage}%
\begin{minipage}{.7\linewidth}
\begin{align}
=&4ig^2 (f^{eab}f^{ecd}+f^{ead}f^{ecb}) \epsilon_{1\perp} \cdot \epsilon_{3\perp} \epsilon_{2\perp} \cdot \epsilon_{4\perp} \nn \\
&+4ig^2 (f^{eac}f^{ebd} +f^{ead}f^{ebc} ) \epsilon_{1\perp} \cdot \epsilon_{4\perp} \epsilon_{2\perp} \cdot \epsilon_{3\perp} \nn \\
&+4ig^2 ( f^{eab}f^{edc}+f^{eac}f^{edb}) \epsilon_{1\perp} \cdot \epsilon_{2\perp} \epsilon_{3\perp} \cdot \epsilon_{4\perp} \,, 
\end{align}
\end{minipage}\\
\newline
from a splitting off of the three gluon vertex,\\
\noindent\begin{minipage}{.3\linewidth}
\begin{equation}
\hspace{1cm}\left.\left( \fd{2.15cm}{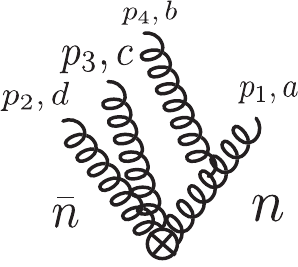}+\text{perms} \right)\right|_{\cO(\lambda^2)} \nn
\end{equation}
\end{minipage}%
\begin{minipage}{.7\linewidth}
\begin{align}
=&2ig^2 \left( \frac{\omega_3-\omega_2}{\omega_4}   \right) f^{abe}f^{cde} \epsilon_{1\perp} \cdot \epsilon_{4\perp} \epsilon_{3\perp} \cdot \epsilon_{2\perp} \nn \\
&  +[(2,d)\leftrightarrow (4,b)] +[(3,c)\leftrightarrow (4,b)] \,,
\end{align}
\end{minipage}\\
\newline
and from multiple emissions off of the two gluon vertex, either using the four gluon vertex with a single off-shell propagator\\
\noindent
\raisebox{1cm}{\begin{minipage}{.3\linewidth}
\begin{equation}
\hspace{0.25cm}\left.\left(\fd{2.15cm}{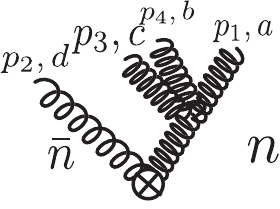}+\text{perms} \right) \right|_{\cO(\lambda^2)} \nn
\end{equation}
\end{minipage}}%
\begin{minipage}{.7\linewidth}
\begin{align}
\hspace{1cm}=& 2ig^2\left( \frac{\omega_2}{\omega_3+\omega_4}\right)\nn \\
& \left[   f^{bae} f^{cde} (\epsilon_{3\perp} \cdot \epsilon_{4\perp} \epsilon_{1\perp} \cdot \epsilon_{2\perp} -\epsilon_{4\perp} \cdot \epsilon_{2\perp} \epsilon_{3\perp} \cdot \epsilon_{1\perp}) \right.\nn \\
&+f^{bce}f^{ade} (\epsilon_{1\perp} \cdot \epsilon_{4\perp} \epsilon_{3\perp} \cdot \epsilon_{2\perp} -\epsilon_{4\perp} \cdot \epsilon_{2\perp} \epsilon_{1\perp} \cdot \epsilon_{3\perp})\nn \\
&\left. +f^{bde}f^{ace}(\epsilon_{1\perp} \cdot \epsilon_{4\perp} \epsilon_{3\perp} \cdot \epsilon_{2\perp}-\epsilon_{3\perp} \cdot \epsilon_{4\perp} \epsilon_{1\perp} \cdot \epsilon_{2\perp}) \right] \nn \\
&\hspace{-0cm}  +[(2,d)\leftrightarrow (3,c)] +[(2,d)\leftrightarrow (4,b)] \,,
\end{align}
\end{minipage}\\
\newline
\noindent or sequential emissions with two off-shell propagators

\noindent\begin{minipage}{.3\linewidth}
\begin{equation}
\left.\left(\fd{2.15cm}{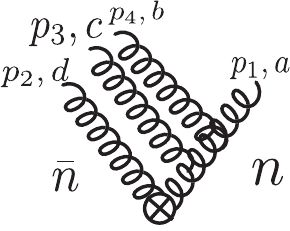}+\text{perms}\right) \right|_{\cO(\lambda^2)} \nn
\end{equation}
\end{minipage}%
\begin{minipage}{.7\linewidth}
\begin{align}
=&2ig^2 \frac{\omega_2 \omega_3}{\omega_4(\omega_3+\omega_4)}   \epsilon_{2\perp} \cdot \epsilon_{3\perp} \epsilon_{4\perp} \cdot \epsilon_{1\perp} f^{abe}f^{ecd}\nn \\
&  +[\text{perms}]   \,.
\end{align}
\end{minipage}\\
In the last case we have not explicitly listed the permutations, since all possible permutations are required.

We now consider the expansion of the full theory diagrams involving on-shell propagators. These will generically involve both local and non-local pieces. The non-local pieces will be directly reproduced by $T$-products in the effective theory.  The first class of diagrams involving on-shell propagators are those with all propagators on-shell. Here, at tree level, the dynamics occurs entirely within a single collinear sector. The two relevant QCD diagrams expanded to $\cO(\lambda^2)$ are
\begin{align}
\left.\fd{2.45cm}{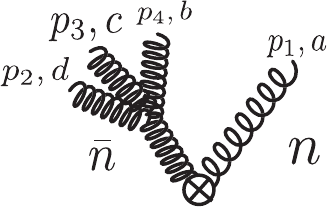} \right|_{\cO(\lambda^2)} &=0\,, \qquad
\left.\left(\fd{2.00cm}{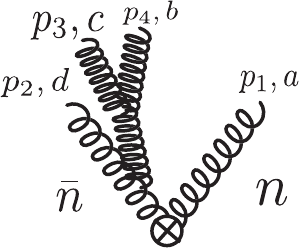} +\text{perms} \right)\right|_{\cO(\lambda^2)} &=0\,,
\end{align}
both of which have vanishing subleading power contributions. 

Next, we consider diagrams involving both on-shell and off-shell propagators. To simplify the results, we will often use the relation
\begin{align}
\frac{p_\perp^2}{(p_2+p_3)^2}=-\frac{\omega_3}{\omega_2}\,,
\end{align}
which will allow us to write the result in terms of a local term, which is just a rational function of the label momenta, and a non-local term, which explicitly contains the on-shell propagator. These non-local terms will be cancelled by the $T$-product diagrams in SCET.  For a first class of diagrams, where we have a nearly on-shell splitting in the $\bar n$-collinear sector, we have both a local term
\begin{align}
\left.\left(\fd{2.25cm}{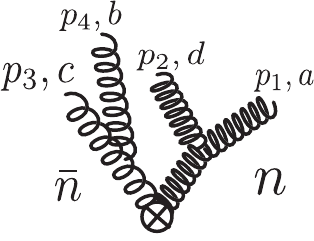}\right) \right|_{\cO(\lambda^2)}& =4ig^2 f^{aed}f^{bce}\frac{(\omega_3 - \omega_4)}{(\omega_3 + \omega_4)}\epsilon_{1\perp} \cdot \epsilon_{2\perp} \epsilon_{3\perp} \cdot \epsilon_{4\perp}\,,
\end{align}
when the splitting is into the particles $3$ and $4$, as well as a term that has both local and non-local pieces
\begin{align}
\left.\left(~\fd{2.25cm}{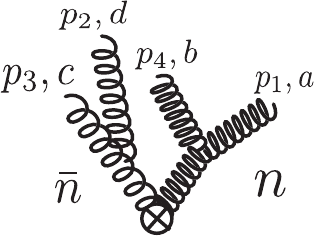}~ +~ \fd{2.25cm}{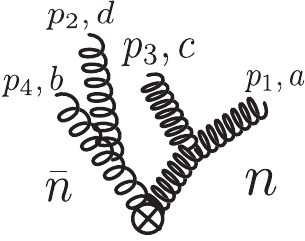}~\right) \right|_{\cO(\lambda^2)}& \\
&\hspace{-6cm} =\frac{4ig^2 f^{aeb}f^{dce}}{\omega_4}\left[ \frac{2(\omega_2 + \omega_3)}{(p_2 + p_3)^2}  p_\perp \cdot \epsilon_{1\perp}   p_\perp \cdot \epsilon_{2\perp}  \epsilon_{3\perp}\cdot \epsilon_{4\perp}  \right.\nn\\
&\hspace{-3.2cm}\left. -(2\omega_3 + \omega_4) \epsilon_{1\perp} \cdot \epsilon_{4\perp} \epsilon_{2\perp} \cdot \epsilon_{3\perp} \vphantom{\frac{(a)}{(b)}}\right]  + [3 \leftrightarrow 4, b \leftrightarrow c , p_\perp \to -p_\perp]\,. \nn
\end{align}
As will be discussed in more detail when we consider the corresponding diagrams in the EFT, the first permutation is purely local, since there is no corresponding $T$-product term in the effective theory, and thus it must be fully reproduced by a hard scattering operator. This particular splitting allows a slight simplification in the calculation of the SCET diagrams. For a second class of diagrams, where we have an on-shell splitting emitted from an off-shell leg, we again have a purely local term
\begin{align}
\left.\left(\fd{2.05cm}{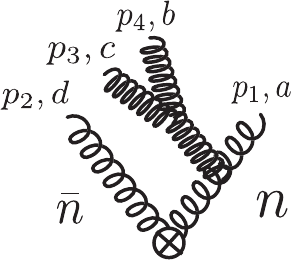}\right) \right|_{\cO(\lambda^2)}& =0\,, 
\end{align}
as well as non-local contributions,
\begin{align}
\left.\left(~\fd{2.05cm}{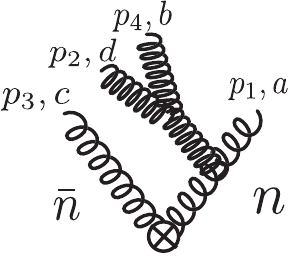}~ +~ \fd{2.05cm}{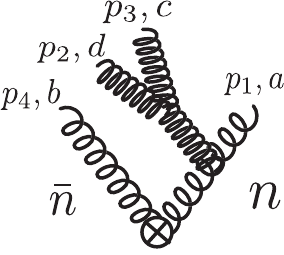}~\right) \right|_{\cO(\lambda^2)}& \\
&\hspace{-6.5cm} =2ig^2 f^{aeb}f^{dce}\left[ \left(\frac{4\omega_4}{(\omega_2 + \omega_3)(p_2 + p_3)^2}\right)   p_\perp \cdot \epsilon_{1\perp} p_\perp \cdot \epsilon_{2\perp} \epsilon_{3\perp}\cdot \epsilon_{4\perp} \right.\nn\\
&\hspace{-4.5cm}\left. -\frac{\omega_3(\omega_2 - \omega_3)(\omega_2 + \omega_3 + \omega_4)^2}{\omega_2\omega_4(\omega_2 + \omega_3)^2} \epsilon_{1\perp} \cdot \epsilon_{4\perp} \epsilon_{2\perp} \cdot \epsilon_{3\perp}  \right]  + [3 \leftrightarrow 4, b \leftrightarrow c , p_\perp \to -p_\perp]\,.\nn
\end{align}
Again, we see the same pattern, that the first permutation gives rise to a purely local term, while the second two permutations give rise to both local and non-local terms.

Finally, we have the diagrams involving the three gluon vertex in the Higgs effective theory. We again have a local contribution
\begin{align}
\left.\left(\fd{2.05cm}{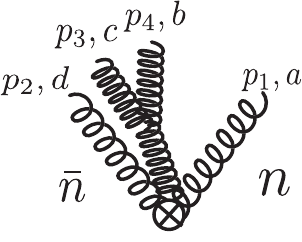}\right) \right|_{\cO(\lambda^2)}& =-2ig^2 f^{ade}f^{ebc} \frac{\omega_2(\omega_3-\omega_4)}{(\omega_3 + \omega_4)^2} \epsilon_{1\perp} \cdot \epsilon_{2\perp} \epsilon_{3\perp} \cdot \epsilon_4\,, 
\end{align}
and a non-local contribution
\begin{align}
\left.\left(~\fd{2.05cm}{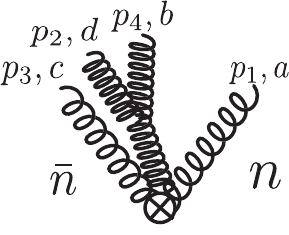}~ +~ \fd{2.05cm}{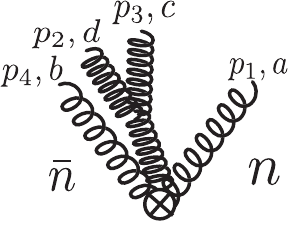}~\right) \right|_{\cO(\lambda^2)}&  \\
&\hspace{-6.5cm} =2ig^2 f^{aeb}f^{dce}\left[ \frac{8}{(p_2 + p_3)^2}  p_\perp \cdot \epsilon_{1\perp}  p_\perp \cdot \epsilon_{2\perp}   \epsilon_{3\perp}\cdot \epsilon_{4\perp}  \right.\nn\\
&\hspace{-4.cm}\left. -\biggl\{\frac{ (\omega_3 + \omega_4)^2 - \omega_2\omega_3}{\omega_2\omega_4}\biggr\}   \epsilon_{1\perp} \cdot \epsilon_{4\perp}   \epsilon_{2\perp} \cdot \epsilon_{3\perp}  \right]  + [3 \leftrightarrow 4, b \leftrightarrow c , p_\perp \to -p_\perp]\,. \nn
\end{align}

The non-local terms in the above expansions must be reproduced by $T$-product terms in the effective theory. First, there are potential contributions from $\cO^{(2)}_{\cP\cB}$, with the two gluon Feynman rule for $\cB_{\bar n,\perp}$, which is given in \App{app:expand_gluon}. Such contributions give vanishing overlap for our choice of $\perp$ polarizations. There are however $T$-product contributions arising from the  three gluon $\cO^{(2)}_{\cP\cB}$ operator, with an  $\cL^{(0)}$ insertion.
The three gluon Feynman rule for the $\cO^{(2)}_{\cP\cB}$ vertex was given in \Eq{eq:feynrule_ggg3}.
Since the $\cO^{(2)}_{\cP\cB}$ operator has an explicit $\cP_\perp$ insertion, it vanishes in the case that either of the particles in the $\bar n$ sector has no perpendicular momentum. This is why our particular choice of momenta for the matching simplifies the structure of the $T$-products. The two non-vanishing permutations are given by
\begin{align}\label{eq:ggggSCET}
    & \fd{2.00cm}{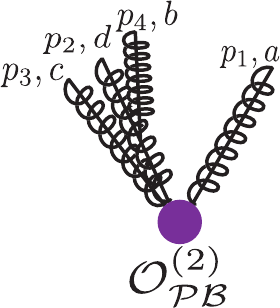} + \fd{2.00cm}{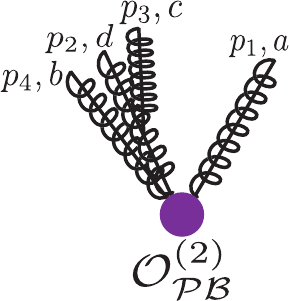}  \\
	=&-8ig^2 f^{abe}f^{ecd} \frac{(\omega_2 + \omega_3 + \omega_4)^2 }{(\omega_3 + \omega_2)\omega_4}\left[ \frac{\omega_3}{(\omega_2 + \omega_3)} \epsilon_{1\perp}\cdot \epsilon_{4\perp}   \epsilon_{3\perp}\cdot \epsilon_{2\perp} - \frac{ p_\perp\cdot\epsilon_{1\perp} p_\perp\cdot \epsilon_{2\perp} \epsilon_{3\perp} \cdot \epsilon_{4\perp}}{(p_2 + p_3)^2} \right] \nn \\
	&+ [3 \leftrightarrow 4, b \leftrightarrow c , p_\perp \to -p_\perp]\,, \nn 
\end{align}
which consists both of a local and a non-local term. The non-local terms exactly reproduce the ones obtained in the QCD expansion 
\begin{align}
	\hspace{-1cm}
   \left(\fd{1.80cm}{figures/matching_gggg_scetb_low} + \fd{1.80cm}{figures/matching_gggg_scetc_low} \right)_\text{\!\!\!non-loc.} \hspace{-0.5cm}
    & \!\! = \left(\fd{1.90cm}{figures/matching_lam2_gggg4b_low} + \fd{1.90cm}{figures/matching_lam2_gggg6b_low} + \fd{1.90cm}{figures/matching_lam2_gggg8b_low} +\! \text{ perms }\!\!\! \right)_\text{\!\!\!non-loc.} \hspace{-1cm} \nn \\[0.4cm]
	&\hspace{-5cm}=8ig^2    p_\perp\cdot\epsilon_{1\perp}  p_\perp\cdot \epsilon_{2\perp}   \epsilon_{3\perp} \cdot \epsilon_{4\perp} \left( \frac{f^{abe}f^{ecd}} {(p_2 + p_3)^2} \frac{(\omega_2 + \omega_3 + \omega_4)^2 }{(\omega_3 + \omega_2)\omega_4} + [3 \leftrightarrow 4 , b \leftrightarrow c] \right) \,. 
\end{align}
While it is of course necessary that the EFT reproduces all such non-local terms, this is also a highly non trivial cross check of both the three and four gluon matching.

The matching coefficients for the hard scattering operators are given by the remaining local terms. Before presenting the result we briefly comment on the organization of the color structure. All diagrams are proportional to $f^{abe}f^{cde}$, $f^{ace}f^{bde}$ or $f^{ade}f^{bce}$, which are related by the Jacobi identity $f^{abe}f^{cde} = f^{ace}f^{bde} - f^{ade}f^{bce}$. A basis in terms of structure constants can easily be related to the trace basis of \eqref{eq:gggg_color} using
\begin{align}\label{eq:colmatrrel}
f^{ace}f^{bde} &= \tr[abdc] + \tr[acdb] - \tr[acbd] - \tr[adbc] = e_2 - e_3\,,\nn \\
f^{ade}f^{bce} &= \tr[abcd] + \tr[adcb] - \tr[acbd] - \tr[adbc] = e_1 - e_3\,, 
\end{align}
where $e_i$ is the $i$-th element of the basis in \eqref{eq:gggg_color}. We find it most convenient to write the Wilson coefficient in the $(f^{ace}f^{bde},f^{ade}f^{bce})$ basis.
After subtracting the local piece of the SCET $T-$product of \eqref{eq:ggggSCET} from the full theory graphs, and manipulating the result to bring it into a compact form, we find the following operator
\begin{align}
\cO^{(2)}_{4g}= 16 \pi \alpha_s f^{ade}f^{bce}  (\cB^a_{n\perp,\omega_i} \cdot \cB^b_{\bar n \perp, \omega_j})(\cB^c_{\bar n \perp, \omega_k} \cdot \cB_{\bar n \perp, \omega_\ell}^d) \left( 3 + \dfrac{\omega_j^3 + \omega_k^3 + \omega_\ell^3 + \omega_j\omega_k\omega_\ell}{(\omega_j + \omega_k)(\omega_j + \omega_\ell)(\omega_k + \omega_\ell)}\right) \,.
\end{align}
The Wilson coefficient is manifestly RPI-III invariant. When the matrix element of this operator is taken we are forced to sum over permutations which gives the proper Bose symmetric result, as well as inducing terms with other color structures. In terms of the helicity operators of \Eq{eq:H_basis_gggg_2}, we have
\begin{align}
\cO^{(2)}_{4g} 
 &= 16 \pi \alpha_s f^{ade}f^{bce}  \left( 3 + \dfrac{\omega_j^3 + \omega_k^3 + \omega_\ell^3 + \omega_j\omega_k\omega_\ell}{(\omega_j + \omega_k)(\omega_j + \omega_\ell)(\omega_k + \omega_\ell)}\right) \nn \\
 &\ \ \times \Big[ \cB^a_{n+,\omega_i}   \cB^b_{\bar n +, \omega_j} \cB^c_{\bar n +, \omega_k}   \cB_{\bar n -, \omega_\ell}^d + \cB^a_{n+,\omega_i}   \cB^b_{\bar n +, \omega_j} \cB^c_{\bar n -, \omega_k}   \cB_{\bar n +, \omega_\ell}^d  
 \nn \\
 &\qquad +\cB^a_{n-,\omega_i}   \cB^b_{\bar n -, \omega_j} \cB^c_{\bar n +, \omega_k}   \cB_{\bar n -, \omega_\ell}^d + \cB^a_{n-,\omega_i}   \cB^b_{\bar n -, \omega_j} \cB^c_{\bar n -, \omega_k}   \cB_{\bar n +, \omega_\ell}^d \Big] 
 \nn \\
 &= 16 \pi \alpha_s  
  \biggl[ 3 + \dfrac{\omega_j^3 \!+\! \omega_k^3 \!+\! \omega_\ell^3 \!+\! \omega_j\omega_k\omega_\ell}{(\omega_j \!+\! \omega_k)(\omega_j \!+\! \omega_\ell)(\omega_k \!+\! \omega_\ell)}\biggr]
  \Big[ (f^{ade}f^{bce}\! +\! f^{ace}f^{bde}) \cB^a_{n+,\omega_i}   \cB^b_{\bar n +, \omega_j} \cB^c_{\bar n +, \omega_k}   \cB_{\bar n -, \omega_\ell}^d  \nn \\
& \qquad\qquad 
  - (f^{ade}f^{bce} + f^{abe}f^{cde} )\cB^a_{n-,\omega_i}   \cB^b_{\bar n +, \omega_j} \cB^c_{\bar n -, \omega_k}   \cB_{\bar n -, \omega_\ell}^d \Big] 
 \,.
\end{align}
We see that all the helicity selection rules are satisfied in the tree level matching, as expected. We have also checked the result using the automatic FeynArts \cite{Hahn:2000kx} and FeynRules implementation of the HiggsEffectiveTheory \cite{Alloul:2013bka}. For more complicated calculations at subleading power in SCET it would be interesting to fully automate the computation of Feynman diagrams involving power suppressed SCET operators and Lagrangians.

The four gluon operators derived in this section can be used to study $\cO(\alpha_s^2)$  collinear contributions at $\cO(\lambda^2)$. It would be interesting to understand in more detail the universality of collinear splittings at subleading power, as well as collinear factorization properties. For some recent work in this direction from a different perspective, see \cite{Stieberger:2015kia,Nandan:2016ohb}.
The behavior of these Wilson coefficients is also quite interesting. They exhibit a singularity as any pair of collinear particles simultaneously have their energy approach zero. This was also observed in the Wilson coefficients for operators describing the subleading collinear limits of two gluons emitted off of a $q\bar q$ vertex \cite{Feige:2017zci}.

\section{Conclusions}\label{sec:conclusions}

In this paper we have presented a complete basis of operators at $\cO(\lambda^2)$ in the SCET expansion for color singlet production of a scalar through gluon fusion, as relevant for $gg\to H$. To derive a minimal basis we used operators of definite helicities, which allowed us to significantly reduce the number of operators in the basis. This simplification is due to helicity selection rules which are particularly constraining due to the scalar nature of the produced particle. We also classified all possible operators which could contribute to the cross section at $\cO(\lambda^2)$. In performing this classification the use of a helicity basis again played an important role, allowing us to see from simple helicity selection rules which operators could contribute.  While the total number of subleading power operators is large, the number that contribute at the cross section level is smaller.  We compared the structure of the contributions to the case of a quark current, $\bar q \Gamma q$, finding interesting similarities, despite a slightly different organization in the effective theory.

A significant portion of this paper was devoted to a tree level calculation of the Wilson coefficients of the subleading power operators which can contribute to the cross section at $\cO(\lambda^2)$. The Wilson coefficients obtained in this matching will allow for a study of the power corrections at NLO and for the study of the leading logarithmic renormalization group structure at subleading power.  An initial investigation of the renormalization group properties of several subleading power operators relevant for the case of $e^+e^-\to \bar q q$ was considered in \cite{Freedman:2014uta}. 

A number of directions exist for future study, with the goal of understanding factorization at subleading power. In particular, one would like to combine the hard scattering operators derived in this paper with the subleading SCET Lagrangians to derive a complete factorization theorem at subleading power for a physical event shape observable.  Combined with the operators in \cite{Feige:2017zci}, all necessary ingredients are now available to construct such a subleading factorization for thrust for $\bar q q$ or $gg$ dijets in $e^+e^-$ collisions. This would also allow for a test of the universality of the structure of the subleading factorization. The operators of this paper can also be used to study threshold resummation, where power corrections of $\mathcal{O}((1-z)^0)$ have received considerable attention \cite{Dokshitzer:2005bf,Grunberg:2007nc,Laenen:2008gt,Laenen:2008ux,Grunberg:2009yi,Laenen:2010uz,Almasy:2010wn,Bonocore:2014wua,White:2014qia,deFlorian:2014vta,Bonocore:2015esa,Bonocore:2016awd}, particularly for the $q\bar q$ channel, but it would be interesting to extend this to the $gg$ case.

An interesting application of current relevance of the results presented in this paper is to the calculation of fixed order power corrections for NNLO event shape based subtractions. Gaining analytic control over power corrections can significantly improve the performance and stability of such subtraction schemes. This has been studied for $q\bar q$ initiated Drell Yan production to NNLO in \cite{Moult:2016fqy} using a subleading power operator basis in SCET (see also \cite{Boughezal:2016zws} for a direct calculation in QCD). Combined with the results for the operator basis and matching for $q\bar q$ initiated processes given in \cite{Feige:2017zci}, the operator basis presented in this paper will allow for the systematic study of power corrections for color singlet production and decay.

\begin{acknowledgments}
We thank the Erwin Schr\"odinger Institute and the organizers of the ``Challenges and Concepts for Field Theory and Applications in the Era of LHC Run-2'' workshop for hospitality and support while portions of this work were completed. This work was supported in part by the Office of Nuclear Physics of the U.S. Department of Energy under the Grant No.~DE-SC0011090, by the Office of High Energy Physics of the U.S. Department of Energy under Contract No. DE-AC02-05CH11231, and the LDRD Program of LBNL. I.S. was also supported by the Simons Foundation through the Investigator grant 327942.
\end{acknowledgments}

\appendix

\section{Generalized Basis with $\cP_{\perp n},~\cP_{\perp \bar n}\neq 0$}\label{app:gen_pt}

In the main text we presented a complete basis of operators to $\cO(\lambda^2)$ in a frame where the total $\cP_\perp$ in each collinear sector is restricted to be zero. In this section we extend the basis, giving the additional operators present when the individual collinear sectors have non-vanishing $\cP_\perp$. We then perform a tree level matching calculation to those operators which can contribute to the cross section at $\cO(\lambda^2)$. While all these operators are fixed by RPI, we choose to find their coefficients by simply performing the tree level matching with more general kinematics.

\subsection{Operators}\label{app:gen_pt_op}

We begin by noting that operators involving two collinear gluon fields with a single insertion of the $\cP_\perp$ operator are eliminated by the helicity selection rules. Operators involving two collinear gluon fields must therefore have two insertions of the $\cP_\perp$ operator. A basis of helicity operators involving two insertions of the $\cP_\perp$ operator, where one $\cP_\perp$ operator acts in each collinear sector, is given by
\begin{align}
 \boldsymbol{(\cP_\perp g_n) (\cP_\perp g_{\bn}):}   {\vcenter{\includegraphics[width=0.18\columnwidth]{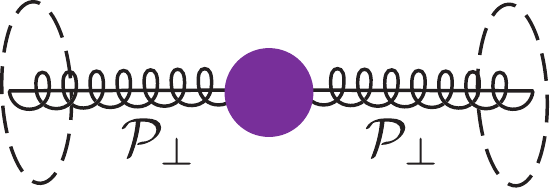}}} \nn
\end{align}
\vspace{-0.4cm}
\begin{alignat}{2}\label{eq:hgg_perp}
 &O_{\cP\cB\cP++[-:-]}^{(0)ab}
=[\cP_\perp^-  \cB_{n +}^a] \, [\cP^-_\perp \cB_{\bar n+}^b]\,   H
\,, \qquad &&O_{\cP\cB\cP--[+:+]}^{(0)ab}
=  [\cP^+_\perp \cB_{n-}^a]\, [\cP^+_\perp \cB_{ \bn -}^b]\, H\,, \nn \\
 &O_{\cP\cB\cP+-[-:+]}^{(0)ab}
=[\cP_\perp^-  \cB_{n +}^a] \, [\cP^+_\perp \cB_{\bar n-}^b]\,   H
\,, \qquad &&O_{\cP\cB\cP-+[+:-]}^{(0)ab}
=  [\cP^+_\perp \cB_{n-}^a]\, [\cP^-_\perp \cB_{ \bn +}^b]\, H\,, \nn \\
 &O_{\cP\cB\cP++[+:+]}^{(0)ab}
=  [\cP^+_\perp \cB_{n +}^a]\, [\cP^+_\perp \cB_{\bar n+}^b]\,   H
\,, \qquad &&O_{\cP\cB\cP--[-:-]}^{(0)ab}
=   [\cP^-_\perp\cB_{n-}^a]\,  [\cP^-_\perp\cB_{ \bn -}^b]\, H
 \,.
\end{alignat}
When both $\cP_\perp$ operators act on the same collinear sector, which we take to be the $n$-collinear sector, then we have
\begin{align}
 \boldsymbol{(\cP_\perp \cP_\perp g_n) g_{\bn}:}   {\vcenter{\includegraphics[width=0.18\columnwidth]{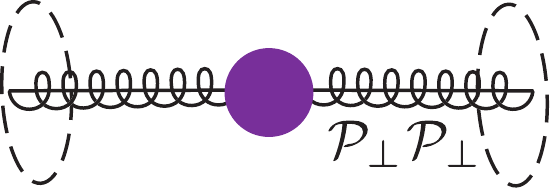}}} \nn
\end{align}
\vspace{-0.4cm}
\begin{alignat}{2}\label{eq:hgg_perp2}
 &O_{\cB\cP\cP++[-+]}^{(0)ab}
=[\cP_\perp^- \cP^+_\perp  \cB_{n +}^a] \, [ \cB_{\bar n+}^b]\,   H
\,, \qquad &&O_{\cB\cP\cP--[-+]}^{(0)ab}
=  [\cP^-_\perp \cP^+_\perp  \cB_{n-}^a]\, [ \cB_{ \bn -}^b]\, H\,, \nn \\
 &O_{\cB\cP\cP++[++]}^{(0)ab}
=  [\cP^+_\perp \cP^+_\perp \cB_{n -}^a]\, [ \cB_{\bar n+}^b]\,   H
\,, \qquad &&O_{\cB\cP\cP--[--]}^{(0)ab}
=   [\cP^-_\perp  \cP^-_\perp \cB_{n+}^a]\,  [\cB_{ \bn -}^b]\, H\,.
\end{alignat}
Note that we have used up our freedom to integrate by parts by never having the $\cP_\perp$ operator act on the $H$ field. The color basis for all these operators before and after the BPS field redefinition is the same as given in \eq{leading_color}.

We also must consider the generalization of the operators involving three gluon or quark fields to generic $\perp$ momentum in the collinear sectors.
As discussed in the text surrounding \Eq{eq:Hqqgpperp_basis_same}, in the case that the $\cP_\perp$ operator is inserted into an operator involving two quark fields and a gluon field, the helicity structure of the operator is highly constrained. In particular, the quark fields must be in a helicity zero configuration, and also have the same chirality. This implies that all operators must involve only the currents $J_{\bar n\, 0}^{\balpha\bt}$ or $J_{\bar n\, \bar 0}^{\balpha\bt}$. Here we have taken without loss of generality that the two quarks are in the $\bar n$-collinear sector. The basis of $\cO(\lambda^2)$ operators for the case that the $\cP_\perp$ operator acts on the $\bar n$ sector, is then given by
\begin{align}
&   \boldsymbol{(g)_n (q\bar q\, \cP_\perp)_{\bn}:}{\vcenter{\includegraphics[width=0.18\columnwidth]{figures/Subleading_qqbarg_orderlam_perp_low}}}  \nn
\end{align}
\vspace{-0.4cm}
\begin{alignat}{2}\label{eq:Hqqgpperp_basis_same_generalized1}
&O_{\cP \chi + (0)[+]}^{(2)a\,\balpha\bt}
= \cB_{n+}^a\, \big\{ \cP_{\perp}^{+} J_{\bar n\, 0}^{\balpha\bt} \big\}\,  H
\,,\qquad &
&O_{\cP\chi - (0)[-]}^{(2)a\,\balpha\bt}
= \cB_{n-}^a\, \big\{ \cP_{\perp}^{-} J_{\bar n\, 0    }^{\balpha\bt} \big\} \, H 
\,,\\
&O_{\cP\chi + (\bar 0)[+]}^{(2)a\,\balpha\bt}
=  \cB_{n+}^a\,  \big\{ \cP_{\perp}^{+} J_{\bar n\, \bar0    }^{\balpha\bt} \big\}\,  H
\,,\qquad &
&O_{\cP\chi - (\bar 0)[-]}^{(2)a\,\balpha\bt}
= \cB_{n-}^a \, \big\{ \cP_{\perp}^{-} J_{\bar n\, \bar0    }^{\balpha\bt} \big\}  \, H
\,,\nn \\
&O_{\cP \chi + (0)[+]}^{(2)a\,\balpha\bt}
= \cB_{n+}^a\, \big\{  J_{\bar n\, 0}^{\balpha\bt} \cP_{\perp}^{\dagger+} \big\}\,  H
\,,\qquad &
&O_{\cP\chi - (0)[-]}^{(2)a\,\balpha\bt}
= \cB_{n-}^a\, \big\{  J_{\bar n\, 0    }^{\balpha\bt} \cP_{\perp}^{\dagger-} \big\} \, H 
\,,\nn \\
&O_{\cP\chi + (\bar 0)[+]}^{(2)a\,\balpha\bt}
=  \cB_{n+}^a\,  \big\{ J_{\bar n\, \bar0    }^{\balpha\bt} \cP_{\perp}^{\dagger+}  \big\}\,  H
\,,\qquad &
&O_{\cP\chi - (\bar 0)[-]}^{(2)a\,\balpha\bt}
= \cB_{n-}^a \, \big\{  J_{\bar n\, \bar0    }^{\balpha\bt}  \cP_{\perp}^{\dagger-} \big\}  \, H
\,,\nn
\end{alignat}
which replaces the four operators in \eq{Hqqgpperp_basis_same}.  For the case that the $\cP_\perp$ operator acts on the $n$ sector the basis is
\begin{align}
&   \boldsymbol{(\cP_\perp g)_n (q\bar q)_{\bn}:}{\vcenter{\includegraphics[width=0.18\columnwidth]{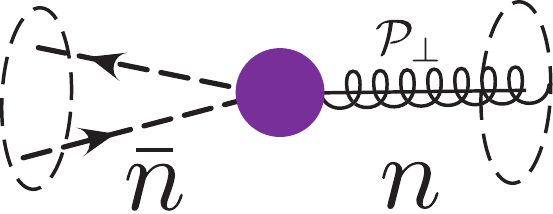}}}  \nn
\end{align}
\vspace{-0.4cm}
\begin{alignat}{2}\label{eq:Hqqgpperp_basis_same_generalized2}
&O_{\cP \chi + (0)[+]}^{(2)a\,\balpha\bt}
= [\cP_{\perp}^{-} \cB_{n+}^a]\,   J_{\bar n\, 0}^{\balpha\bt} \,  H
\,,\qquad &
&O_{\cP\chi - (0)[-]}^{(2)a\,\balpha\bt}
= [\cP_{\perp}^{+} \cB_{n-}^a]\,   J_{\bar n\, 0    }^{\balpha\bt}  \, H 
\,,\\
&O_{\cP\chi + (\bar 0)[+]}^{(2)a\,\balpha\bt}
= [\cP_{\perp}^{-} \cB_{n+}^a]\,   J_{\bar n\, \bar0    }^{\balpha\bt} \,  H
\,,\qquad &
&O_{\cP\chi - (\bar 0)[-]}^{(2)a\,\balpha\bt}
=[\cP_{\perp}^{+} \cB_{n-}^a] \,  J_{\bar n\, \bar0    }^{\balpha\bt}  \, H
\,.\nn
\end{alignat}
The color basis for all these operators (before and after the BPS field redefinition) is the same as given in \eqs{nnlp_color_quark_perp}{nnlp_color_quark_perpBPS}.

The final case we must consider are the generalized versions of \Eq{eq:Hgggpperp_basis}, which involve the insertion of a single $\cP_\perp$ operator into an operator involving three collinear gluon fields. In this case a basis of  $\cO(\lambda^2)$ operators for the case that the $\cP_\perp$ operator acts in the $\bar n$-collinear sector is given by
\begin{align}
&  \boldsymbol{(g)_n (gg\, \cP_\perp)_{\bn}:}{\vcenter{\includegraphics[width=0.18\columnwidth]{figures/Subleading_3g_perp_low}}}  \nn
\end{align}
\vspace{-0.4cm}
\begin{alignat}{2} \label{eq:Hgggpperp_basis_general1}
&O_{\cP\cB +++[-]}^{(2)abc}
=  \cB_{n+}^a\, \cB_{\bar n+}^b\, \left [\cP_{\perp}^{-} \cB_{\bar n+}^c \right ] \,H
\,,\qquad && O_{\cP\cB ---[+]}^{(2)abc}
= \cB_{n-}^a\, \cB_{\bar n-}^b\, \left [\cP_{\perp}^{+} \cB_{\bar n-}^c \right ] \,H\,,
 \nn \\
&O_{\cP\cB ++-[+]}^{(2)abc}
= \, \cB_{n+}^a\, \cB_{ \bar n+}^b\, \left [\cP_{\perp}^{+} \cB_{\bar n-}^c \right ] \,H
\,,\qquad 
&&O_{\cP\cB --+[-]}^{(2)abc}
= \cB_{n-}^a\, \cB_{ \bn -}^b\, \left [\cP_{\perp}^{-} \cB_{\bar n +}^c \right ] \,H
\,, \nn \\
&O_{\cP\cB +-+[+]}^{(2)abc}
= \, \cB_{n+}^a\, \cB_{ \bar n-}^b\, \left [\cP_{\perp}^{+} \cB_{\bar n+}^c \right ] \,H
\,,\qquad 
&&O_{\cP\cB -+-[-]}^{(2)abc}
= \cB_{n-}^a\, \cB_{ \bn +}^b\, \left [\cP_{\perp}^{-} \cB_{\bar n -}^c \right ] \,H
\,,
\end{alignat}
where these six operators replace the four in \eq{Hgggpperp_basis}. In addition we have operators for the case that the $\cP_\perp$ acts in the $n$-collinear sector,\\[-10pt]
\vbox{
\begin{align}
&  \boldsymbol{(\cP_\perp g)_n (gg)_{\bn}:}{\vcenter{\includegraphics[width=0.18\columnwidth]{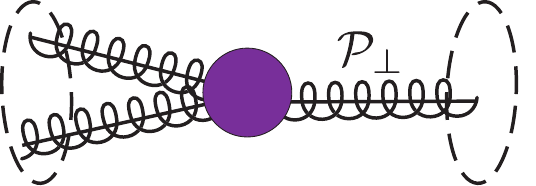}}}  \nn
\end{align}
\vspace{-0.4cm}
\begin{alignat}{2} \label{eq:Hgggpperp_basis_general2}
&O_{\cP\cB +++[-]}^{(2)abc}
=  [\cP_{\perp}^{+} \cB_{n+}^a]\, \cB_{\bar n+}^b\,  \cB_{\bar n+}^c  \,H
\,,\qquad && O_{\cP\cB ---[+]}^{(2)abc}
=[ \cP_{\perp}^{-} \cB_{n-}^a]\, \cB_{\bar n-}^b\,  \cB_{\bar n-}^c  \,H\,,
 \nn \\
&O_{\cP\cB ++-[+]}^{(2)abc}
=[\cP_{\perp}^{-}  \, \cB_{n+}^a\,] \cB_{ \bar n+}^b\,  \cB_{\bar n-}^c \,H
\,,\qquad 
&&O_{\cP\cB --+[-]}^{(2)abc}
= [\cP_{\perp}^+ \cB_{n-}^a\,] \cB_{ \bn -}^b\,  \cB_{\bar n +}^c  \,H
\,.
\end{alignat}
}
The color basis for all of these operators is the same as given in \eq{ggg_perp_color}.

\subsection{Matching}\label{app:gen_pt_match}

We now consider the matching to these operators. We begin with the matching to the operators involving two $\cP_\perp$ insertions into the leading power operator. We use a two gluon final state and take the kinematics as
\begin{align}
p_1^\mu =\omega_1 \frac{n^\mu}{2}+p_{1\perp}^\mu+p_{1r}\frac{\bar n^\mu}{2} , \qquad  p_2 =\omega_2 \frac{\bar n^\mu}{2}+p_{2\perp}^\mu +p_{2r} \frac{n^\mu}{2}\,.
\end{align}
We find at $\cO(\lambda^2)$
\begin{align}
\left. \fd{2.0cm}{figures/matching_gg_low}\right|_{\cO(\lambda^2} &= -4i \delta^{ab} p_{1\perp}\cdot p_{2\perp} \epsilon_{3\perp} \cdot \epsilon_{4\perp}\,.
\end{align}
This is recognized as the tree level matrix element of the operator
\begin{align}\label{eq:ggpp_matched}
\cO^{(2)}_{\cP \cB  \cP}= -4 \delta^{ab} g_{\mu \nu}g_{\delta \rho} [\cP^\delta_{\perp} \cB^{a\nu}_{n\perp,\omega_1}] [\cP^\rho_{\perp} \cB^{b\mu}_{\bar n \perp,\omega_2}]\,,
\end{align}
or in terms of helicity operators,
\begin{alignat}{2}
 &O_{\cP\cB\cP++[-:-]}^{(0)ab}
=-4[\cP_\perp^-  \cB_{n +}^a] \, [\cP^-_\perp \cB_{\bar n+}^b]\,   H
\,, \qquad &&O_{\cP\cB\cP--[+:+]}^{(0)ab}
=  -4[\cP^+_\perp \cB_{n-}^a]\, [\cP^+_\perp \cB_{ \bn -}^b]\, H\,, \nn \\
 &O_{\cP\cB\cP++[+:+]}^{(0)ab}
=  -4[\cP^+_\perp \cB_{n +}^a]\, [\cP^+_\perp \cB_{\bar n+}^b]\,   H
\,, \qquad &&O_{\cP\cB\cP--[-:-]}^{(0)ab}
=   -4[\cP^-_\perp\cB_{n-}^a]\,  [\cP^-_\perp\cB_{ \bn -}^b]\, H\,,
\end{alignat}
We see that not all possible helicity combinations appear in the tree level matching. Furthermore, the operators of \Eq{eq:hgg_perp2} where both $\cP_\perp$ insertions are in the same collinear sector do not appear at this order.

We now consider the matching to the operators of \Eqs{eq:Hqqgpperp_basis_same_generalized1}{eq:Hqqgpperp_basis_same_generalized2}. We can simplify the matching by performing it in two steps. First, to extract the Wilson coefficient of the operator involving the action of the $\cP_\perp$ on the collinear gluon field we take our kinematics as
\begin{align}
\hspace{-0.3cm} p_1^\mu=\omega_1 \frac{n^\mu}{2}+p_\perp^\mu +p_{1r} \frac{\bar n^\mu}{2}\,, \quad  p_2^\mu=\omega_2 \frac{n^\mu}{2}-p_\perp^\mu +p_{2r} \frac{\bar n^\mu}{2}\,, \quad p_3^\mu=\omega_3 \frac{\bar n^\mu}{2}+p_{3\perp}^\mu+p_{3r} \frac{n^\mu}{2}\,.
\end{align}
With this choice, all subleading Lagrangian insertions vanish, for similar reasons as for the $gq\bar q$ matching discussed in the text, as do insertions of the operator of \Eq{eq:ggpp_matched}, so that the result must be reproduced by hard scattering operators. Expanding the QCD result, we find
\begin{align}
\left. \fd{2.25cm}{figures/matching_lam2_qqg_low} \right|_{\cO(\lambda^2)}&= 0\,.
\end{align}

To extract the operators where the $\cP_\perp$ acts in the $n$-collinear sector we simplify the matching by taking
\begin{align}
p_1^\mu=\omega_1 \frac{n^\mu}{2}+p_{1\perp}^\mu +p_{1r} \frac{\bar n^\mu}{2}\,, \qquad  p_2^\mu=\omega_2 \frac{n^\mu}{2}+p_{2\perp}^\mu +p_{2r} \frac{\bar n^\mu}{2}\,, \qquad p_3^\mu=\omega_3 \frac{\bar n^\mu}{2}\,,
\end{align}
where, unlike in the text, we have allowed for a generic $\perp$ momentum in the $n$-collinear sector. Note that for this configuration it is still true that subleading $T$-products vanish,f or similar reasons as for the $gq\bar q$ matching discussed in the text, at least at this order. Only the operator of \Eq{eq:ggpp_matched} appeared in the matching, however its contribution vanishes for this matching configuration. Expanding the full theory result we find
\begin{align}
\left. \fd{2.25cm}{figures/matching_lam2_qqg_low} \right|_{\cO(\lambda^2)}&=0 \,,
\end{align}
just as was the case when the $\perp$ momenta in each sector were restricted to vanish. 

Finally, we must consider the matching with general $\perp$ momenta to the three gluon operators. Again, we can perform the matching in two steps. In the first step we take the momenta as
\begin{align}
\hspace{-0.3cm} p_1^\mu=\omega_1 \frac{n^\mu}{2}+p_{1\perp}^\mu +p_{1r} \frac{\bar n^\mu}{2}\,, \quad  p_2^\mu=\omega_2 \frac{n^\mu}{2}-p_\perp^\mu +p_{2r} \frac{\bar n^\mu}{2}\,, \quad p_3^\mu=\omega_3 \frac{\bar n^\mu}{2}+p_{\perp}^\mu+p_{3r} \frac{n^\mu}{2}\,.
\end{align}
to isolate the action of the operator with an insertion of the $\cP_\perp$ operator in the $n$-collinear sector. The QCD amplitudes expanded to this order are 
\begin{align}
&\left. \left(\fd{2.25cm}{figures/matching_lam2_ggg4_low} + \fd{2.25cm}{figures/matching_lam2_ggg1_low}\right) \right |_{\cO(\lambda^2)}\hspace{-0.4cm}= -4 g f^{abc}\left[\frac{\omega_3}{\omega_2} (\epsilon _1 \cdot \epsilon _3) (p_{1\perp}\cdot \epsilon_2) -\frac{\omega _2}{\omega _3}(\epsilon _1\cdot\epsilon _2) (p_{1\perp}\cdot\epsilon _3) \right] \,, \nn \\
&\left. \fd{2.25cm}{figures/matching_lam2_ggg2_low}\right|_{\cO(\lambda^2)} =0\,, \nn \\
&\left.\fd{2.25cm}{figures/matching_lam2_ggg3_low}\right|_{\cO(\lambda^2)} = 4 g f^{abc} \left[(\epsilon_1 \cdot \epsilon _2) (p_{1\perp}\cdot\epsilon_3)-(\epsilon _1 \cdot \epsilon _3) (p_{1\perp}\cdot \epsilon_2) \right] 
\,. 
\end{align}
There are no SCET contributions at this order, since for our choice of kinematics there is no perpendicular momentum flowing in the $\bar n$ leg.
Therefore, the hard scattering operators which appear in the tree level matching are
\begin{align}
	\cO_{\cP\cB}^{(2)}&=4 g if^{abc}\left(1+\frac{\omega_2}{\omega _3}\right) \mathcal{B}^b_{\bar n \perp,\omega_2}\cdot[\mathcal{B}^a_{n \perp,\omega_1} \mathcal{P}^\dagger_\perp]\cdot \mathcal{B}^c_{\bar n \perp,\omega_3}\,.
\end{align}

In the second step of the matching we can take the kinematics as
\begin{align}
p_1^\mu=\omega_1 \frac{n^\mu}{2}\,, \qquad  p_2^\mu=\omega_2 \frac{n^\mu}{2}+p_{2\perp}^\mu +p_{2r} \frac{\bar n^\mu}{2}\,, \qquad p_3^\mu=\omega_3 \frac{\bar n^\mu}{2}+p_{3\perp}^\mu+p_{3r} \frac{n^\mu}{2}\,,
\end{align}
which allows us to determine the Wilson coefficients of the operators with a $\cP_\perp$ acting in the $\bar n$-collinear sector.
Expanding the relevant QCD diagrams to $\cO(\lambda^2)$, we find
\begin{align}
&\left. \left(\fd{2.25cm}{figures/matching_lam2_ggg4_low} + \fd{2.25cm}{figures/matching_lam2_ggg1_low}\right) \right |_{\cO(\lambda^2)}\hspace{-0.8cm}  \\
&= 4 g f^{abc}  \left[\frac{\omega_3}{\omega_2} \left[(\epsilon _2 \cdot \epsilon_3) (p_{2,\perp} \cdot\epsilon_1)-(\epsilon _1 \cdot \epsilon _2) (p_{2,\perp} \cdot\epsilon _3)\right]+(\epsilon _1\cdot\epsilon _2) (p_{3,\perp}\cdot\epsilon _3) - (2 \leftrightarrow 3)\right]\,, \nn \\
&\left. \fd{2.25cm}{figures/matching_lam2_ggg2_low}\right|_{\cO(\lambda^2)} =0\,, \nn \\
&\left.\fd{2.25cm}{figures/matching_lam2_ggg3_low}\right|_{\cO(\lambda^2)} = 4 g f^{abc}\left[-\frac{\omega _3}{\omega_2} (\epsilon _1\cdot\epsilon _3) (p_{2,\perp} \cdot\epsilon_2) - (\epsilon _1\cdot\epsilon _2) \left(p_{2,\perp} \cdot\epsilon _3\right)+(\epsilon _2\cdot\epsilon _3) (p_{2,\perp} \cdot\epsilon _1) - (2\leftrightarrow 3) \right]\,. \nn 
\end{align}
There are no SCET $T$-product contributions, so that these must be exactly reproduced by hard scattering operators in the effective theory. We therefore find the following operators
\begin{align}
\cO^{(2)}_{\cP \cB1}&=-4g \left(  1+\frac{\omega_3}{\omega_2} \right)if^{abc} \cB^a_{n\perp,\omega_1}\cdot \left[  \cP_\perp \cB^b_{\bar n \perp,\omega_2}\cdot  \right] \cB_{\bar n \perp,\omega_3}^c    H\,, \nn \\
\cO^{(2)}_{\cP \cB2}&=4g\left( 2+\frac{\omega_3}{\omega_2}  \right)if^{abc}  \cB^a_{n\perp,\omega_1} \cdot \left[ \cB_{\perp \bar n, \omega_2}^c    \cP^\dagger_\perp \right]  \cdot \cB_{\bar n \perp,\omega_3}^b    H\,, \nn \\
\cO^{(2)}_{\cP \cB3}&=4g\left( 2+\frac{\omega_3}{\omega_2} \right)if^{abc}   \cB^a_{n\perp,\omega_1} \cdot \cB_{\perp \bar n, \omega_3}^c   \left[ \cP_\perp \cdot \cB_{\bar n \perp,\omega_2}^b \right]  H\,.
\end{align}
These can be projected onto definite helicities following \Eq{eq:3g_hel_match}.

\section{Useful Feynman Rules}\label{app:expand_gluon}

In this appendix we summarize for convenience several useful Feynman rules used in the text, both from the Higgs effective theory, and from SCET. 

The Feynman rules in the Higgs effective theory with
\begin{align}
O^{\text{hard}}=G^{\mu \nu}G_{\mu \nu} H\,,
\end{align}
are well known, and are given by
\begin{align}\label{eq:feynrule_eft}
\fd{2.25cm}{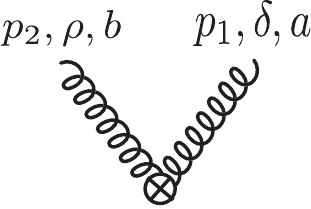} &=-4i\delta^{ab} (p_1 \cdot p_2 g^{\rho \delta}  -p_1^\rho p_2^\delta )\,,  
\end{align}

\noindent\begin{minipage}{.5\linewidth}
\begin{align}
\hspace{3cm}\fd{3cm}{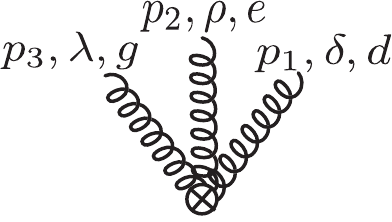} \nn
\end{align}
\end{minipage}%
\hspace{-3cm}
\begin{minipage}{.6\linewidth}
\begin{align}
=&-4g f^{deg}(p_1^\rho g^{\delta \lambda} -p_1^\lambda g^{\rho \delta}) \nn \\
& -4g  f^{ged}(p_3^\rho g^{\delta \lambda} -p_3^\delta g^{\lambda \rho}) \nn \\
&-4g  f^{egd}(p_2^\lambda g^{\rho\delta} -p_2^\delta g^{\lambda \rho})\,, 
\end{align}
\end{minipage}

\noindent\begin{minipage}{.1\linewidth}
\begin{align}
\hspace{3.5cm}\fd{3.8cm}{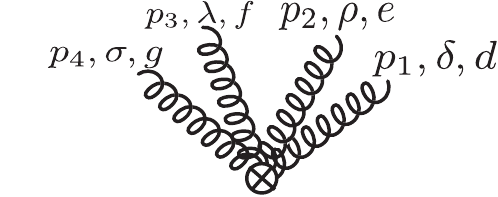} \nn
\end{align}
\end{minipage}%
\hspace{-2cm}
\begin{minipage}{.7\linewidth}
\begin{align}
=&4ig^2 ( f^{adf}f^{aeg}+f^{aef}f^{adg} )g^{\delta \rho} g^{\lambda \sigma} \nn \\
&  +4ig^2 (f^{ade}f^{afg}+f^{adg}f^{afe}) g^{\delta \lambda} g^{\rho \sigma} \nn \\
&+4ig^2 (f^{ade} f^{agf}+ f^{adf} f^{age})  g^{\delta \sigma} g^{\rho \lambda}\,. 
\end{align}
\end{minipage}\\

Before presenting the subleading power Feynman rules in SCET, we begin by briefly reviewing the Lagrangian, and gauge fixing for the collinear gluons. The gauge covariant derivatives that we will use to write the Lagrangian are defined by
\begin{align}
iD^\mu_n &= i\partial^\mu_n +g A^\mu_n\,, \qquad
 i \partial^\mu_n = \frac{\bn^\mu}{2} n \cdot \partial + \frac{n^\mu}{2} \overline{\cP} + \cP_\perp^\mu\,, \nn \\
 iD^\mu_{ns} &=i D^\mu_n +\frac{\bn^\mu}{2}gn \cdot A_{us}\,,\qquad
i\partial^\mu_{ns}=i \partial^\mu_n +\frac{\bn^\mu}{2} gn\cdot A_{us}\,,
\end{align}
and
\begin{align}
iD_{us}^\mu=i\partial^\mu+gA^\mu_{us}\,,
\end{align}
and their gauge invariant versions are given by
\begin{align}
i\cD^\mu_{n}=W_n^\dagger iD^\mu_{n} W_n\,, \nn \\
i\cD^\mu_{n\perp}=W_n^\dagger iD^\mu_{n\perp} W_n= \cP^\mu_{n\perp}+gB^\mu_{n\perp}\,, \nn \\
i \cD^\mu_{ns}=W_n^\dagger iD^\mu_{ns} W_n\,.
\end{align}
The leading power SCET Lagrangian can be written as
\begin{align} \label{eq:leadingLag_2}
\cL^{(0)} &= \cL^{(0)}_{n \xi} + \cL^{(0)}_{n g} +  \cL^{(0)}_{us}\,, 
\end{align}
where~\cite{Bauer:2001yt}
\begin{align}
\cL^{(0)}_{n \xi} &= \bar{\xi}_n\big(i n \cdot D_{ns} + i \slashed{D}_{n \perp} W_n \frac{1}{\overline{\cP}_n} W_n^\dagger i \slashed{D}_{n \perp} \big)  \frac{\slashed{\bar{n}}}{2} \xi_n\,, \\
\cL^{(0)}_{n g} &= \frac{1}{2 g^2} \tr \big\{ ([i D^\mu_{ns}, i D^\nu_{ns}])^2\big\} + \frac{1}{\alpha} \tr \big\{ ([i \partial^\mu_{ns},A_{n \mu}])^2\big\}+2 \tr \big\{\bar{c}_n [i \partial_\mu^{ns}, [i D^\mu_{ns},c_n]]\big\} \,, \nn
\end{align}
and the ultrasoft Lagrangian, $\cL^{(0)}_{us}$, is simply the QCD Lagrangian. We have used a covariant gauge with gauge fixing parameter $\alpha$ for the collinear gluons.

The $\mathcal{O}(\lambda)$ Lagrangian can be written 
\begin{align}
\cL^{(1)}={\cal L}_{\chi_n}^{(1)}+{\cal L}_{A_n}^{(1)}+{\cal L}_{\chi_n q_{us}}^{(1)} \,,
\end{align}
where \cite{Chay:2002vy,Pirjol:2002km,Manohar:2002fd,Bauer:2003mga}
\begin{align}\label{eq:sublagcollq_2}
{\cal L}_{\chi_n}^{(1)} &= \bar \chi_n \Big(
i \slashed{D}_{us\perp}\frac{1}{ \bar \cP} 
i \slashed{\cal D}_{n\perp}
+i \slashed{\cal D}_{n\perp}\, \frac{1}{\bar \cP} 
i \slashed{D}_{us\perp} 
\Big)\frac{\slashed{\bar{n}}}{2} \chi_n \ ,
 \\ 
{\cal L}_{A_n}^{(1)}&= \frac{2}{g^2}\text{Tr}\Big(
\big[ i {\cal D}_{ns}^\mu,i {\cal D}_{n\perp}^\nu \big]\big[
i {\cal D}_{ns\mu},iD^\perp_{us\,\nu} 
\big]
\Big)  
+ 2 \frac{1}{\alpha} \text{Tr} \left(   [i D_{us\perp}^\mu, A_{n\perp \mu}] [i\partial^\nu_{ns},A_{n\nu}] \right) \nn \\
&+2 \text{Tr}   \left( \bar c_n [iD_{us\perp}^\mu, [iD^\perp_{n\mu}, c_n  ]]  \right)          +2 \text{Tr}   \left( \bar c_n [\cP_\perp^\mu,[ W_n iD^\perp_{us\,\mu} W_n^\dagger, c_n   ]]  \right)    \,, \nn \\
{\cal L}_{\chi_n q_{us}}^{(1)} 
&= \bar{\chi}_n  g \slashed{\cB}_{n\perp} q_{us}+\text{h.c..}\nn
\end{align}

Finally, the $\cO(\lambda^2)$ Lagrangian can be written as \cite{Pirjol:2002km,Manohar:2002fd,Bauer:2003mga}
\begin{align}
	{\cal L}^{(2)} &= {\cal L}_{\chi_n}^{(2)}  + {\cal L}_{A_n}^{(2)} 
	+ {\cal L}_{\chi_n q_{us}}^{(2)} \,,
\end{align}
where
\begin{align}
	\cL_{\xi_n q_{us}}^{(2)}&=  \bar \chi_n \frac{\Sl \bn}{2} [ W_n^\dagger in\cdot D W_n]  q_{us}    + \bar \chi_n \frac{\Sl \bn}{2}  i\Sl \cD_{n\perp}    \frac{1}{\overline{\cP}}   ig \Sl \cB_{n\perp}  q_{us}+ \text{h.c.} \,, \\
	\cL_{n\xi}^{(2)}&=  \bar \chi_n   \left( i\Sl D_{us\perp} \frac{1}{\overline{\cP}}  i\Sl D_{us\perp}-  i\Sl \cD_{n\perp}   \frac{i\bn \cdot D_{us}}{(\overline{\cP})^2}   i\Sl \cD_{n\perp}   \right)    \frac{\Sl \bn}{2} \chi_n\,, \nn\\
	\cL_{ng}^{(2)}&=\frac{1}{g^2} \text{Tr} \left(  [  i\cD^\mu_{ns} , iD_{us}^{\perp \nu}   ]    [ i \cD_{ns\mu}   ,i D^\perp_{us\nu}   ] \right)   +\frac{1}{g^2} \text{Tr} \left(  [ iD^\mu_{us\perp}  ,i D^\nu_{us\perp}   ]    [ i\cD^\perp_{n\mu}  , i\cD^\perp_{n\nu}   ] \right) \nn \\
	&+ \frac{1}{g^2} \text{Tr} \left(  [ i\cD_{ns\mu}  , i n\cdot \cD_{ns}  ]    [  i\cD_{ns\mu} , i\bn \cdot D_{us}   ] \right)+\frac{1}{g^2} \text{Tr} \left(  [ iD^\mu_{us\perp}  ,  i\cD^\perp_{n\nu}  ]    [  i\cD^\perp_{n\mu}  ,  iD_{us \perp}^\nu  ] \right)\,, \nn\\
	\cL_{gf}^{(2)}&= \frac{1}{\alpha} \text{Tr} \left(   [ i D^\mu_{us\perp}, A_{n\perp \mu}]   [i D^\nu_{us\perp}, A_{n\perp \nu}] \right)   + \frac{1}{\alpha} \text{Tr} \left(   [i \bn \cdot D_{us},n \cdot A_n]   [i \partial^\mu_{ns}, A_{n\mu}] \right) \nn \\
	&+ 2 \text{Tr} \left(  \bar c_n [iD^\mu_{us\perp},[ W_n iD^\perp_{us\mu}W_n^\dagger ,c_n]] \right)+ \text{Tr} \left( \bar c_n [i \bn \cdot D_{us} ,[ i n\cdot D_{ns},c_n]] \right) \nn \\
	&+ \text{Tr} \left( \bar c_n [\overline{\cP},[W_n i \bn \cdot D_{us} W_n^\dagger,c_n]] \right)\,. \nn
\end{align}

Using these Lagrangians, one can derive the required Feynman rules for the calculations described in the text. The $\mathcal{O}(\lambda)$ Feynman rule for the emission of a ultrasoft gluon from a collinear gluon in a general covariant gauge, specified by a gauge fixing parameter $\alpha$, is given by
\begin{align}
\fd{2.5cm}{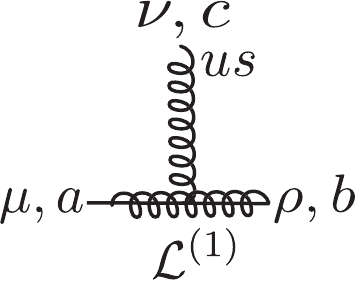}=&-g f^{abc}\left[   g_{\perp}^{\nu\rho}\left(  \left(  1-\frac{1}{\alpha} \right)p^\mu_n-   \left(  1+\frac{1}{\alpha} \right)n\cdot p_s \frac{\bar n^\mu}{2}  -\frac{p_n^2 \bar n^\mu}{\bar n \cdot p_n}  \right)  \right. \nn \\
&-2g^{\mu \nu}p^\rho_{n\perp} +g^{\mu \rho}_\perp\left(   \left(  1-\frac{1}{\alpha} \right)p_n^\nu    -\frac{p_n^2 \bar n^\nu}{\bar n \cdot p_n}  \right) \nn \\
&\left.+\left( \bar n^\mu p^\nu_n+ \bar n^\nu p^\mu_n +\frac{1}{2}\bar n^\mu \bar n^\nu n\cdot p_s    \right)   \frac{p^\rho_{n\perp}}{\bar n \cdot p_n}\right]\,,
\end{align}
and the $\mathcal{O}(\lambda)$ propagator correction to the gluon propagator is given by
\begin{align}
\fd{2.0cm}{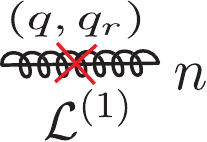}&=-4i \delta^{ab}g^{\mu \nu} q_\perp \cdot q_{r\perp}+2i(1-\frac{1}{\alpha}) \delta^{ab}\left[  q_{r\perp}^\mu q^\nu +q^{\mu}q_{r\perp}^\nu  \right]\,.
\end{align}

For the matching calculation for the operators involving an ultrasoft derivative in \Sec{sec:us_deriv}, we also needed the $\mathcal{O}(\lambda^2)$ corrections to the propagator, which is given by
\begin{align}\label{eq:lam2_gluon_prop}
\fd{2.0cm}{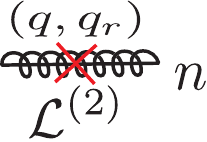}&=-i \delta^{ab} q_r^\perp \cdot q_r^\perp g^{\mu \nu}_\perp +i \delta^{ab} \left(  1-\frac{1}{\alpha}  \right) q_{r\perp}^\mu q_{r\perp}^\nu \nn \\
&+\frac{i}{2}\delta^{ab}\left(  1-\frac{1}{\alpha}  \right)(q_\perp^\mu n^\nu \bar n \cdot q_r +q_\perp^\nu n^\mu \bar n \cdot q_r ) +\cdots \,, 
\end{align}
where the dots indicate the other tensor components in the light cone basis, which are not relevant for the current discussion.
For simplicity, the matching was performed using a $\perp$ polarized gluon. In the $n$-collinear sector, the leading power hard scattering operator produces only $\bar n$, and $\perp$ polarized gluons. Therefore, only the $\perp-\perp$ and $n-\perp$ components of the propagator are needed.
In the matching, the $\perp-\perp$ term vanishes since it proportional to the residual $\perp$ momentum, which is set to zero, and the $n-\perp$ term vanishes for a $\perp$ polarized gluon, due to the gluons equation of motion, $q_\perp \cdot \epsilon_\perp=0$.

At $\mathcal{O}(\lambda^2)$, the individual propagator and emission factors are sufficiently complicated that it is also convenient to give the complete result for the matrix element
\begin{align} 
\fd{3cm}{figures/soft_lam2_scet_low}&=\left .\langle 0| T \{\cB^\nu_{n\perp}(0), \cL^{(2)}_{A_n}\} | \epsilon_n, p_n; \epsilon_s,p_s \rangle \right |_{\alpha=1}=\nn\\
&=-i f^{abc} \epsilon_{n\mu} \frac{2\epsilon_{s\rho} p_{s\sigma}}{\nbar \cdot p_n\, n\cdot p_s} \left(  g^{\mu \rho}_\perp g^{\sigma \nu}_\perp -    g^{\mu \sigma}_\perp g^{\rho \nu}_\perp  \right)\,,
\end{align}
where we have restricted to $\alpha=1$ for simplicity.

Since we have also matched to operators involving collinear quarks, we also summarize the subleading power  Feynman rules involving collinear quark. The Feynman rules for the correction to a collinear quark propagator are given by
\be
\fd{3cm}{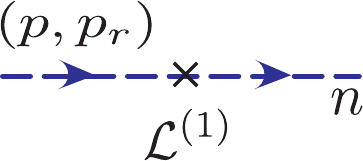}=i \frac{\Sl{\bar n}}{2}\frac{2p_\perp \cdot p_{r\perp}}{\bar n \cdot p}\,,
\ee
\be
\fd{3cm}{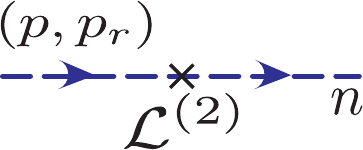}=i \frac{\Sl{\bar n}}{2}\frac{p_{r\perp}^2}{\bar n\cdot p}\,,
\ee
and the Feynman rules for the emission of a collinear gluon are given by
\be
\fd{2cm}{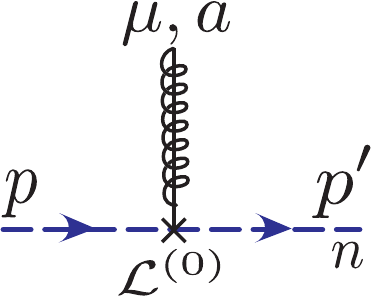}=i g T^a \left( n_\mu +\frac{\gamma^\perp_\mu \Sl{p}_\perp}{\bar n \cdot p}+\frac{\Sl{p}_\perp^{'} \gamma_\mu^\perp}{\bar n \cdot p'}- \frac{\Sl{p}_\perp \Sl{p}_\perp^{'}}{\bar n\cdot p \bar n \cdot p'}\bar n_\mu   \right) \frac{\Sl{\bar n}}{2}\,,
\ee
\begin{align}
&\fd{2cm}{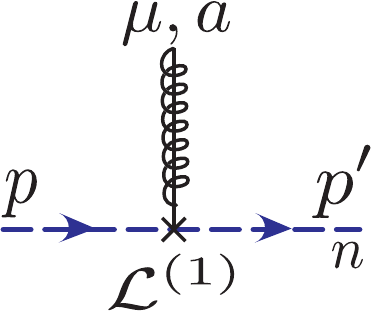}= \\
 &i g T^a \left( \frac{\gamma^\perp_\mu \Sl{p}_{r\perp}}{\bar n \cdot p}+\frac{\Sl{p}_{r\perp}^{'} \gamma_\mu^\perp}{\bar n \cdot p'}+        \frac{\Sl{p}_{r\perp} \Sl{p}_{\perp}}{\bar n\cdot q \bar n \cdot p}\bar n_\mu                    -\frac{\Sl{p}_\perp^{'} \Sl{p}_{r\perp}^{'}}{\bar n\cdot q \bar n \cdot p'}\bar n_\mu-                    \frac{\Sl{p}_{r\perp}^{'} \Sl{p}_\perp}{\bar n\cdot q \bar n \cdot p'}\bar n_\mu+                     \frac{\Sl{p}_\perp^{'} \Sl{p}_{r\perp}}{\bar n\cdot q \bar n \cdot p'}\bar n_\mu \right) \frac{\Sl{\bar n}}{2}\,,\nn
\end{align}
\begin{align}
&\fd{2cm}{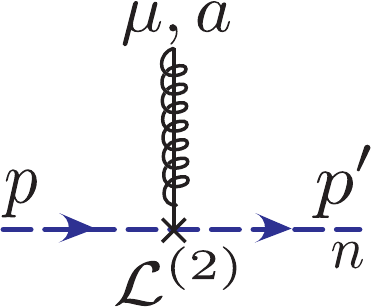}=\\
	&ig T^a \left(\frac{\bar n ^\mu p_{r\perp}^2}{\bar n \cdot p}     -\frac{\bar n ^\mu p_{r\perp}^{'2}}{\bar n \cdot p'}     -\frac{\gamma^\mu_\perp \Sl{p}_\perp  \bar n \cdot p_r }{(\bar n \cdot p)^2}     -\frac{ \Sl{p}^{'}_\perp  \gamma^\mu_\perp \bar n \cdot p_r}{(\bar n \cdot p')^2}     -\frac{\bar n^\mu \Sl{p}^{'}_\perp \Sl{p}_\perp \bar n \cdot p_r}{\bar n \cdot q(\bar n \cdot p)^2}     +\frac{\bar n^\mu \Sl{p}^{'}_\perp \Sl{p}_\perp \bar n \cdot p_r}{\bar n \cdot q(\bar n \cdot p')^2}     \right)    \frac{\Sl{\bar n}}{2}\,.\nn
\end{align}
We can see that each term in the power suppressed collinear Lagrangian insertions are proportional to either $p_{r\perp}$, or $\bar n \cdot p_r$. At tree level, and in the absence of ultrasoft particles, one can use RPI to set all these terms to zero. This was used extensively to simplify our matching calculations.

For convenience we also give the expansion of the Wilson lines and collinear gluon field to two emissions. The collinear Wilson lines are defined by
\begin{align}
W_n =\left[  \sum\limits_{\text{perms}} \exp \left(  -\frac{g}{\bar \cP} \bar n \cdot A_n(x)  \right) \right]\,.
\end{align}
Expanded to two gluons with incoming momentum $k_1$ and $k_2$, we have
\begin{align}
W_n &=1-\frac{gT^a \bar n \cdot A_{nk}^a}{\bar n \cdot k}  +g^2 \left[   \frac{T^a T^b}{\bar n \cdot k_1 (\bar n \cdot k_1+\bar n \cdot k_2)} +\frac{T^b T^a}{\bar n \cdot k_2(\bar n \cdot k_1+\bar n \cdot k_2)}  \right]    \frac{\bar n \cdot A_{nk1}^a   \bar n \cdot A_{nk2}^b  }{2!}\,, \nn \\
W^\dagger_n& =1+\frac{gT^a \bar n \cdot A_{nk}^a}{\bar n \cdot k}  +g^2 \left[   \frac{T^a T^b}{\bar n \cdot k_1 (\bar n \cdot k_1+\bar n \cdot k_2)} +\frac{T^b T^a}{\bar n \cdot k_2(\bar n \cdot k_1+\bar n \cdot k_2)}  \right]    \frac{\bar n \cdot A_{nk1}^a   \bar n \cdot A_{nk2}^b  }{2!}\,.
\end{align}
The collinear gluon field is defined as
\begin{align}
\cB^\mu_{n\perp}=\frac{1}{g}\left[   W_n^\dagger i D^\mu_{n\perp}W_n \right]\,.
\end{align}
Expanded to two gluons, both with incoming momentum, we find
\begin{align}
g\cB^\mu_{n\perp}&=g\left(   A^{\mu a}_{\perp k} T^a -k^\mu_\perp \frac{\bar n \cdot A^a_{nk} T^a}{\bar n \cdot k} \right)+g^2(T^a T^b-T^b T^a) \frac{\bar n \cdot A^a_{nk1} A^{\mu b}_{\perp k2} }{\bar n \cdot k_1}  \\
&+g^2 (k^\mu_{1\perp} +k^\mu_{2\perp}) \left(   \frac{T^a T^b}{\bar n \cdot k_1 (\bar n \cdot k_1+\bar n \cdot k_2)}+\frac{T^b T^a}{\bar n \cdot k_2 (\bar n \cdot k_1+\bar n \cdot k_2)} \right)\frac{\bar n \cdot A_{n k1}^a \bar n \cdot A^b_{n k2}}{2!} \,. \nn
\end{align}
In both cases, at least one of the gluons in the two gluon expansion is not transversely polarized. Such terms can therefore be eliminated in matching calculations by choosing particular polarizations, as was done in the text.

\bibliography{../../overallbib}{}
\bibliographystyle{jhep}

\end{document}